\documentclass[prd,
superscriptaddress,
preprintnumbers,
twocolumn,
 amsmath,amssymb,
 aps,longbibliography,nofootinbib,
floatfix
]{revtex4-2}

\usepackage{graphicx}
\usepackage{dcolumn}
\usepackage{bm}
\usepackage{xcolor}
\usepackage{orcidlink}
\usepackage{bbold}
\usepackage[T1]{fontenc}
\usepackage{tabularray}

\newcommand\Sec[1]{Sect.~\ref{Sec:#1}}
\newcommand\App[1]{Appendix~\ref{Sec:#1}}
\newcommand\Tab[1]{Table~\ref{tab:#1}}
\newcommand\Fig[1]{Fig.~\ref{fig:#1}}
\newcommand\Eq[1]{Eq.~(\ref{eq:#1})}

\newcommand{\beq}{\begin{equation}}
\newcommand{\eeq}{\end{equation}}

\newcommand{\beqs}{\begin{eqnarray}}
\newcommand{\eeqs}{\end{eqnarray}}
\newcommand{\nn}{\nonumber}
\newcommand{\dd}{\mbox{d}}

\newcommand{\orcidauthorPIAI}{0000-0002-2251-0111}
\newcommand{\orcidauthorBENNETT}{0000-0002-1678-6701}
\newcommand{\orcidauthorLUCINI}{0000-0001-8974-8266}
\newcommand{\orcidauthorVADACCHINO}{0000-0002-5783-5602}
\newcommand{\orcidauthorHONG}{0000-0002-3923-4184}
\newcommand{\orcidauthorLIN}{0000-0003-3743-0840}
\newcommand{\orcidauthorLEE}{0000-0002-4616-2422}
\newcommand{\orcidauthorHSIAO}{0000-0002-8522-5190}

\newcommand{\PlaquettePhaseDiagramBetas}{$\beta = 6.4$, $6.45$, $6.5$}

\newcommand{\MHatVFHatPSRatio}{6.04(13)}

\newcommand{\MVFPSExtrapolationRatio}{6.59(11)}
\newcommand{\MVFPSExtrapolationCL}{0.069(15)}
\newcommand{\MVFPSExtrapolationCW}{-0.10(21)}
\newcommand{\MVFPSExtrapolationROverRootTwo}{4.656(80)}

\newcommand{\ChiPTBetaValues}{$\beta = 6.6$, $6.65$, $6.7$, $6.75$, $6.8$, and $6.9$}

\newcommand{\WZeroIncompleteEnsembles}{ASB5M1}

\newcommand{\SpectrumPlotTargetBeta}{6.7}

\newcommand{\FiniteVolumeStudyBeta}{6.8}
\newcommand{\FiniteVolumeStudyMass}{-1.03}
\newcommand{\FiniteVolumeStudyDeltaTraj}{12}

\newcommand{\HeavyPSMassLimit}{0.45}

\newcommand{\DEFTCommonYValue}{2.6}

\begin{document}

\preprint{CTPU-PTC-24-32, UTHEP-794, UTCCS-P-160}

\title{Meson spectroscopy in the $Sp(4)$ gauge theory with three antisymmetric fermions}

\author{Ed Bennett\,\orcidlink{\orcidauthorBENNETT}}%
\email{E.J.Bennett@swansea.ac.uk}
\affiliation{Swansea Academy of Advanced Computing, Swansea University (Bay Campus), Fabian Way, SA1 8EN Swansea, Wales, United Kingdom}

\author{Deog Ki Hong\,\orcidlink{\orcidauthorHONG}}
\email{dkhong@pusan.ac.kr}
\affiliation{Department of Physics, Pusan National University, Busan 46241, Korea}
\affiliation{Extreme Physics Institute, Pusan National University, Busan 46241, Korea}

\author{Ho Hsiao\,\orcidlink{\orcidauthorHSIAO}}
 \email{hohsiao@ccs.tsukuba.ac.jp}
\affiliation{Center for Computational Sciences, University of Tsukuba, 1-1-1 Tennodai, Tsukuba, Ibaraki 305-8577, Japan}
\affiliation{Institute of Physics, National Yang Ming Chiao Tung University, 1001 Ta-Hsueh Road, Hsinchu 30010, Taiwan}

\author{Jong-Wan Lee\,\orcidlink{\orcidauthorLEE}}
 \email{j.w.lee@ibs.re.kr}
\affiliation{%
Particle Theory  and Cosmology Group, Center for Theoretical Physics of the Universe, Institute for Basic Science (IBS), Daejeon, 34126, Korea 
}%

\author{C.-J. David Lin\,\orcidlink{\orcidauthorLIN}}
\email{dlin@nycu.edu.tw}
\affiliation{Institute of Physics, National Yang Ming Chiao Tung University, 1001 Ta-Hsueh Road, Hsinchu 30010, Taiwan}
\affiliation{Centre for High Energy Physics, Chung-Yuan Christian University, Chung-Li 32023, Taiwan}

\author{Biagio Lucini\,\orcidlink{\orcidauthorLUCINI}}
\email{B.Lucini@Swansea.ac.uk}
\affiliation{Swansea Academy of Advanced Computing, Swansea University (Bay Campus), Fabian Way, SA1 8EN Swansea, Wales, United Kingdom}
\affiliation{Department of Mathematics, Faculty of Science and Engineering, Swansea University (Bay Campus), Fabian Way, SA1 8EN Swansea, Wales, United Kingdom}

\author{Maurizio Piai\,\orcidlink{\orcidauthorPIAI}}
\email{m.piai@swansea.ac.uk}
\affiliation{Department of Physics, Faculty  of Science and Engineering, Swansea University, Singleton Park, SA2 8PP, Swansea, Wales, UK}

\author{Davide Vadacchino\,\orcidlink{\orcidauthorVADACCHINO}}
\email{davide.vadacchino@plymouth.ac.uk}
\affiliation{Centre for Mathematical Sciences, University of Plymouth, Plymouth, PL4 8AA, United Kingdom}

\date{\today}

\begin{abstract}
We report the results of an extensive numerical study of 
the $Sp(4)$ lattice gauge theory coupled to fermion matter content consisting of three (Dirac) flavors,
transforming  in the two-index antisymmetric representation of the gauge group.
In the presence of (degenerate) fermion masses, the theory has an enhanced global $SU(6)$ symmetry,
broken explicitly and spontaneously to its $SO(6)$ subgroup. 
This symmetry breaking pattern  makes the theory interesting for applications in the context of composite Higgs models, as well as for the
implementation of top partial compositeness.  Alternatively, it can also provide a dynamical realisation of the strongly interacting massive particle 
paradigm for the origin of dark matter. 
We adopt the standard plaquette gauge action, along with the Wilson-Dirac formulation for the fermions, 
and apply the (rational) hybrid Monte Carlo algorithm in our ensemble generation process. 
We monitor the autocorrelation and topology of the ensembles.
We explore the bare parameter space, and identify the weak and strong coupling regimes, which are
separated by a line of first-order bulk phase transitions.

We  measure two-point correlation functions between meson operators
that transform as non-trivial representations of $SO(6)$, and extract the ground-state
masses,  in all accessible spin and parity channels.
We assess the size of finite volume effects, and restrict attention to measurements in which
these systematic effects are negligibly small compared to the statistical uncertainties.  
 The accuracy of our data enables us to extract the decay constants of the composite particles in the pseudoscalar, vector and axial-vector channels. 
 In addition, we measure the mass of the
 first excited state for one of the channels, the vector meson,
  by performing a generalised eigenvalue problem analysis involving two different meson operators.
Spectral quantities show a mass dependence that is compatible with the expectation that,
at long distances, the theory undergoes confinement, accompanied by the spontaneous breaking of the approximate global symmetries acting on the matter fields.
Finally, we discuss the continuum and massless extrapolations within the framework of Wilson chiral perturbation theory, after setting the physical scale using the gradient flow method, 
and compare the results to those of existing studies in the quenched approximation, as well as to the
literature on closely related theories. 

\end{abstract}

\maketitle

\tableofcontents

\section{\label{Sec:intro}Introduction}

New  physics systems made of composite particles,  arising from novel strong coupling dynamics, yield promising scenarios 
within which to address the big open questions in contemporary particle physics and astrophysics. 
They form the basis of
 Composite Higgs Models (CHMs)~\cite{Kaplan:1983fs,
Georgi:1984af,Dugan:1984hq} (see also the reviews in Refs.~\cite{Panico:2015jxa,Witzel:2019jbe,
Cacciapaglia:2020kgq}, the tables in Refs.~\cite{Ferretti:2013kya,Ferretti:2016upr,Cacciapaglia:2019bqz}, a selection of publications in Refs.~\cite{Katz:2005au,Barbieri:2007bh,
Lodone:2008yy,Gripaios:2009pe,Mrazek:2011iu,Marzocca:2012zn,Grojean:2013qca,Cacciapaglia:2014uja,
Ferretti:2014qta,Arbey:2015exa,Cacciapaglia:2015eqa,Feruglio:2016zvt,DeGrand:2016pgq,Fichet:2016xvs,
Galloway:2016fuo,Agugliaro:2016clv,Belyaev:2016ftv,Csaki:2017cep,Chala:2017sjk,Golterman:2017vdj,
Csaki:2017jby,Alanne:2017rrs,Alanne:2017ymh,Sannino:2017utc,Alanne:2018wtp,Bizot:2018tds,
Cai:2018tet,Agugliaro:2018vsu,Cacciapaglia:2018avr,Gertov:2019yqo,Ayyar:2019exp,
Cacciapaglia:2019ixa,BuarqueFranzosi:2019eee,Cacciapaglia:2019dsq,Cacciapaglia:2020vyf,
Dong:2020eqy,Cacciapaglia:2021uqh,Banerjee:2022izw,Ferretti:2022mpy, Cai:2022zqu, Cacciapaglia:2024wdn,Banerjee:2024zvg} 
and the holographic models in
Refs.~\cite{Contino:2003ve,Agashe:2004rs,Agashe:2005dk,Agashe:2006at,
Contino:2006qr,Falkowski:2008fz,Contino:2010rs,Contino:2011np,Erdmenger:2020lvq,Erdmenger:2020flu,Elander:2021kxk,Elander:2023aow,Erdmenger:2023hkl,Elander:2024lir}), 
 models of Top Partial Compositeness (TPC)~\cite{Kaplan:1991dc} (see also Refs.~\cite{Grossman:1999ra,Gherghetta:2000qt,Chacko:2012sy}, for useful discussions),
 models of dark sectors~\cite{Strassler:2006im,Cheung:2007ut,Hambye:2008bq,Feng:2009mn,
Cohen:2010kn,Foot:2014uba,
Bertone:2016nfn}, composite dark 
matter~\cite{DelNobile:2011je,
Hietanen:2013fya,Cline:2016nab,Cacciapaglia:2020kgq,Dondi:2019olm,
Ge:2019voa,Beylin:2019gtw,Yamanaka:2019aeq,Yamanaka:2019yek,Cai:2020njb,
Pomper:2024otb} (see also Ref.~\cite{Appelquist:2024koa}),
and strongly interacting massive particles (SIMPs)~\cite{Hochberg:2014dra,Hochberg:2014kqa,Hochberg:2015vrg,Bernal:2017mqb,Berlin:2018tvf,
Bernal:2019uqr,Tsai:2020vpi,Kondo:2022lgg,Bernal:2015xba,Chu:2024rrv} (see also the review~\cite{Cirelli:2024ssz} and references therein). Their presence might even affect the thermal history of the early universe, giving rise to inflation~\cite{Arkani-Hamed:2003wrq,Cacciapaglia:2023kat,Liu:2024xrh}. 
 A stochastic relic density of gravitational waves might arise if 
 such systems underwent a  first-order phase transition 
 during the times our universe was hot~\cite{Witten:1984rs,Kamionkowski:1993fg,Allen:1996vm,Schwaller:2015tja, 
Croon:2018erz,Christensen:2018iqi}. Such effects are
detectable in present and future experiments~\cite{Seto:2001qf,
 Kawamura:2006up,Crowder:2005nr,Corbin:2005ny,Harry:2006fi,
 Hild:2010id,Yagi:2011wg,Sathyaprakash:2012jk,Thrane:2013oya,
 Caprini:2015zlo,
 LISA:2017pwj,
 LIGOScientific:2016wof,Isoyama:2018rjb,Baker:2019nia,
 Brdar:2018num,Reitze:2019iox,Caprini:2019egz,Maggiore:2019uih} (see the discussions in  Refs.~\cite{Huang:2020crf,Halverson:2020xpg,Kang:2021epo,Reichert:2021cvs,
 Reichert:2022naa,
 Pasechnik:2023hwv,Banerjee:2024puz,Bruno:2024dha} and~\cite{
 Mason:2022trc,
 Mason:2022aka,
 Springer:2021liy,
 Springer:2022qos,
 Springer:2023wok,
 Lucini:2023irm,
 Mason:2023ixv,
 Springer:2023hcc,Bennett:2024bhy}).

Motivated by compositeness scenarios,
an extensive programme of Theoretical Explorations on the Lattice 
with Orthogonal and Symplectic groups (TELOS), in particular in the case of  $Sp(2N)$ groups,
has led to significant recent advancements in their understanding~\cite{Bennett:2017kga,
Bennett:2019jzz,Bennett:2019cxd,Bennett:2020hqd,Bennett:2020qtj,
Bennett:2022yfa,Bennett:2022gdz,Bennett:2022ftz,
Bennett:2023wjw,Bennett:2023gbe,Bennett:2023mhh,
Bennett:2023qwx,Bennett:2024cqv,Bennett:2024wda}---see also 
Refs.~\cite{Lee:2018ztv,Lucini:2021xke,Bennett:2021mbw,Hsiao:2022gju,Hsiao:2022kxf,Bennett:2024orw,Zierler:2024ivh,Hsiao:2024tjf} as well as 
Refs.~\cite{Maas:2021gbf,Zierler:2021cfa,
Kulkarni:2022bvh,Zierler:2022qfq,Zierler:2022uez,Bennett:2023rsl,Dengler:2023szi,Dengler:2024maq} and the  pioneering work in Ref.~\cite{Holland:2003kg}. 
Field theories with symplectic gauge group, $Sp(N_c=2N)$, 
lead to enhanced global symmetry patterns, distinctive from those emerging in $SU(N_c)$ gauge theories.
Nevertheless, in the large-$N_c$ limit, $Sp(N_c)$ and $SU(N_c)$ gauge theories are expected to share the same physics in a common sector. Interesting ideas have been 
 put forward to describe the associated non-perturbative phenomena~\cite{Bochicchio:2016toi,Bochicchio:2013sra,
Hong:2017suj,Bennett:2020hqd,Bennett:2022gdz}, hence providing more
general reasons to gain understanding of non-perturbative properties of $Sp(2N)$ theories.
Finally, dedicated studies of these theories have also appeared
within bottom-up holographic models ~\cite{Elander:2020nyd,Elander:2021bmt,
Erdmenger:2024dxf}---an example of top-down holographic models is found in 
 Ref.~\cite{Imoto:2009bf}.

 In the CHM context,
the minimal model amenable to lattice studies
exploits the global symmetry patterns described by the $SU(4)/ Sp(4)$ coset. 
It can be realised by the $Sp(2N)$ gauge theory 
coupled to $N_{\rm f}=2$ matter fermions transforming in the fundamental representation, (f) \cite{Peskin:1980gc}.
For $N>1$, the addition of $N_{\rm as}=3$ fermions transforming in the
two-index antisymmetric representation, (as), leads to
 the minimal realisation of a model combining
CHM and TPC~\cite{Barnard:2013zea}.\footnote{The $SU(4)/ Sp(4)$ coset
appears also
in $SU(2)$ lattice theories~\cite{Hietanen:2014xca,Detmold:2014kba,
Arthur:2016dir,Arthur:2016ozw,Pica:2016zst,Lee:2017uvl,Drach:2017btk,Drach:2020wux,Drach:2021uhl, 
Drach:2021uhl,Bowes:2023ihh}. In this case, though,  the antisymmetric representation is trivial, and TPC cannot be realised in the same way.} 
The
enhanced $SU(6)$ global symmetry acting on the $N_{\rm as}=3$ fermions is broken to $SO(6)$, 
both by the presence of degenerate fermion masses, and by the formation of a fermion condensate \cite{Peskin:1980gc}. The $SU(3)$ group
associated with Quantum Chromodynamics (QCD) is identified with a subgroup of the
unbroken $SO(6)$, and fermion bound states involving one (as)-type and two (f)-type fermions
have the right quantum numbers to act as partners of the top quark.

In this paper, we study a closely related $Sp(4)$ lattice theory,
 coupled to $N_{\rm as}=3$  (as)-type dynamical Dirac fermions, 
but  in which  $N_{\rm f}=0$.
Because of the large multiplicity of the antisymmetric representation, 
this single-representation lattice theory  is expected to approximate well the two species theory, 
  without the significant complications due to the mixed matter field content.
  We refer to the
 handful of existing lattice studies in theories with mixed representations: see for instance Refs.~\cite{Ayyar:2017qdf,Ayyar:2018zuk,Ayyar:2018ppa,
 Ayyar:2018glg,Cossu:2019hse,Lupo:2021nzv,Hasenfratz:2023sqa} for the $SU(4)$ theory
 with fundamental and two-index antisymmetric fermions,  Refs.~\cite{Bennett:2022yfa,Bennett:2024wda,Bennett:2024cqv} for the $Sp(4)$ theory with $N_{\rm as}=3$ and $N_{\rm f}=2$, and Refs.~\cite{Bergner:2020mwl, Bergner:2021ivi} for the $SU(2)$ gauge theories with fundamental and adjoint matter.

This lattice theory is of interest in itself,
as the strongly coupled  origin of effective field theories (EFTs) 
based on the $SU(6)/SO(6)$ coset. Such EFTs can be used to
 provide an alternative  CHM, which includes a composite dark matter candidate~\cite{Cacciapaglia:2019ixa}.
The SIMP mechanism could also be realised in this theory, by generalising the $SU(4)/SO(4)$ analysis in Ref.~\cite{Pomper:2024otb}.

A further reason why this lattice theory is interesting {\it per se} pertains the mapping of the phase space of gauge theories at zero temperature.
Given the large multiplicity of its fermion matter fields, it is important to check with an {\it ab initio} calculation how far   the $Sp(4)$ theory with $N_{\rm as}=3$ sits from the lower edge of the conformal window~\cite{Appelquist:1988yc,Cohen:1988sq,
 Sannino:2004qp,Dietrich:2006cm,Ryttov:2007cx,
Pica:2010mt,Pica:2010xq,Kim:2020yvr,Lee:2020ihn}---see also the higher-loop analyses in Refs.~\cite{Ryttov:2016ner,Ryttov:2016hdp,
Ryttov:2016asb,Ryttov:2016hal,Ryttov:2017toz,
Ryttov:2017kmx,Ryttov:2017dhd,Gracey:2018oym,Ryttov:2018uue,Ryttov:2020scx,Ryttov:2023uzc},
which generalise the Banks-Zaks (BZ) fixed point~\cite{Caswell:1974gg,Banks:1981nn} to higher-loop orders using the results in Refs.~\cite{Chetyrkin:1997dh,Vermaseren:1997fq,Baikov:2016tgj,Herzog:2017ohr}.
Non-perturbative hints of near conformal dynamics  might
manifest themselves in the slow running of the coupling (walking~\cite{Holdom:1984sk,Yamawaki:1985zg,Appelquist:1986an}), 
and unconventional scaling of composite operators~\cite{Cohen:1988sq,Kaplan:2009kr}. We are going to test these possibilities, and provide evidence  that this theory confines and breakes its global, approximate, continuous symmetries, in a way not dissimilar from QCD, the theory of strong nuclear forces. The  literature on lattice studies of 
candidate theories with unconventional, near conformal, strongly coupled  dynamics---see, for instance, Refs.~\cite{Catterall:2007yx,DelDebbio:2008zf,Hietanen:2008mr,Appelquist:2009ty,Hietanen:2009az,
DelDebbio:2009fd,LSD:2009yru,Bursa:2009we,Fodor:2009ff,DelDebbio:2010hx,DeGrand:2010na,
Patella:2010dj,Hayakawa:2010yn,Fodor:2011tu,Bursa:2011ru,Appelquist:2011dp,DeGrand:2011cu,
Karavirta:2011zg,DeGrand:2012qa,Appelquist:2012nz,Lin:2012iw,Aoki:2012eq,Cheng:2013eu,
DeGrand:2013uha,Hasenfratz:2013eka,LSD:2014nmn,Lombardo:2014pda,Hasenfratz:2014rna,Athenodorou:2014eua,
Fodor:2015baa,
Fodor:2015zna,Rantaharju:2015yva,Rantaharju:2015cne,Fodor:2016zil,
Athenodorou:2016ndx,Hasenfratz:2016dou,
Leino:2017lpc,Leino:2017hgm,Amato:2018nvj,Fodor:2018tdg,Hasenfratz:2019dpr,
Hasenfratz:2020ess,LatticeStrongDynamics:2020uwo,Lopez:2020van,
Athenodorou:2021wom,Bennett:2021ivn,Bergner:2022hoo,Bennett:2022bhc,Bergner:2022trm,Hasenfratz:2023wbr,Athenodorou:2024rba}---is reviewed  in Ref.~\cite{Rummukainen:2022ekh}.

For this work, we adopt the standard  (unimproved) Wilson gauge action and Wilson-Dirac fermions. Their
dynamical  implementation is achieved 
 through the rational hybrid Monte-Carlo (RHMC) algorithm~\cite{Duane:1987de,Clark:2006fx}. 
Preliminary results have been presented in Ref.~\cite{Hsiao:2022gju}.
We use the Wilson flow~\cite{Luscher:2010iy,Luscher:2011bx,Luscher:2013vga} to set the scale,
 and also as a smoothening procedure in computing the topology of the ensembles,
 with which we monitor their autocorrelations.
To measure the meson mass spectra, we complement the use of stochastic wall sources~\cite{Boyle:2008rh} 
by the implementation of  Wuppertal smearing of sources and 
sinks~\cite{Gusken:1989qx,Roberts:2012tp,Alexandrou:1990dq}, supplemented by APE smearing of the link variables~\cite{APE:1987ehd,Falcioni:1984ei}.  The spectrum of the vector meson sector is
obtained from a generalised eigenvalue problem (GEVP)---see also Ref.~\cite{Bennett:2024wda}.
We compute also  (renormalised) meson decay constants~\cite{Martinelli:1982mw,Lepage:1992xa}, 
where this is possible with our available data.
 We perform an extensive study of finite-volume effects.
We extrapolate our numerical results towards the massless and continuum limits,
relying on  Wilson chiral perturbation theory (W$\chi$PT)~\cite{
Sheikholeslami:1985ij,Rupak:2002sm} by borrowing ideas from Ref.~\cite{Sharpe:1998xm}, and from the literature on improvement~\cite{Symanzik:1983dc,Luscher:1996sc}.

This is the first systematic,  dedicated calculation, in this lattice gauge theory,
to allow for an extrapolation towards the continuum limit.
We benchmark our results against existing measurements, obtained either 
in the quenched approximation of the same theory, or in other related theories with dynamical fermions: the $Sp(4)$ theory coupled to $N_{\rm f}=2$ fermions, and the $SU(3)$ theory with three 
quarks---we remind the reader that, in $SU(3)$, the two-index 
antisymmetric representation coincides with the conjugate of the fundamental one.
Interesting trends emerge from these critical comparisons, which we highlight later in the paper.

The paper is organised as follows.
We introduce the (continuum and lattice)  theories of interest in Sect.~\ref{sec:model}.
We devote Sect.~\ref{Sec:latticemodel} to studying the properties of the lattice theory. We
identify its bulk phase transitions in (lattice) parameter space.
We perform a finite-volume study, to identify criteria that allow us to ensure that such systematic effects can be neglected, in comparison with existing statistical uncertainties.
We introduce at this stage our treatment of the Wilson flow, and its complementary uses to set the scale and to compute the  topology in our configurations.
Sect.~\ref{Sec:meson} details our definitions of meson operators, correlation functions, masses, and decay constants, and the processes we follow in our measurements. 
 The numerical results are summarised in Sect.~\ref{Sec:results}, 
 which describes also our approach to the continuum and massless limits. We discuss our main results and 
 outline  future avenues for investigation in Sect.~\ref{sec:conclusion}. 
Appendices~\ref{Sec:data}, \ref{Sec:autocorrelation}, \ref{Sec:largemass}, and~\ref{Sec:mpcac} contain tables of intermediate numerical results
 and technical details that may be helpful in reproducing our analysis---we  provide access to our data and the analysis code
 in Refs.~\cite{data_release}  and~\cite{analysis_release}.

Appendix~\ref{Sec:deft} presents a preliminary, alternative analysis of the spectral measurements we collected,
performed in the light of dilaton effective field theory (dEFT)~\cite{Matsuzaki:2013eva,Golterman:2016lsd,Kasai:2016ifi,Hansen:2016fri,Golterman:2016cdd,Appelquist:2017wcg,Appelquist:2017vyy,Golterman:2018mfm,Cata:2019edh,Cata:2018wzl,Appelquist:2019lgk,Golterman:2020tdq,Golterman:2020utm,Appelquist:2020bqj,Appelquist:2022qgl,Appelquist:2022mjb}.\footnote{The idea that a 
light scalar particle, the dilaton, associated with the spontaneous breaking of scale invariance,
might appear in proximity of the lower edge of the conformal window is quite old~\cite{Leung:1985sn,Bardeen:1985sm,Yamawaki:1985zg}, and so is the first effective field theory (EFT)
description of its behaviour~\cite{Migdal:1982jp,Coleman:1985rnk}. 
The striking phenomenological implications for new physics
of the emergence of a dilaton~\cite{Goldberger:2007zk} are the subject of a vast literature---an incomplete selection of interesting work includes Refs.~\cite{Hong:2004td,Dietrich:2005jn,Hashimoto:2010nw,Appelquist:2010gy,Vecchi:2010gj,Chacko:2012sy,Bellazzini:2012vz,Bellazzini:2013fga,Abe:2012eu,Eichten:2012qb,Hernandez-Leon:2017kea,CruzRojas:2023jhw} and references therein.  Examples of application of dEFT  in the CHM context can be found in Refs.~\cite{Appelquist:2020bqj,Appelquist:2022qgl}---see also related earlier work in Refs.~\cite{Vecchi:2015fma,Ma:2015gra,BuarqueFranzosi:2018eaj}.}
It provides an unconventional interpretation and analysis tool, with respect  to the Wilson chiral perturbation theory adopted in the main body of the paper. 
This new instrument has given interesting results in dynamical theories with near conformal dynamics, such as the case of  $SU(3)$ theories with $N_{\rm f}=8$ fundamental fermions~\cite{LatKMI:2014xoh,Appelquist:2016viq,LatKMI:2016xxi,Gasbarro:2017fmi,LatticeStrongDynamics:2018hun,LatticeStrongDynamicsLSD:2021gmp,Hasenfratz:2022qan,LSD:2023uzj,LatticeStrongDynamics:2023bqp},
or $N_{(s)}=2$ fermions transforming in the two-index symmetric representation~\cite{Fodor:2012ty,
	Fodor:2015vwa,
	Fodor:2016pls,Fodor:2017nlp,Fodor:2019vmw,Fodor:2020niv}.
We hence decided to perform and report  this exercise, although we anticipate  that
 our results are  inconclusive.
Such an analysis promises to become important for future  higher statistics calculations, 
performed at smaller fermion masses and 
closer to the continuum limit, particularly if flavour-singlet scalar mesons are accessible.

\section{\label{sec:model}The $Sp(4)$ theory of interest}
The field content of the $Sp(4)$ gauge theory of  interest consists of the gauge fields, $V_\mu\equiv \sum_A V_{{\mu}}^A T^A$ (where $T^A$, for $A=1,\,\cdots,\,10$, are the hermitian generators of the group, normalised so that ${\rm Tr}\,T^AT^B=\frac{1}{2}\delta^{AB}$),
and three flavors of  massive (degenerate) hyperquarks, $\Psi$. These Dirac fermions,
that are described by antisymmetric $4\times 4$ matrices in $Sp(4)$,
obey the condition  ${\rm Tr}\left[ \Omega \Psi\right]=0$, where the symplectic matrix, $\Omega$, is
\beqs
\Omega&\equiv&
\begin{pmatrix} 0 & \mathbb{1}_{2\times 2} \\ -\mathbb{1}_{2\times 2} & 0\end{pmatrix}\,.
\eeqs
The fermions transform in the antisymmetric, two-index representation, as $\Psi \rightarrow U \Psi U^T$, under the action of a group element, $U\in Sp(4)$. 
The continuum Lagrangian density is 
\beq
\mathcal{L}= -\frac{1}{2} \textrm{Tr}\,V_{\mu\nu}  V^{\mu\nu} + \overline{\Psi^j} (i D_\mu \gamma^\mu-m) \Psi^j,
\label{eq:action}
\eeq
where $\mu,\nu=1,\,2,\,3,\,4$ are space-time indexes and $j=1,\,2,\,3$ flavour indexes, while summations over  repeated indices are understood. 
In our convention  the Minkowski metric has signature $(+,-,-,-)$. 
The field strength tensor, $V_{\mu\nu}$, and the covariant derivative, $D_\mu$, are defined by
\beqs
V_{\mu\nu} &\equiv& \partial_\mu V_\nu -\partial_\nu V_\mu + ig  [V_\mu,V_\nu ],\\
D_\mu \Psi &\equiv& \partial_\mu \Psi + ig V_\mu \Psi +i g \Psi V_\mu^T,
\eeqs
where $g$ is the gauge coupling.

The fermion degrees of freedom populating the entries of the antisymmetric matrices, 
$\Psi^{ab}$, with $a,\,b=1,\,\cdots,\,4$,  span an
 irreducible representations of $Sp(4)$,
which, as the group is locally isomorphic to $SO(5)$, coincides with
the vectorial representation, $\bold 5$, of $SO(5)$,
after imposing the aforementioned $\Omega$-traceless requirement on $\Psi$.
Because such a representation is real, the  global (flavor) symmetry acting on $N_{(\rm as)}=3$
Dirac fermions  is enhanced from $U(3)_L\times U(3)_R$ to $U(1)\times SU(6)$. In this paper, we ignore the (anomalous) $U(1)$, focusing on non-trivial representations of the non-Abelian, $SU(6)$, factor. This symmetry is  broken to the maximal  $SO(6)$ 
subgroup by the non-zero, degenerate fermion mass, $m$. 
In the massless case, the $SU(6)$ global symmetry is also spontaneously  broken along the same pattern, if a non-vanishing  condensate, $\langle \overline{\Psi}\Psi\rangle \neq 0$, emerges.


\subsection{\label{Sec:latticeaction}Lattice action}

The lattice theory is written by applying a Wick rotation to the four-dimensional space-time,
and discretising the Euclidean space on hypercubic lattice of  size
 $N_t \times N_s^3= (T/a) \times (L/a)^3$, with $a$ the lattice spacing, $T$ and $L$
 the temporal and spatial extents, respectively.
 As the discretised version of the action based on the Lagrangian density in \Eq{action}, 
we adopt the standard plaquette action for the gauge fields, supplemented by
 the Wilson-Dirac fermion formulation for hyperquarks, $\Psi$, and write 
\begin{widetext}
\beqs
S &\equiv& \beta \sum_{n,\,\mu<\nu} \left(1-\frac{1}{4}{\rm Re}{\rm Tr}\mathcal{P}_{\mu\nu}(n)\right)
+a^4  \sum_n \overline{\Psi}_j(n) D^{(as)}  \Psi_j(n)\,,
\label{eq:lattice_action}
\eeqs
\end{widetext}
where $\mu,\nu=1,\,2,\,3,\,4$ are space-time indexes, $j=1,\,2,\,3$ the flavour indexes, and $n$ denotes the lattice sites. 
The lattice bare coupling, $\beta$, is related with the gauge coupling, $g$, by $\beta=8/g^2$. 
The elementary plaquette, $\mathcal{P}_{\mu\nu}$, is defined as
\beq
\mathcal{P}_{\mu\nu}(n)\equiv U_\mu (n) U_\nu (n+\hat{\mu}) U_\mu^\dagger(n+\hat{\nu}) U_\nu^\dagger(n)\,,
\label{eq:plaquette}
\eeq
where $U_\mu \in Sp(4)$ denotes the gauge link, which satisfies the condition
$U^*=\Omega U \Omega^\dagger$. 
We impose periodic boundary conditions on all the fields, except for the temporal directions for the fermion fields, $\Psi$, which obey anti-periodic boundary conditions.

We exploit the aforementioned properties of the irreducible representation of the fermions, $\bold 5$, 
to  write the Wilson-Dirac operator  in terms of 
the link variable for the antisymmetric representation, $U^{\rm{(as)}}_\mu$, 
and the bare hyperquark mass, $m_0$, 
as
\begin{widetext}
\beqs
D^{(as)} \Psi_j(x) \equiv (4/a+m_0) \Psi_j(x)-\frac{1}{2a}\sum_\mu
\left\{\frac{}{}(1-\gamma_\mu)U^{\rm{(as)}}_\mu(x) \Psi_j(x+\hat{\mu})
+(1+\gamma_\mu)U^{\rm{(as)}\, \dagger}_\mu(x-\hat{\mu}) \Psi_j(x-\hat{\mu})\frac{}{}\right\}\,.
\label{Eq:DiracF}
\eeqs
\end{widetext}

The relation to the fundamental link, $U_\mu$, reads 
\beq
U_{\mu,\,AB}^{(as)}(x) \equiv {\rm Tr}\left[
e^{(as)\, \dagger}_A U_\mu (x) e^{(as)}_B U_\mu^{\rm T}(x)
\right],
\eeq
where the multi-indexes, $A=(ab)$ and $B=(cd)$,  denote
 the ordered pairs with $1\leq a (c) < b (d) \leq 4$. The basis matrices, $e_A^{(as)}$, are antisymmetric, and obey  the 
 defining relations ${\rm Tr} \left[ \Omega\, e_A^{(as)}\right] =0$
 (they are $\Omega$-traceless)
and
\beq
(e^{(as)}_{A=(ab)} )_{cd} = \frac{1}{\sqrt{2}}(\delta_{ad}\delta_{bc}-\delta_{ac}\delta_{bd})\,,
\eeq
except for $A=(24)$, which is given by
\beq
e^{(as)}_{A=(24)}  = \begin{pmatrix}
0 & 0 & \frac{1}{2} & 0\\
0 & 0 & 0 & -\frac{1}{2}\\
-\frac{1}{2} & 0 & 0 & 0\\
0 & \frac{1}{2} & 0 & 0
\end{pmatrix}\,.
\eeq
These five basis matrices satisfy also the orthonormalisation condition ${\rm Tr}\left[ e^{(as)\,T}_A e^{(as)}_B \right]=\delta_{AB}$.

\begin{table*}[t]
\caption{%
\label{tab:ensemble}
Ensembles of dynamical $Sp(4)$ lattice gauge theories coupled to $N_{\rm as}=3$ Wilson-Dirac hyperquarks transforming in the two-index antisymmetric representation of the gauge group. For each ensemble analysed, we report the lattice extent,  
$N_t\times N_s^3$, the bare coupling, $\beta$, the fermion mass, $am_0$, the number of  configurations, $N_{\rm cfg}$, 
and the expectation value of average plaquette, $\langle P \rangle$. 
We also include a comment in the last column,  discussed in the body of the paper.
}
\begin{center}
\begin{tblr}{width=\textwidth,colspec=|c|c|c|c|c|c|c|}
\hline\hline
Ensemble & $N_t \times N_s^3$ & $\beta$ & $am_0$ & $N_{\mathrm{cfg}}$ & $\langle \mathcal{P} \rangle$ & Comment \\
\hline
\hline
ASB0M1 & $48 \times 18^3$ & 6.6 & -1.075 & 100 & 0.580347(15) & heavy \\
ASB0M2 & $48 \times 24^3$ & 6.6 & -1.08 & 140 & 0.582733(10) & heavy \\
ASB0M3 & $48 \times 32^3$ & 6.6 & -1.085 & 130 & 0.5849929(62) &  \\
\hline
ASB1M1 & $48 \times 18^3$ & 6.65 & -1.05 & 128 & 0.579896(12) & heavy \\
ASB1M2 & $48 \times 18^3$ & 6.65 & -1.06 & 120 & 0.583984(12) & heavy \\
ASB1M3 & $48 \times 18^3$ & 6.65 & -1.063 & 135 & 0.585150(14) & heavy \\
ASB1M4 & $48 \times 24^3$ & 6.65 & -1.07 & 137 & 0.5877893(86) &  \\
ASB1M5 & $48 \times 28^3$ & 6.65 & -1.075 & 215 & 0.5896176(54) &  \\
ASB1M6 & $48 \times 32^3$ & 6.65 & -1.08 & 180 & 0.5914586(56) &  \\
\hline
ASB2M1 & $48 \times 16^3$ & 6.7 & -1.02 & 200 & 0.578740(28) & heavy \\
ASB2M2 & $48 \times 16^3$ & 6.7 & -1.03 & 110 & 0.582264(35) & heavy \\
ASB2M3 & $48 \times 18^3$ & 6.7 & -1.04 & 100 & 0.585678(14) & heavy \\
ASB2M4 & $48 \times 24^3$ & 6.7 & -1.045 & 120 & 0.5873337(84) & heavy \\
ASB2M5 & $48 \times 24^3$ & 6.7 & -1.05 & 110 & 0.5889499(90) & heavy \\
ASB2M6 & $48 \times 24^3$ & 6.7 & -1.055 & 180 & 0.5905740(86) &  \\
ASB2M7 & $54 \times 28^3$ & 6.7 & -1.06 & 201 & 0.5921942(59) &  \\
ASB2M8 & $54 \times 28^3$ & 6.7 & -1.063 & 150 & 0.5931513(75) &  \\
ASB2M9 & $54 \times 32^3$ & 6.7 & -1.065 & 150 & 0.5937583(54) &  \\
ASB2M10 & $54 \times 36^3$ & 6.7 & -1.067 & 195 & 0.5944477(41) &  \\
ASB2M11 & $54 \times 36^3$ & 6.7 & -1.069 & 218 & 0.5950649(38) &  \\
\hline
ASB3M1 & $54 \times 18^3$ & 6.75 & -1.03 & 180 & 0.590439(12) & heavy \\
ASB3M2 & $54 \times 24^3$ & 6.75 & -1.041 & 120 & 0.5935363(84) &  \\
ASB3M3 & $54 \times 24^3$ & 6.75 & -1.046 & 180 & 0.5949951(72) &  \\
ASB3M4 & $54 \times 28^3$ & 6.75 & -1.051 & 196 & 0.5963914(66) &  \\
ASB3M5 & $54 \times 32^3$ & 6.75 & -1.055 & 225 & 0.5975680(48) &  \\
\hline
ASB4M1 & $56 \times 16^3$ & 6.8 & -1.01 & 171 & 0.592230(14) & heavy \\
ASB4M2 & $54 \times 16^3$ & 6.8 & -1.02 & 165 & 0.594777(13) & heavy \\
ASB4M3 & $54 \times 24^3$ & 6.8 & -1.03 & 180 & 0.5972763(73) &  \\
ASB4M4 & $56 \times 24^3$ & 6.8 & -1.035 & 275 & 0.5985597(78) &  \\
ASB4M5 & $54 \times 32^3$ & 6.8 & -1.04 & 170 & 0.5998238(50) &  \\
ASB4M6 & $54 \times 32^3$ & 6.8 & -1.043 & 251 & 0.6006178(38) &  \\
ASB4M7 & $54 \times 36^3$ & 6.8 & -1.046 & 219 & 0.6014000(34) &  \\
\hline
ASB5M1 & $54 \times 24^3$ & 6.9 & -1.01 & 391 & 0.6045303(54) &  \\
ASB5M2 & $54 \times 32^3$ & 6.9 & -1.017 & 216 & 0.6060103(43) &  \\\hline\hline
\end{tblr}

\end{center}

\end{table*}

\subsection{\label{Sec:simulation}Simulation details}

Throughout this work, we employ standard bootstrap methods in our statistical analyses. 
We perform numerical calculations using a branch of the HiRep code~\cite{DelDebbio:2008zf}, in which some of us implemented the $Sp(2N)$ gauge groups~\cite{Bennett:2017kga}. 
Dynamical gauge ensembles are generated with an admixture of both the Hybrid Monte Carlo (HMC) 
and Rational HMC (RHMC)~\cite{Clark:2003na, Clark:2006fx}
 algorithms\footnote{
 Using HiRep commit IDs \texttt{d3ab8d8}~\cite{hirep:sa2c_2018} and \texttt{9e66e56}~\cite{hirep:sa2c}.
 };
we introduce two pseudofermions, one in the HMC (for two flavor species)
and the other (for the third species) in the RHMC evolutions, 
and dial their masses to reproduce the presence of 
 three mass-degenerate Wilson-Dirac fermions---we verified elsewhere that this choice yields compatible result with using the RHMC algorithm for all three species~\cite{Bennett:2022yfa,Bennett:2023gbe}. 
The determinant of the Dirac operator is real and positive for fermions in the antisymmetric representation (even with an odd number of them), thus numerical simulations are  free from sign problems. 
 The molecular dynamics (MD) evolution is implemented using a second-order Omelyan integrator~\cite{Takaishi:2005tz}.

The main properties of all our ensembles are summarised in Table~\ref{tab:ensemble}.
Each ensemble consists of $N_{\rm cfg}$ thermalised configurations. 
The lattice coupling, $\beta$, and  the bare hyperquark mass,  $a m_0$, as well as the lattice volume, 
 $N_t\times N_s^3$, are chosen 
in such a way that our numerical results do not suffer from two typical artefacts due to the implementation of
dynamical fermions in lattice calculations:
the presence of  first-order bulk transition and finite volume 
effects. We will discuss them in details in the next section.

For each ensemble we present the expectation value, $\langle \cal P \rangle$, of the average plaquette, $ \cal P$,  defined as
\beqs
{\cal P} &\equiv& \frac{1}{6N_t N_s^3}\sum_n\sum_{{\mu<\nu}}\left[\frac{1}{4}
{\rm Re}{\rm Tr}\mathcal{P}_{\mu\nu}(n)\right]\,. 
\eeqs
We also investigate the history of the average plaquette along the trajectories to ensure that the system is thermalised. 
We discard a few hundreds of initial trajectories, 
the precise number of which, for each ensemble, is chosen by monitoring the value of $\mathcal{\cal P}$. 
We compute the autocorrelation length of the average plaquette, $\tau^{\cal{P}}_{\rm exp}$, over the thermalised configurations, and separate them by $\delta_{\rm traj}$ trajectories, with $\delta_{\rm traj} \gtrsim \tau^{\cal{P}}_{\rm exp}$. The definition and the resulting values of $\tau^{\cal{P}}_{\rm exp}$, as well as our choices of $\delta_{\rm traj}$, can be found in \App{autocorrelation}. 

In the last column of \Tab{ensemble}, we also anticipate some qualitative assessment of the 
 properties of the ensembles, that will be discussed in more details later in the body of the paper.  We denote as `heavy'  those ensembles in which the bare mass is so large that an unexpected behaviour (to be discussed in Sect.~\ref{Sec:results}) shows up in the measured spectral quantities; we exclude such ensembles from the continuum extrapolation.


\section{\label{Sec:latticemodel}Characterisation of the lattice theory}

We devote this section to the characterisation of the lattice theory of interest. 
This part of the numerical investigation  determines the range of lattice theory bare parameters and volumes used in the generation of gauge ensembles and physics measurements.
We reveal the existence of a first-order bulk phase transition separating the range of lattice parameters that is connected to the continuum theory from the lattice strong coupling regime. We assess the size of finite-volume effects on spectroscopic observables. 
We introduce the Wilson flow as a scale setting procedure, and as a tool to compute the topological properties of the configurations.

\subsection{\label{Sec:phase}Phase space}

\begin{figure}
\begin{center}
\includegraphics{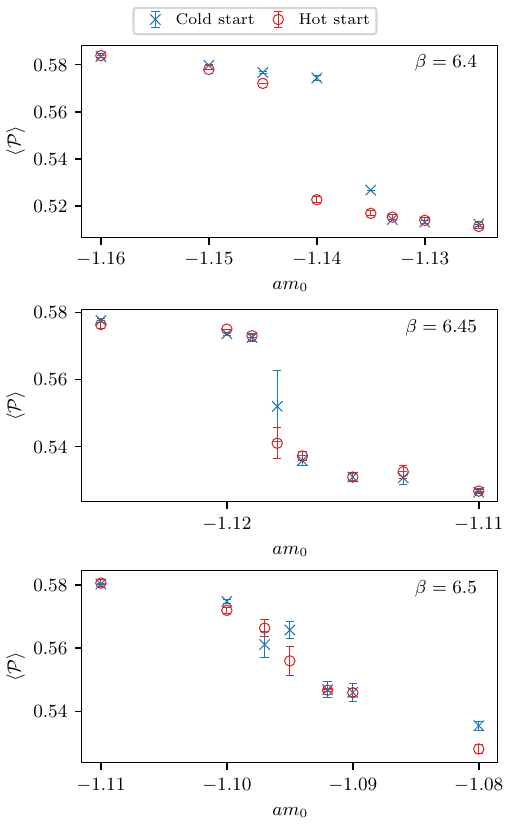}
\caption{%
\label{fig:mass_scan}%
Numerical measurement of the ensemble average of the plaquette, $\langle {\cal P} \rangle$,
as a function of the (degenerate)  bare mass, $am_0$, of the fermions transforming as the antisymmetric representation of the $Sp(4)$ gauge group,  for three representative values of the lattice coupling, \PlaquettePhaseDiagramBetas ~(top to bottom panels, respectively). 
Results are presented for isotropic lattices with extent $N_t N_s^3=8^4$. 
Average plaquette values are measured using thermalised configurations with both cold (unit) and hot (random) starts; we denote  the former by blue crosses and the latter by red empty circles. 
}
\end{center}
\end{figure}

The lattice spacing, $a$, serves as an ultraviolet (UV) regulator. 
The  continuum theory of interest is recovered in the limit  $a\rightarrow 0$, corresponding to the quantum critical point of the lattice theory. 
In the space of lattice parameters, the continuum theory can be approached by extrapolating physical 
measurements towards the limit
 $\beta\rightarrow \infty$. 
In the presence of fermions, there is also a second free parameter, the bare mass,  $a m_0$,
which must be dialed towards the limit of interest.
In the Wilson-Dirac formulation of the fermions, this parameter  is affected by 
additive renormalisation, which further complicates extrapolations towards physically interesting regions of parameter space.
Yet, the main concern is the existence of potential bulk phase transitions, restricting the 
basin of attraction of the Gaussian fixed point in lattice parameter space.

Preliminary studies have shown the existence of a first-order bulk phase transition in the strong coupling regime~\cite{Lee:2018ztv,Bennett:2023gbe}, highlighted by evidence  of hysteresis in  the average plaquette, $\langle \mathcal{P} \rangle$. 
Phase transition points appear to lie along a line in the plane of lattice parameters,  $(am_0,\,\beta)$, 
with the line ending at a critical point.
For this publication, we conducted an additional study, which refines earlier findings 
by narrowing down the range of $\beta$ values to \PlaquettePhaseDiagramBetas, near the critical point, 
and increasing the statistics. The results are shown in \Fig{mass_scan}.  
The transition is first order for $\beta=6.4$, 
but becomes a smooth crossover at $\beta \simeq 6.45$. We therefore restrict our computations
to the weak coupling regime, defined by the constraint $\beta > 6.45$.

\subsection{\label{Sec:FV}Finite volume effects}

The finiteness of the  volume introduces a spurious discretisation of all physical spectra measured on the lattice. The volume itself appears as an unphysical scale, acting as an infrared (IR) regulator, and affects all spectral quantities.
For the purpose of assessing (and minimising) the size of these  finite volume (FV) effects,
in this brief subsection we anticipate some preliminary results of our analysis of
 spectral quantities. We identify general criteria that
we later apply to select the ensembles we retain in the physical analysis in the body of the paper.
Having done so,  FV effects will be ignored in subsequent sections.

In confining theories, spectral observables are expected to
receive exponentially suppressed corrections due to
 FV effects, as long as 
the spatial length, $L=a N_s$, is larger than the longest intrinsic length scale in the physical system, 
 the Compton wavelength of the lightest composite state. 
As can be confirmed a posteriori, the pseudoscalar (ps) meson 
is the lightest among the composite states considered in this work. 
We hence focus attention on the mass of the ground state in the ps channel,
and study its dependence on the spatial volume---we return to describing this measurement
 in the next section.

\begin{figure}
\begin{center}
\includegraphics{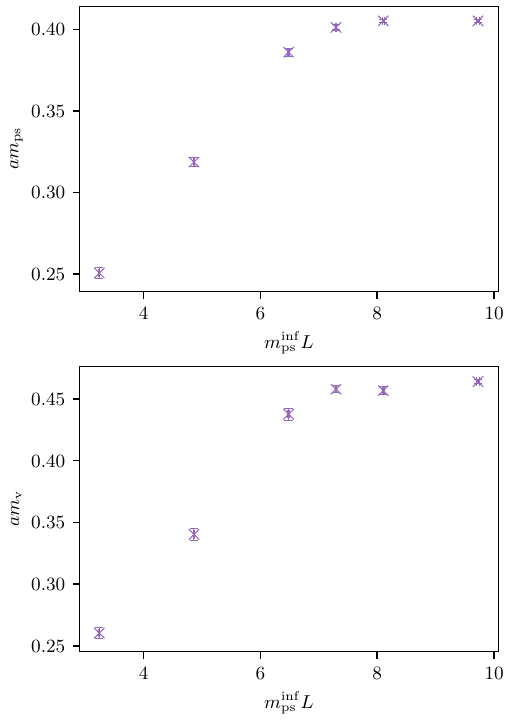}
\caption{%
\label{fig:mpsL}%
Pseudoscalar meson mass, $am_{\rm ps}$ (top panel), and  vector meson mass, $am_{\rm v}$ (bottom panel), as a function of the infinite-volume
 pseudoscalar mass, $a m_{\rm ps}^{\rm inf}\,$, multiplied by the spatial extent $N_s=L/a$.
The lattice coupling and the bare hyperquark mass used for the calculations are $\beta=\FiniteVolumeStudyBeta$ and $m_0=\FiniteVolumeStudyMass$, respectively. 
The pseudoscalar mass at  infinite volume, $m_{\rm ps}^{\rm inf}$, is estimated by taking the one measured in the largest available lattice, with $N_t\times N_s^3=54\times 24^3$. 
}
\end{center}
\end{figure}

\begin{table*}
\caption{%
\label{tab:FV}%
Ensembles generated for the study of finite volume effects. 
The lattice coupling and the bare mass are fixed to $\beta=\FiniteVolumeStudyBeta$ and $am_0=\FiniteVolumeStudyMass$, respectively, while
$N_{\rm cfg}$ is the number of configurations, separated by $\delta_{\rm traj}=\FiniteVolumeStudyDeltaTraj$ trajectories between adjacent configurations, and $\langle {\cal P} \rangle$ is the average plaquette.
The measured PCAC hyperquark and meson masses are expressed in lattice units, while $m_{\rm ps}^{\rm inf}$ denotes the pseudoscalar mass at infinite volume 
which is estimated as  the mass measured at the largest available lattice. 
}
\begin{center}
\begin{tblr}{width=\textwidth,colspec=|c|c|c|c|c|c|c|}
\hline\hline
$N_t \times N_s^3$ & $N_{\mathrm{cfg}}$  & $\langle\mathcal{P}\rangle$ & $am_{\mathrm{PCAC}}$  & $am_{\mathrm{ps}}$ & $am_{\mathrm{v}}$  & $m_{\mathrm{ps}}^{\mathrm{inf}} L$ \\
\hline
$54 \times 8$ & 200 & 0.596973(37) & 0.05742(64) & 0.2506(36) & 0.2605(47) & 3.2413(60) \\
$54 \times 12$ & 300 & 0.597306(14) & 0.05837(24) & 0.3186(28) & 0.3403(49) & 4.8619(91) \\
$54 \times 16$ & 209 & 0.597278(13) & 0.05926(18) & 0.3860(25) & 0.4374(48) & 6.483(12) \\
$54 \times 18$ & 200 & 0.5972729(99) & 0.05975(14) & 0.4012(16) & 0.4577(30) & 7.293(14) \\
$54 \times 20$ & 200 & 0.5972551(91) & 0.05987(12) & 0.4052(10) & 0.4565(30) & 8.103(15) \\
$54 \times 24$ & 180 & 0.5972763(73) & 0.059877(85) & 0.40516(75) & 0.4639(19) & 9.724(18) \\\hline\hline
\end{tblr}

\end{center}
\end{table*}

In \Tab{FV}, 
we present an example of the numerical  results we collected for the purpose of FV effects study.
We fix  the lattice parameters, $\beta=6.8$ and $a m_0=-1.03$, 
the length of the time direction in the lattice, $N_t=54$.
 We also provide some details about the characterisation of the ensembles.
 We consider six different volumes, with $N_s=L/a=8,\,12,\,16,\,18,\,20,\,24$,  
and for each measure the mass of the lightest ps state, $a m_{\rm ps}$. 
For completeness, in the table we report the (bare) partially-conserved-axial-current (PCAC) mass, 
$am_{\rm PCAC}$, which can be used as a more physical assessment of the fermion mass,
being free of additive renormalisation. Finally, we show the value of the mass of the lightest vector meson, $a m_{\rm v}$.

In the upper panel of \Fig{mpsL}, we show the measured mass of the ps mesons, $am_{\rm ps}$, 
as a function  of $m_{\rm ps}^{\rm inf} L$.
We estimate $m_{\rm ps}^{\rm inf}$ by assuming it coincides with the mass measured at the largest 
available lattice, $N_t\times N_s^3=54\times 24^3$.\footnote{
We remark that the commonly used exponential fit in the mass of the pseudoscalar to determine the latter in the infinite volume limit does not appear to work in the regime of our calculation. 
}  As expected, the ${\rm ps}$ mass quickly converges to its infinite-volume asymptotic value as the volume increases.
The results for the two largest-volume lattices are  indistinguishable, given present statistical uncertainties, 
indicating that  FV effects can be ignored, when  $m_{\rm ps}^{\rm inf} L \gtrsim 7.5$ .
We show also  $a m_{\rm v}$ as a function of $m_{\rm ps}^{\rm inf} L$,
 in the lower panel of \Fig{mpsL}.
In the case of the three largest lattice volumes, the size of FV effects affecting $a m_{\rm v}$ is comparable to the statistical uncertainty,
which confirms  that FV effects are negligible for $m_{\rm ps}^{\rm inf} L \gtrsim 7.5$.

This requirement is somewhat stronger than the analogous bound on the lattice volume applied to typical calculations in QCD, but significantly weaker than in theories known to be in the conformal window such as $SU(2)$ with two adjoint fermions \cite{DelDebbio:2015byq}. Assuming that our theory is deep inside the confined and chirally broken phase, like is the case in QCD, this effect might be due to the fact that our dynamical ensembles probe a region of parameter space with relatively large mass, as will be discussed in \Sec{results}, where FV effects are enlarged by a factor of $m_{\rm ps}^2$.
The observed, negative contribution of  FV effects to $am_{\rm ps}$ is consistent with next-to-leading-order chiral perturbation theory (NLO $\chi$PT) expectations, as the sign of the coefficient of the contribution due to FV corrections, 
at the one-loop level, solely depends on the chiral symmetry breaking pattern~\cite{Bijnens:2009qm}. 
We refer the reader to Refs.~\cite{Bennett:2021mbw,Bennett:2022yfa,Bennett:2023wjw} for further details and discussion of FV effects in the $Sp(4)$ theory with dynamical fermions in different representations.

\subsection{\label{Sec:scale} Wilson flow and scale setting procedure}

We adopt the gradient flow and its lattice implementation, the Wilson flow~\cite{Luscher:2010iy,Luscher:2011bx,Luscher:2013vga}, 
 to set the scale in our lattice results and  extrapolate them
towards the continuum limit.
This scale-setting procedure  relies only on 
theoretical input,
 and
amounts to the study of a diffusion process, in five dimensions, along a 
fictitious time, $t$. Gradient flow and
 Wilson flow can also be 
used to study 
the non-perturbative evolution of renormalised couplings~\cite{Fodor:2012td,Hasenfratz:2019hpg}.
Furthermore, they provide a smoothening procedure, suppressing short-distance fluctuations and
optimising the computation of long-distance properties, such as the topological charge, $Q$.
We follow the ideas exposed in Ref.~\cite{BMW:2012hcm} to further reduce discretisation effects.

\begin{figure}
\begin{center}
\includegraphics{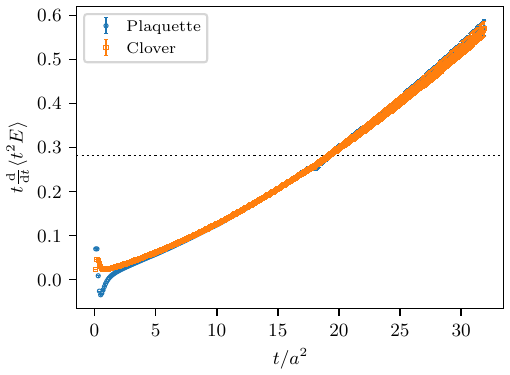}
\caption{%
\label{fig:wflow}%
The derivative of the expectation value of the energy density, $\mathcal{W}(t)\equiv t \,d\langle t^2 E \rangle/dt$, defined in Eq.~(\ref{Eq:WW}), and built from the field strength tensor, $G_{\mu \nu}$, obtained either with the plaquette (blue circle) or the alternative,  clover-leaf (orange square) definitions, as a function of  flow time, $t$, in the ensemble ASB4M5. 
The black dashed line denotes our choice of reference value, 
$\mathcal{W}_0=0.28125$.
Qualitatively equivalent plots can be obtained with other available ensembles.
}
\end{center}
\end{figure}

The diffusion equation defining the gradient flow is 
\beqs
\frac{d}{dt}V_\mu(x,t) = - V_\mu(x,t) \frac{\partial S(V_\mu)}{\partial V_\mu(x,t)}\,,
\label{Eq:wflow}
\eeqs
and the flow variable, $V_{\mu}$, at $t=0$ obeys the initial condition $V_\mu(x,0)=U_\mu(x)$. 
The flow equation can be solved via  numerical integration, with infinitesimal flow time, $\delta t$. 
For the gauge action in Eq.~(\ref{Eq:wflow}), $S$, one can use the Wilson plaquette action, the first term of \Eq{lattice_action}, which defines the  {\it Wilson flow}. 
(One can replace, in $S$,  the plaquette $\mathcal{P}_{\mu\nu}$, with
a different object, so that the results differ only by terms that vanish in the continuum, $a\rightarrow 0$,  limit.) 
As the fictitious time, $t$, flows, the gauge fields diffuse 
and  high momentum fluctuations (discretisation effects) 
are suppressed, while long-distance  physics is preserved. 

The gradient (Wilson) flow acts as Gaussian smoothening operator on the 
fields (configurations), removing short-distance singularities.
The diffusion radius associated with this smoothening process has a characteristic length scale  $\sim 1/\sqrt{8t}$~\cite{Luscher:2010iy}.
For $t > 0$,  furthermore, it has been shown that correlation functions are finite to all orders in perturbation theory, 
in the sense described in Ref.~\cite{Luscher:2011bx}.
Hence, one can fix a physical scale by assigning a
reference value to one, conveniently chosen, physical observable.
Following the proposal in Ref.~\cite{Luscher:2010iy}, one can consider the
energy density
\beq
E(t)\equiv \frac{1}{2} {\rm Tr}\, G_{\mu\nu}(t) G_{\mu\nu}(t)\,,
\eeq
where $G_{\mu\nu}$ is the field strength tensor associated with $V_{\mu}$. 
As the expectation value of $E(t)$ has dimension of a mass to the fourth power, while $t$ scales as the inverse of a mass squared,
one defines the following dimensionless quantities, defined for any $t>0$~\cite{Luscher:2010iy,BMW:2012hcm}:
\beq
\mathcal{E}(t) \equiv t^2 \langle E(t)\rangle\,,\\
\eeq
and its (logarithmic) derivative, 
\beq
\label{Eq:WW}
\mathcal{W}(t) \equiv t \frac{\dd}{\dd t} \mathcal{E}(t)\,.
\eeq
One then defines two alternative scales, $t_0$ and $w_0$, by imposing the following conditions, respectively:
\beq
\label{Eq:flow_refs}
\mathcal{E}(t)|_{t=t_0} = \mathcal{E}_0\,,~~~\textrm{or}~~~\mathcal{W}(t)|_{t=w_0^2} = \mathcal{W}_0\,.
\eeq

The references values, $\mathcal{E}_0$ or $\mathcal{W}_0$, are chosen empirically, by aiming at minimising systematic effects, such as lattice spacing and finite volume artefacts. 
To assess the size of discretisation effects, we adopt two distinct, alternative definitions for the field-strength tensor, $G_{\mu\nu}$, obtained with either the plaquette in \Eq{plaquette}, $\cal{P}_{\mu\nu}$, or the clover-leaf, $\mathcal{C}_{\mu\nu}$~\cite{Sheikholeslami:1985ij,Hasenbusch:2002ai}, 
\begin{widetext}
\beqs
\mathcal{C}_{\mu\nu}(x)&\equiv&\frac{1}{8}\left\{ \frac{}{}U_\mu(x)U_\nu(x+\hat{\mu})U^\dagger_\mu(x+\hat{\nu})U^\dagger_\nu(x)\right.
+U_\nu(x)U^\dagger_\mu(x+\hat{\nu}-\hat{\mu})U^\dagger_\nu(x-\hat{\mu})U_\mu(x-\hat{\mu})\\
&&+U^\dagger_\mu(x-\hat{\mu})U^\dagger_\nu(x-\hat{\nu}-\hat{\mu})U_\mu(x-\hat{\nu}-\hat{\mu})U_\nu(x-\hat{\nu})\nn
+U^\dagger_\nu(x-\hat{\nu})U_\mu(x-\hat{\nu})U_\nu(x-\hat{\nu}+\hat{\mu})U^\dagger_\mu(x)
\left.-h.c.\frac{}{}\right\}\,,
\eeqs
\end{widetext}
built out of the discretised lattice link variables by replacing $U_\mu(x)$ with $V_\mu (x,t)$, at $t>0$. 
With our ensembles, we could demonstrate the fact that, for $\mathcal{E}(t)$, the plaquette and clover definitions 
 are not in complete agreement, yet show little discernible difference if we choose $\mathcal{E}_0 \gtrsim 0.25$. 
As illustrated in the example in \Fig{wflow}, though, we
 find that the two definitions yield  statistically consistent
results, if one instead uses $\mathcal{W}(t)$, and  $\mathcal{W}_0\gtrsim  0.1$.

In view of these considerations, in our numerical study we use $w_0$ as the gradient flow 
scale, measured with the choice $\mathcal{W}_0=0.28125$, applied to the clover-leaf definition of the energy density.
This numerical choice for $\mathcal{W}_0$ is
derived by first adopting $\mathcal{W}_0=0.3$ for the $SU(3)$ Yang-Mills theory, and 
then  applying plausible assumptions to the large-$N_c$ scaling of the relevant observables, to
compare to $Sp(4)$~\cite{Bennett:2020qtj}.\footnote{
The renormalised coupling at the scale $\mu=1/\sqrt{8t}$ is given by $\alpha(\mu)=k_\alpha \mathcal{E}(t)$ at the leading order in the perturbative expansion, with $k_\alpha^{-1}=3N_c C_2(G)/16\pi$, 
and $C_2(G)$  the quadratic Casimir operator of the gauge group, $G$. 
Combined with the numerical value of $k_\alpha=16\pi/15$ for the $Sp(4)$ theory, 
our choice of $\mathcal{W}_0=0.28125$ yields that $\alpha(\mu)\simeq 0.94$, 
suggesting that we are indeed testing the theory at the scale of hadronic physics.
}
For this choice of $\mathcal{W}_0$, we find that
 finite volume effects are at most ${\cal O}(5\%)$, as $\sqrt{8 t_0}/L\lesssim 0.35$, according to the estimates in Ref.~\cite{Fodor:2012td}, except for a few heavy ensembles and for $\beta=6.9$.

 \begin{table*}[t]
\caption{%
\label{tab:topology}
Measurement of the gradient flow scale, $w_0/a$, and the measurement of the topological charge, $Q$, with Gaussian fit of the resulting distributions, in the ensembles used in this work. 
For each ensemble, the central value, $Q_0$, and standard deviation, $\sigma_Q$, are tabulated. 
Regarding the measurement of $w_0/a$, the number of uncorrelated configurations, $N_{\rm traj}^{\rm GF}$, 
is also presented. 
In the case of \WZeroIncompleteEnsembles, the gradient flow scale is not available, because for some of the configurations we cannot reach the reference scale $\mathcal{W}_0$ in the measurement of $\mathcal{W}(t)$.
}
\begin{center}
\begin{tblr}{width=\textwidth,colspec=|c|c|c|c|c|}
\hline\hline
Ensemble & $N_{\mathrm{traj}}^{\mathrm{GF}}$ & $w_0 / a$ & $Q_0$ & $\sigma_Q$  \\
\hline
\hline
ASB0M1 & 32 & $1.855(12)$ & $-0.37(48)$ & $3.61(42)$ \\
ASB0M2 & 46 & $2.1435(70)$ & $1.84(46)$ & $4.14(38)$ \\
ASB0M3 & 18 & $2.575(12)$ & $-3.37(46)$ & $4.04(39)$ \\
\hline
ASB1M1 & 42 & $1.6260(80)$ & $-1.31(60)$ & $4.69(47)$ \\
ASB1M2 & 23 & $1.982(17)$ & $-1.21(45)$ & $3.71(34)$ \\
ASB1M3 & 26 & $2.148(19)$ & $-0.67(34)$ & $3.19(29)$ \\
ASB1M4 & 33 & $2.600(19)$ & $0.10(26)$ & $2.40(23)$ \\
ASB1M5 & 17 & $3.063(29)$ & $0.13(16)$ & $2.00(11)$ \\
ASB1M6 & 8 & $3.654(56)$ & $0.03(11)$ & $1.386(86)$ \\
\hline
ASB2M1 & 66 & $1.4303(39)$ & $0.06(38)$ & $4.51(32)$ \\
ASB2M2 & 36 & $1.6262(70)$ & $0.20(38)$ & $3.17(32)$ \\
ASB2M3 & 16 & $1.925(21)$ & $0.28(39)$ & $2.91(33)$ \\
ASB2M4 & 39 & $2.1132(86)$ & $-1.31(39)$ & $3.34(32)$ \\
ASB2M5 & 27 & $2.342(17)$ & $0.61(48)$ & $3.93(39)$ \\
ASB2M6 & 14 & $2.630(36)$ & $0.37(25)$ & $2.86(19)$ \\
ASB2M7 & 9 & $3.125(68)$ & $1.75(32)$ & $2.33(24)$ \\
ASB2M8 & 18 & $3.405(41)$ & $0.42(23)$ & $2.22(17)$ \\
ASB2M9 & 12 & $3.614(56)$ & $-0.814(89)$ & $1.001(70)$ \\
ASB2M10 & 21 & $4.183(41)$ & $0.426(77)$ & $0.963(65)$ \\
ASB2M11 & 10 & $4.330(56)$ & $1.462(76)$ & $0.714(59)$ \\
\hline
ASB3M1 & 25 & $2.206(16)$ & $-0.79(20)$ & $2.34(17)$ \\
ASB3M2 & 19 & $2.634(24)$ & $0.61(31)$ & $2.74(26)$ \\
ASB3M3 & 22 & $3.059(37)$ & $1.82(11)$ & $1.302(87)$ \\
ASB3M4 & 9 & $3.628(51)$ & $0.21(23)$ & $2.08(16)$ \\
ASB3M5 & 12 & $4.079(55)$ & $-2.46(16)$ & $1.49(11)$ \\
\hline
ASB4M1 & 33 & $2.101(14)$ & $-0.429(86)$ & $1.839(62)$ \\
ASB4M2 & 23 & $2.437(36)$ & $-1.38(15)$ & $1.60(12)$ \\
ASB4M3 & 19 & $2.958(28)$ & $-1.08(11)$ & $1.296(86)$ \\
ASB4M4 & 21 & $3.365(36)$ & $0.814(80)$ & $1.075(47)$ \\
ASB4M5 & 28 & $3.828(22)$ & $-1.30(13)$ & $1.310(81)$ \\
ASB4M6 & 35 & $4.378(38)$ & $-0.025(54)$ & $0.697(39)$ \\
ASB4M7 & 13 & $4.692(52)$ & $3.506(55)$ & $0.591(46)$ \\
\hline
ASB5M1 & $\cdots$ & $\cdots$ & $0.053(60)$ & $0.781(33)$ \\
ASB5M2 & 35 & $4.304(27)$ & $-0.239(43)$ & $0.407(30)$ \\
\hline\hline
\end{tblr}

\end{center}
\end{table*}

 Another potential source of sizable systematic errors, particularly in ensembles with large $\beta$ and small $a m_0$,
is the presence of autocorrelation in the measurements of the Wilson flow, which can be quantified by the autocorrelation time $\tau^{w_0}_{\rm exp}$ of the action density, $\mathcal{E}(t)$, evaluated at $t=w_0^2$. 
We refer the reader to \App{autocorrelation} for the details of the determination of $\tau^{w_0}_{\rm exp}$ and to \Tab{autocorr} for the result. 
We find that $\tau^{w_0}_{\rm exp}$ is substantially larger than $\delta_{\rm traj}$ in any given ensemble. We therefore enlarge the separation between adjacent configurations to $\delta^{\rm GF}_{\rm traj}\gtrsim\tau^{w_0}_{\rm exp}$ in the measurements of the gradient flow scale. 
We report in \Tab{topology} the measured values of $w_0/a$, obtained from $N_{\rm traj}^{\rm GF}$ configurations, separated by $\delta_{\rm traj}^{\rm GF}$ trajectories. 
In some cases, we are left  with only a  handful of independent configurations, hence we alert the reader to use caution
in assessing the size of  the statistical errors.
Similar considerations will reappear later in the paper, when discussing the topological charge.

\begin{figure}
\begin{center}
\includegraphics{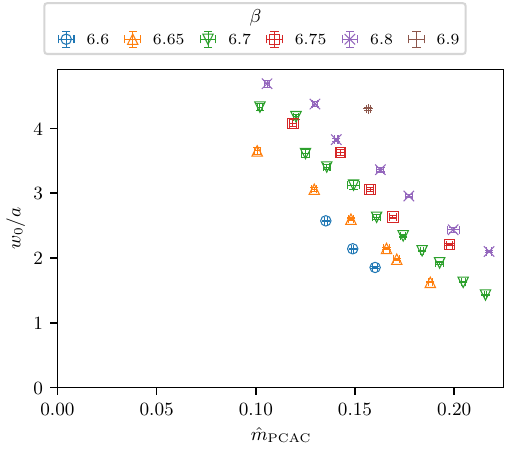}
\caption{%
\label{fig:w0vsmpcac}%
The Wilson flow scale, $w_0/a$, measured in all the ensembles considered in this work, as a function of $\hat{m}_{\rm PCAC}\equiv m_{\rm PCAC}\, w_0$, The clover-leaf definition has been adopted in the energy density, while the reference value $\mathcal{W}_0=0.28125$ has been used in the scale-setting exercise. 
The value of $\beta$ is indicated by the colour and marker, as shown in the legend.
}
\end{center}
\end{figure}

We rewrite all dimensionful quantities
in terms of the Wilson flow scale, $w_0$. We denote the resulting quantities as
$\hat{m}=m w_0$ and $\hat{f}=f w_0$, for masses and decay constants, respectively. 
As shown in \Fig{w0vsmpcac}, we find that the values of $w_0/a$ grow as $\beta$ grows,
 over the range $\beta \in [6.6,\,6.9]$ available for  this work. 
We also find evidence of a significant mass dependence in $w_0/a$, 
which is different to lattice QCD~\cite{BMW:2012hcm}, but has been observed in lattice calculations for other gauge theories with dynamical fermions~\cite{Ayyar:2017qdf,Bennett:2019cxd,Athenodorou:2021wom}. 
Following Refs.~\cite{Ayyar:2017qdf,Bennett:2019cxd}, 
throughout this work we
adopt a mass-dependent scale-setting scheme, and rescale all  dimensional quantities using the value of $w_0/a$ as measured in the individual ensembles. 
The implications 
for our 
spectral measurementss are discussed in Sect.~\ref{Sec:mesonN} and Appendix~\ref{Sec:largemass}.

\subsection{\label{Sec:topology}Topology}

  \begin{figure*}[t]
\begin{center}
\includegraphics{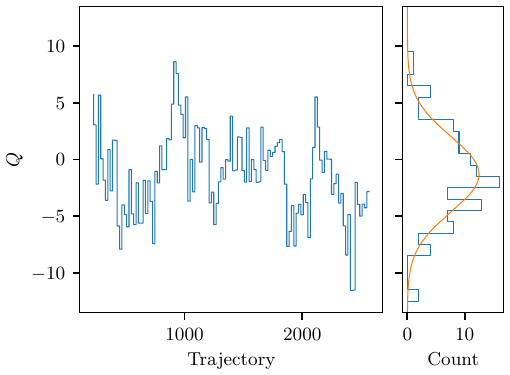}\hfill
\includegraphics{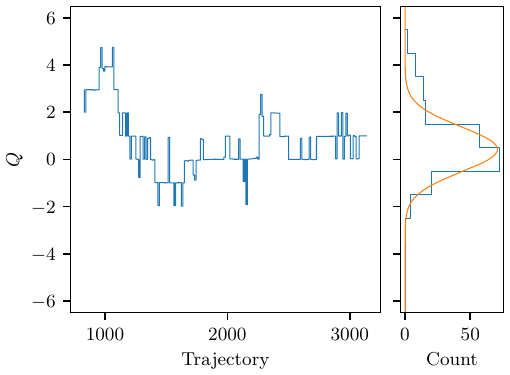}\\
\includegraphics{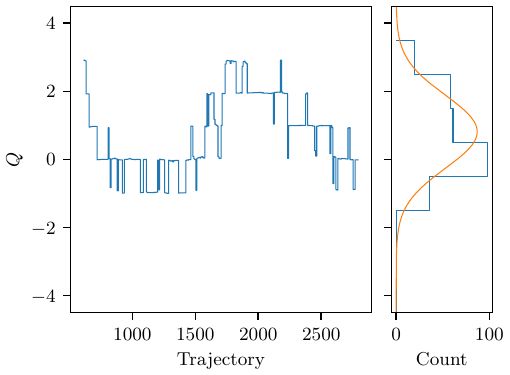}\hfill
\includegraphics{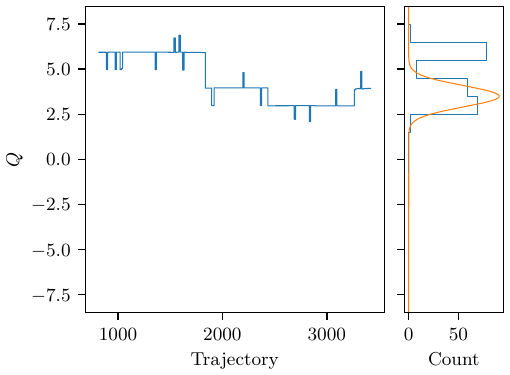}
\caption{%
\label{fig:Q_history}%
Examples of the history of the topological charge, $Q$, along the (R)HMC trajectories.
The lattice parameters, $(\beta,\,a m_0)$, used for the figures are $(6.7,-1.045)$ (top-left), $(6.7,-1.067)$ (top-right), $(6.8,-1.035)$ (bottom-left), and $(6.8,-1.046)$ (bottom-right). For each figure we also present an histogram of
the distribution of  $Q$, superimposed onto  a Gaussian fit of the results. 
}
\end{center}
\end{figure*}

The approach to the continuum limit of lattice calculations is affected by the critical slowing down 
of  the local update algorithms, 
such as the HMC and RHMC used in this work. 
 The topological charge, $Q$, is  particularly sensitive to this problem. Its autocorrelation time 
rapidly grows as the lattice spacing decreases, 
and eventually becomes one of the longest ones in the system, as shown in Ref.~\cite{DelDebbio:2002xa} for $SU(N_c)$ and Ref.~\cite{Bennett:2022ftz} for $Sp(N_c=2N)$ gauge theories. 
If one keeps reducing the lattice spacing naively, asymptotically the algorithm 
 gets trapped into one topological sector,
rather than sampling the space, and violates ergodicity conditions.  

This phenomenon of {\it topological freezing} may  introduce  sizable systematic effects
 in physical observables. 
For instance, 2-point correlation functions of flavor-singlet pseudoscalar mesons  are expected to receive corrections of ${\cal O}(Q^2/V^2)$, with $V$ the lattice volume, at large Euclidean time~\cite{Aoki:2007ka}. 
Since our primary interest is in the spectrum of flavored mesons, the problem might be less severe.
Nevertheless, we do monitor the topological charge of our Monte Carlo trajectories, 
 to assess whether any physical measurements are performed with ensembles that are
affected by  topological freezing.

Because the gradient flow is  a diffusion process which essentially smoothens out the quantum fields and thus suppresses UV fluctuations, 
we take advantage of it to study the topological charge  on the lattice, $Q_L(t)\equiv \sum_x q_L(x,t)$, which is otherwise difficult to compute~\cite{Luscher:2010iy}.
We define the lattice topological charge density, $q_L(x,t)$,  in terms of
the  clover-leaf operator, $\mathcal{C}_{\mu\nu}(x,t)$, obtained from 
the flowed gauge fields, $V_\mu(x,t)$:
\beq
q_L(x,t)\equiv \frac{1}{32\pi^2}\epsilon^{\mu\nu\rho\sigma} \,{\rm Tr} \,\mathcal{C}_{\mu\nu}(x,t) \,\mathcal{C}_{\rho\sigma}(x,t)\,.
\eeq
As the fields, $V_{\mu}$, flow along the flow time, $t$, the values assumed by $Q_L(t)$ approach quasi-integer numbers, and their distribution displays a clear 
separation into different topological sectors. For large enough $t$,
the measurement  of $Q_L(t)$ becomes independent of $t$.
In practice, we evaluate it at $t=t_Q\equiv L^2/32$. 
For notational convenience, we drop the subscript $L$ from $Q_L$, for the rest of this section,
being it understood that our topology measurements are always performed on the lattice.

In \Fig{Q_history}, we show examples of the history of the topological charge, $Q$, 
along the (R)HMC trajectory, for four representative ensembles. In the figure,
going from left to right we reduce the fermion mass, while holding fixed $\beta$ value, and from top to bottom we increase the value of $\beta$. 
We also perform  a Gaussian fit of the histogram distribution of $Q$, and display 
the results of the maximum likelihood analysis, by using the fit function
\beq
n(Q)\equiv A_n {\rm exp}\,\left(-\frac{(Q-Q_0)^2}{2\sigma_Q^2}\right)\,,
\eeq
in which $Q_0$ and $\sigma_Q$ are the mean and the standard deviation, respectively. 
Our results for $Q_0$ and $\sigma_Q$ are summarised in \Tab{topology}. Except for a few light ensembles at larger $\beta$ values, we find that the distribution of topological charge in the ensemble reproduces the expected Gaussian around zero.

The examples in \Fig{Q_history} show long autocorrelation in the history of the topological charge along the trajectories. We estimate the autocorrelation time, $\tau^{Q}_{\rm exp}$, from exponential fit to the autocorrelation function for $Q$. 
In \App{autocorrelation}, we present the results for $\tau^{Q}_{\rm exp}$ in \Tab{autocorr}.  
Because in each ensemble the fermion mass is different, one must use caution in  making comparisons 
between different lattice couplings, 
yet the general trend is that the autocorrelation time, $\tau^{Q}_{\rm exp}$, increases while going towards finer lattices. 
Furthermore,  $\tau^{Q}_{\rm exp}$ rapidly grows, for fixed $\beta$, as the fermion mass decreases.
In the cases of light and fine ensembles, we find that the topological charge is almost frozen and the estimate of $\tau^{Q}_{\rm exp}$ is comparable in size to that of the available ensemble.
The mass dependence of our results agrees with what we found in
 the calculation of the gradient flow scale, $w_0/a$, 
as well as its autocorrelation time, $\tau^{w_0}_{\rm exp}$: 
$\tau^{Q}_{\rm exp}$ and $\tau^{w_0}_{\rm exp}$ rapidly grow as the lattice becomes
 finer and/or the fermions are lighter.

\begin{table}[th!]
\caption{
Interpolating operators, $\mathcal{O}_M$, for mesons with spin $J=0,\,1$, and parity $P=\pm1$, 
built  of Dirac fermions in the antisymmetric representation, $\Psi^{i\,ab}$, of $Sp(4)$.  We show explicitly the
 flavour indices $i\neq j=1,\,2,\,3$, while colour and spinor indices are implicit and summed over.
We also show the spin and parity quantum numbers, $J^P$, the  corresponding QCD
meson sourced by the analogous operator, and
 the irreducible representation  of  the unbroken global
$SO(6)$ spanned by the meson,
ignoring the $SO(6)$-singlets.
}
\label{tab:mesons}
\begin{center}
\begin{tblr}{colspec=|c|c|c|c|c|,colsep=0.5em}
\hline\hline
{\rm Label} & {\rm Interpolating} & {\rm Meson}
& 
&   \\
 & {\rm operator} &   {\rm in}
& {\rm $J^{P}$}
& $SO(6)$  \\
 $M$ & $\mathcal{O}_M$& {\rm QCD}
& &   \\
\hline
ps & $\overline{\Psi^i}\gamma_5 \Psi^j$ & $\pi$ & $0^{-}$ 
& $20^{\prime} 
$\\
s & $\overline{\Psi^i} \Psi^j$ & $a_0$ & $0^{+}$ 
& $20^{\prime}  
$\\
v & $\overline{\Psi^i}\gamma_\mu \Psi^j$ & $\rho$ & $1^{-}$ 
& $15$ \\
t & $\overline{\Psi^i}\gamma_0\gamma_\mu \Psi^j$ & $\rho$ & $1^{-}$ 
& $15 
$\\
av & $\overline{\Psi^i}\gamma_5\gamma_\mu \Psi^j$ & $a_1$ & $1^{+}$ 
& $20^{\prime}  
$\\
at & $\overline{\Psi^i}\gamma_5\gamma_0\gamma_\mu \Psi^j$ & $b_1$ & $1^{+}$ 
&
$15 
$\\
\hline\hline
\end{tblr}
\end{center}
\end{table}

\section{\label{Sec:meson} Meson spectroscopy}

This section is devoted to defining the observables relevant to our study.
We list the meson operators in the theory, the correlation functions we measure, 
and we describe the processes by which we extract  the masses and decay constants of the composite states of interest.

\subsection{\label{Sec:observable} Interpolating operators}

In \Tab{mesons}, we list  the operators considered in this work.
They are gauge invariant hyperquark bilinears, in the form $O_M (x)\equiv \overline{\Psi}^i(x) \Gamma_M \Psi^j(x)$, 
and we refer to them as  {\it mesons}, borrowing terminology from QCD. The index,
$M= {\rm ps, s, v, t , av, at}$, denotes pseudoscalar, scalar, vector, tensor, axial-vector, and axial-tensor mesons, respectively. We restrict attention to combinations with flavor indexes  $i\neq j$, 
and for each of them we specify the corresponding irreducible representations of the global (flavor) symmetry group, $SO(6)$. This choice introduces a simplification, as
we are only required to compute connected diagrams. 
For completeness, the table reports also the spin and parity quantum numbers, $J^P$,  associated to each operator.
(Charge conjugation is trivial.)

We anticipate here an observation that is going to be useful later in the paper.
The vector (v) and tensor (t) operators  carry the same $SO(6)$ quantum numbers, and thus 
 interpolate the same meson states in the continuum theory.  
As will be demonstrated by our numerical results, even on the discretised lattice the masses of these two mesons are statistically compatible. We exploit this property 
 with a GEVP analysis, that allows us to extract the first excited state in the vector meson channel.

\subsection{\label{Sec:measurement}Two-point correlation functions}

We write the zero-momentum correlation functions involving 
source meson operators, $O_{M'}$, located at $x\equiv (t,\vec{x})$,
and sink, $O_M$, at $y\equiv (t_0,\vec{y})$, 
as follows
\beq
C_{M,M'}(t-t_0)=\sum_{\vec{x},\vec{y}}
\langle O_{M}(x) O_{M'}^\dagger(y)\rangle\,.
\eeq
We omit flavour indexes, $i,\,j=1,\,\cdots,\,3$, for which we only consider the non-diagonal terms with $i\neq j$.
 After applying the appropriate Wick contractions, we write
\beq
C_{M,M'}(t-t_0)=-\sum_{\vec{x},\vec{y}}
{\rm Tr}\, [\gamma^5 \Gamma_M S_i(x,y) \overline{\Gamma}_{M'} \gamma^5 S_j^{\dagger} (x,y)]\,,
\eeq
where $\overline{\Gamma}\equiv \gamma^0\Gamma^\dagger \gamma^0$ and the trace is taken over colour and spinor indices. 
The hyperquark propagator, $S_j(x,y)$, carrying flavour, $j$, is defined by
\beq
S^{AB}_{j,\,\alpha\beta}(x,y)=\langle \Psi^{A}_{j\,\alpha}(x) \overline{\Psi}^{B}_{j\,\beta}(y)\rangle,
\eeq
where $A$, $B$ and $\alpha$, $\beta$ are (ordered) pairs of colour and spinor indexes, respectively. 
Without loss of generality, we set the initial time of the source operator to be zero,  $t_0=0$.

The spectral decomposition of two-point functions in  finite (Euclidean) space-time volume, 
$\tilde{V}$, is
 discrete.  Finite energy eigenvalues, $E_n$, are  labelled by an integer, $n=0,\,1,\,2,\,\cdots$.
At sufficiently large Euclidean times, 
 the two-point function is dominated by the contribution of the ground state, with the lowest energy, $E_0$,
while that of excited states has exponentially suppressed.
For $M=M'$, this state is the lightest meson, $|M\rangle$, with  mass $m_M$, interpolated by $O_M$. The correlation function approaches the  functional form:
\beq
C_{M,\,M}(t)\rightarrow \frac{|\langle 0 | O_M | M \rangle|^2}{2m_M} \left(
e^{-m_M t}+e^{-m_M(T-t)}
\right)\,.
\label{eq:corr_fit}
\eeq
The second term arises from the contribution of backward  propagation, with finite temporal extent, $T$. 
We then extract the meson masses and the matrix elements by fitting the lattice correlation function to \Eq{corr_fit}, 
having restricted the fit range to large Euclidean time.

The matrix elements of vector, (${\rm v}$), and axial-vector, (${\rm av}$), operators (currents),
bracketed  between the relevant one-meson state and the vacuum, 
are related to the decay constants of the corresponding mesons, $f_{\rm v}$ and $f_{\rm av}$, respectively. 
 For vanishing three-momentum, $\vec{p}=0$, they are parametrised as follows:
\beqs
\label{eq:fv}
\langle 0 | O^\mu_{\rm v} | {\rm v} \rangle &=& \sqrt{2} f_{\rm v} m_{\rm v} \epsilon^\mu\,,\\
\langle 0 | O^\mu_{\rm av} | {\rm av} \rangle &=& \sqrt{2} f_{\rm av} m_{\rm av} \epsilon^\mu\,,
\label{eq:fav}
\eeqs
where $\epsilon_\mu$ is the polarisation four-vector 
obeying the defining relations $p_\mu \epsilon_\mu=0$ and $\epsilon^{*}_{\mu} \epsilon^{}_\mu=1$.

The wave-functions of the pseudoscalar mesons overlap with the axial-vector current.
We adopt  the following definition for the pseudoscalar decay constant, $f_{\rm ps}$:\footnote{Analogous conventions and normalisations yield, in QCD, 
 the pion decay constant,  $ f_\pi \simeq 93\,{\rm MeV}$, in lieu of $f_{\rm ps}$. }
\beq
\langle 0 | O^0_{\rm av} | {\rm ps} \rangle = \sqrt{2} f_{\rm ps} m_{\rm ps}\,.
\label{eq:fps}
\eeq
To determine $f_{\rm ps}$, we calculate $C_{\rm ps,\,ps}(t)$,  as well as the following correlation function:
\beq
C_{\rm av,\,ps}(t) \rightarrow \frac{f_{\rm ps} \langle 0|O_{\rm ps} | {\rm ps} \rangle^*}{\sqrt{2}} \left(
e^{-m_{\rm ps} t} -e^{-m_{\rm ps}(T-t)}
\right)\,.
\eeq
By combining these correlation functions, $C_{\rm ps,\, ps}$ and $C_{\rm av,\, ps}$, and exploiting
the partially-conserved-axial-current (PCAC) relation, we also define  the  PCAC mass, $a m_{\rm PCAC}$ (see  \App{mpcac}), which provides a useful  
definition of the hyperquark mass, in place of the
(additively renormalised) Wilson bare mass, $a m_0$.

The decay constants, extracted from lattice matrix elements,
must be renormalised and matched to their continuum counterparts. 
We adopt a renormalisation procedure based on 
 the analytical evaluation of $1$-loop integrals in lattice perturbation theory with Wilson fermions, in the ${\overline{\rm MS}}$ scheme~\cite{Martinelli:1982mw}.
The conversion factors
from lattice results, $f^{\rm latt}_M$, to  continuum ones, $f^{\rm con}_M$,  read as follows:
\beqs
f^{\rm con}_{\rm ps\,(av)}&=& \left(1+C_{\rm R}(\Delta_{\Sigma_1}+\Delta_{\gamma_\mu\gamma_5})\frac{\tilde{g}^2}{16\pi^2}
\right)f^{\rm latt}_{\rm ps\,(av)}\,, \\
f^{\rm con}_{\rm v}&=& \left(1+C_{\rm R}(\Delta_{\Sigma_1}+\Delta_{\gamma_\mu})\frac{\tilde{g}^2}{16\pi^2}
\right)f^{\rm latt}_{\rm v}\,.
\eeqs
In these relations, the eigenvalue of the quadratic Casimir operator for antisymmetric fermions in the $Sp(4)$ gauge theory is $C_{\rm R=as}=2$. 
The first numerical factor, $\Delta_{\Sigma_1}=-12.82$, arises from wave-function renormalisation, 
while $\Delta_{\gamma_\mu}=-7.75$ and $\Delta_{\gamma_\mu\gamma_5}=-3.0$ descend from  vertex renormalisation. 
Perturbative matching can further be improved, as discussed in Ref.~\cite{Lepage:1992xa}, by using as definition of the effective coupling  the combination $\tilde{g}^2\equiv g^2/\langle P \rangle$, instead of the bare gauge coupling, where $\langle P \rangle$ is the average plaquette. 
This approach is effectively equivalent to a mean-field approximation, in which the contributions of tadpole diagrams, which are absent in the continuum theory, are subtracted from the gauge links.

Our numerical calculation of  $2$-point correlation functions adopts
 two different and complementary strategies in
the construction of source and sink  operators. Both procedures
 are well established in the lattice QCD literature.
We  have developed and tested these implementations in the present work. 

We first consider  $Z_2\otimes Z_2$ stochastic wall sources, with the one-end trick---see, e.g., Ref.~\cite{Boyle:2008rh}---implemented in the HiRep code. 
In comparison with using naive point sources (delta functions), this choice 
allows to improve the signal,  and determine masses and decay constants with higher precision. 
We set the number of hits to $3$, for all meson correlation functions and for all the ensembles listed in \Tab{ensemble}. 
We verified that this strategy leads to
a reduction of statistical noise and the appearance of a  cleaner plateau in the effective mass plots at large Euclidean time, compared to the point source, but
 for the same computational cost.

Having completed these measurements, we select a subset of our ensembles that 
are useful for 
our first extrapolation of the meson spectroscopy observables towards the continuum and massless limits. 
We repeat the measurement of meson masses in these ensembles, by applying 
 a more sophisticated combination of two noise reduction 
techniques,  referred to as APE smearing~\cite{APE:1987ehd,Falcioni:1984ei} and Wuppertal smearing~\cite{Gusken:1989qx,Roberts:2012tp,Alexandrou:1990dq}, respectively.
Both smearing procedures introduce 
non-nearest-neighbour interactions, and 
some degree of non-locality in the measurement process, as we shall explain.

 APE smearing is a procedure that  smoothens out the UV fluctuations of the gauge links,
  by combining  staples and  gauge link iteratively~\cite{APE:1987ehd}.
The smearing procedure is controlled by two parameters, 
the smearing step-size, $\alpha_{\rm APE}$, 
and the number of iterations, $N_{\rm APE}$, the choices of which have to be optimised.

Gauge covariant Wuppertal smearing, by contrast, acts on the hyperquark fields 
and composite operators,
by replacing point-like sources and sinks with extended configurations, 
defined through an iterative diffusion process~\cite{Gusken:1989qx}. 
Doing so increases the  overlap between 
smeared  interpolating operators and meson eigenstates of interest, 
by suppressing contributions to correlation functions coming from other states. 
This procedure requires optimising (on a channel-by-channel basis, in principle), its
step-size, $\epsilon_W$, 
and the number of iterations, $N_{\rm source}$ and $N_{\rm sink}$,
with the aim of maximising the overlaps, without introducing unwanted systematic errors.

A typical consequence of smearing is that  the plateau in the effective mass plots start at earlier  Euclidean time. 
This is particularly useful in measuring the properties of comparatively heavy states,
for which one loses signal  into noise, at large time. 
The increase in fitting range 
also yields a reduction of
  statistical uncertainties.

A drawback of the adoption of   smeared operators is that their non-local nature 
alters the correlation functions in respect to those built of local operators in \Eq{corr_fit}. 
One can still extract the decay constants in Eqs.~(\ref{eq:fv})-(\ref{eq:fps}) by performing a simultaneous fit 
to  smeared-smeared and  smeared-point correlation functions, but one
expects little to no numerical gain, in terms of noise reduction, in doing so.
 Hence, in our measurement of meson decay constants, we revert to 
 $Z_2\otimes Z_2$
 stochastic wall sources. 

Both APE and Wuppertal smearing have been recently implemented in the HiRep code\footnote{
 Specifically, commit ID \texttt{1b204b6}~\cite{hirep:sa2c_fzieler}.
 }, 
 extensively tested and used for the measurements of spin-$1/2$ chimera baryons in quenched $Sp(4)$
 gauge theories~\cite{Bennett:2023mhh,hirep:sa2c}. 
 Their use in the study of
 connected diagram contributions to flavour-singlet meson correlation functions
  in the $Sp(4)$ theory coupled to $N_{\rm f}=2$ fundamental and $N_{\rm as}=3$ antisymmetric 
fermions dramatically improves noise control, and makes it possible to perform measurements in this sector of the theory~\cite{Bennett:2024wda}. 
Application to the same theory further provides a supporting tool for 
the spectral density approach to lattice spectroscopy~\cite{Bennett:2024cqv}.

 \begin{figure}[t]
\begin{center}
\includegraphics{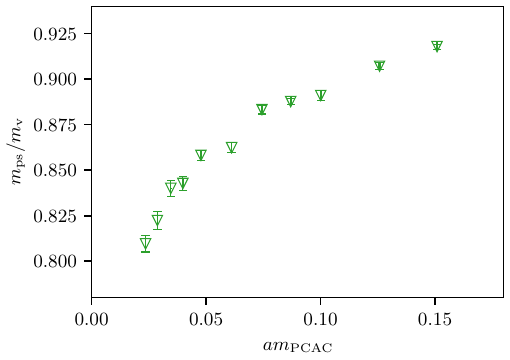}
\caption{%
\label{fig:mvmps_b6p7}%
The ratio between masses of ground-state  pseudoscalar and vector mesons, $m_{\rm ps}/m_{\rm v}$,
extracted using $Z_2 \otimes Z_2$ stochastic wall sources in the definition of the correlation functions. The lattice coupling, $\beta=\SpectrumPlotTargetBeta$,
is held fixed, while the 
bare mass of the hyperquarks, $a m_0$, varies, and so does the PCAC hyperquark mass, $a m_{\rm PCAC}$.
}
\end{center}
\end{figure}

Finally, as anticipated  in \Sec{observable}, 
the vector, ${\rm v}$,  and tensor, ${\rm t}$, operators  interpolate the same physical states.
We exploit this observation to optimise the extraction 
of the first excited state, 
 by formulating (and solving)  a GEVP, taking  the form
\beq
\mathcal{C}(t) v_n(t,t_1) = \lambda_n(t,t_1) \mathcal{C}(t_1) v_n(t,t_1)\,,
\eeq
where the $2\times 2$ matrix-valued correlation function, $\mathcal{C}(t)$, is defined by
\beqs
\mathcal{C}(t) \equiv \left(
\begin{array}{cc} C_{\rm v,v}(t) & C_{\rm v,t}(t) \\
C_{\rm t,v}(t) & C_{\rm t,t}(t) \end{array}
\right)\,.
\eeqs
For fixed $t_1=1$, and choosing $t>t_1$, the resulting eigenvalues, $\lambda_0(t,t_1)$ and $\lambda_1(t,t_1)$,
exhibit single-exponential decays 
at large Euclidean time,  with the decay rates measured by the masses of ground state and first excited state of the system, respectively. 

\section{\label{Sec:results}Numerical results}

In this section we report  our measurements of masses and decay constants, for the flavored mesons sourced by the operators in \Tab{mesons}. 
A first, preliminary analysis of the fermion mass dependence of the relevant observables allows us to identify two distinct regions
of parameter space explored by our ensembles, 
at lower and higher masses, respectively, that
exhibit different dynamical properties.
We then restrict our attention to ensembles in the lower-mass region, 
within which the mass dependence of the observables is compatible with expectations from 
Wilson chiral perturbation theory, truncated at the next-to-leading order (NLO).
In the resulting, restricted set of ensembles, we refine our measurements, fix the scale with the gradient flow method discussed in \Sec{scale},
 and perform the first continuum and massless extrapolations of the spectroscopic observables in the theory. 
 We critically discuss our numerical results, by comparing them to the quenched studies reported in Ref.~\cite{Bennett:2019cxd}.

\begin{figure}[t]
\begin{center}
\includegraphics{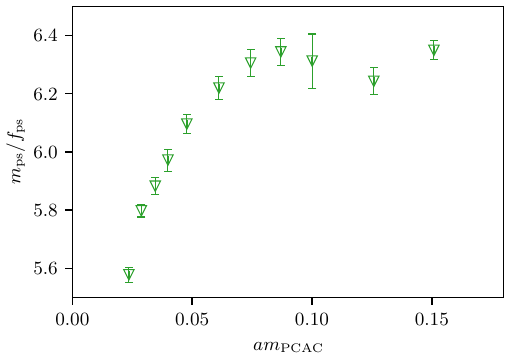}
\caption{%
\label{fig:mpsfps_b6p7}%
The ratio between  mass and decay constant of  ground-state  pseudoscalar  mesons, $m_{\rm ps}/f_{\rm ps}$,
extracted using $Z_2 \otimes Z_2$ stochastic wall sources in the definition of the correlation functions. The lattice coupling, $\beta=\SpectrumPlotTargetBeta$, 
is held fixed, while the 
bare mass of the hyperquarks, $a m_0$, varies, and so does the PCAC hyperquark mass, $a m_{\rm PCAC}$.
}
\end{center}
\end{figure}
 
We deploy a GEVP analysis to extract the mass of the ground and first excited states
of the vector mesons,  the flavored spin-$1$ composite states with negative parity.
We also measure the ratio
$m_{\rm v}/f_{\rm ps}$, and compare our results to other related theories.
If one takes at face value the phenomenological 
Kawarabayashi-Suzuki-Riazuddin-Fayyazuddin (KSRF) relation,
$m_{\rm v}^2=2g_{\rm v\,ps\,ps}^2f_{\rm ps}^2$~\cite{Kawarabayashi:1966kd,Riazuddin:1966sw},
these measurements provide a first,
naive estimate  of the effective coupling, $g_{\rm v\,ps\,ps}$,  associated with the decay of vector into pairs of pseudoscalar mesons.

\begin{figure*}[t]
\begin{center}
\includegraphics{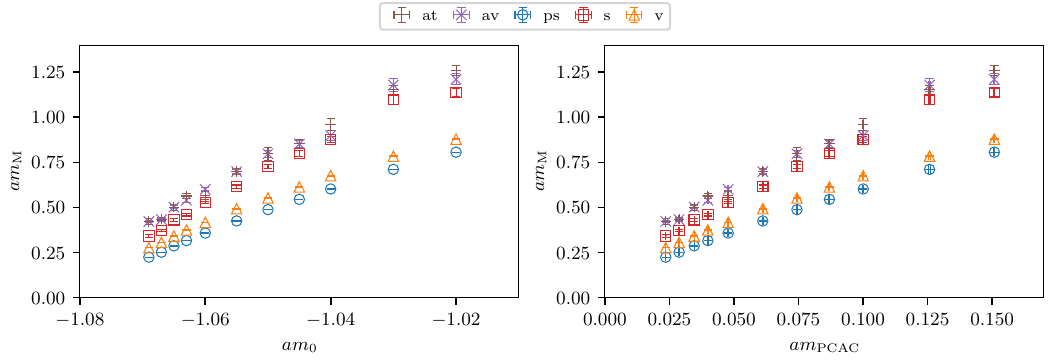}
\caption{%
\label{fig:mass_b6p7}%
Masses of flavored mesons, $a m_{M}$, extracted using $Z_2 \otimes Z_2$ stochastic wall sources in defining the correlation functions, in ensembles with fixed lattice coupling, $\beta=\SpectrumPlotTargetBeta$, and varying  Wilson-Dirac mass of the hyperquarks, $a m_0$ (left panel), or, equivalently,  the PCAC hyperquark mass, $a m_{\rm PCAC}$ (right panel). 
The colour and marker in the legend denote measurements with $M={\rm ps,\, v,\, s,\, av,\,at}$ for pseudoscalar, vector, scalar, axial-vector, and axial-tensor mesons.
}
\end{center}
\end{figure*}

\subsection{\label{Sec:mesonN}Masses and decay constants}

We start by discussing measurements extracted from correlation functions defined 
with $Z_2 \otimes Z_2$ stochastic wall sources.
The complete spectroscopic results are reported in \App{data}, for all the ensembles listed in \Tab{ensemble}. 
In this section, we restrict our discussion to the ensembles with lattice coupling $\beta=\SpectrumPlotTargetBeta$,
for convenience.
Similar considerations to those exposed in the following apply to other ensembles as well.

In  a typical non-Abelian gauge theory, 
in which confinement appears at long distances, and is accompanied by the formation of a bilinear fermion condensate,
with the associated spontaneous breaking of approximate global symmetries,
 one expects chiral perturbation theory  ($\chi$PT) to apply.
The pseudoscalar meson is the lightest particle in the spectrum, and its mass is expected to approach zero with the functional form
 $m_{\rm ps} \propto \sqrt{m_f}$, where $m_f$ stands for a measurement of the fermion mass.
 
 By contrast, in a theory with IR-conformal dynamics, in which the massless theory does not confine, but rather flows into a new, non-trivial fixed point of the renormalisation group flow at long distances,
  all the meson masses are expected to approach zero with the same exponent,  $m_M \propto (m_f)^{1/(1+\gamma^*)}$, with $\gamma^*$ an  anomalous dimension, measured at the IR fixed point~\cite{DelDebbio:2010ze}.

We start by testing the hypothesis of IR conformality.
We focus on the pseudoscalar and vector mesons, 
as the measurements of their mass and decay constant are the most precise available to us.
In \Fig{mvmps_b6p7}, we plot the  ratio between the pseudoscalar and vector meson masses, $m_{\rm ps}/m_{\rm v}$. 
In the available ensembles, the smallest value of this ratio is approximately $m_{\rm ps}/m_{\rm v}\sim 0.8$, for the lightest ensemble, which is quite far away from the opening of the kinematical  threshold for vector decay to two pseudoscalars.
In a IR-conformal theory  this ratio should be independent of the fermion mass; conversely
we find clear evidence of a variation of this ratio, as the hyperquark mass (represented here by
 $a m_{\rm PCAC}$) changes.
The ratio decreases as the hyperquark mass is reduced, and furthermore, the slope of the resulting curve becomes steeper at lower masses.

We show the numerical results for the ratio of the mass and decay constant of the pseudoscalar meson, $m_{\rm ps}/f_{\rm ps}$, 
 in \Fig{mpsfps_b6p7}. We find   evidence of two distinct dynamical regimes. 
For $am_{\rm PCAC}\gtrsim 0.07$, our measurements form 
a plateau at an approximately constant value,
$m_{\rm ps}/f_{\rm ps}\sim 6.3$.
However, this ratio sharply drops for $am_{\rm PCAC}\lesssim0.07$. 
Combined with our evidence of  significant mass dependence for the ratio $m_{\rm ps}/m_{\rm v}$, the results for the lighter ensembles contradict the hyperscaling hypothesis, and we infer that the theory in the limit of vanishing hyperquark mass is not IR conformal.

 \begin{figure*}[t!]
\begin{center}
\includegraphics{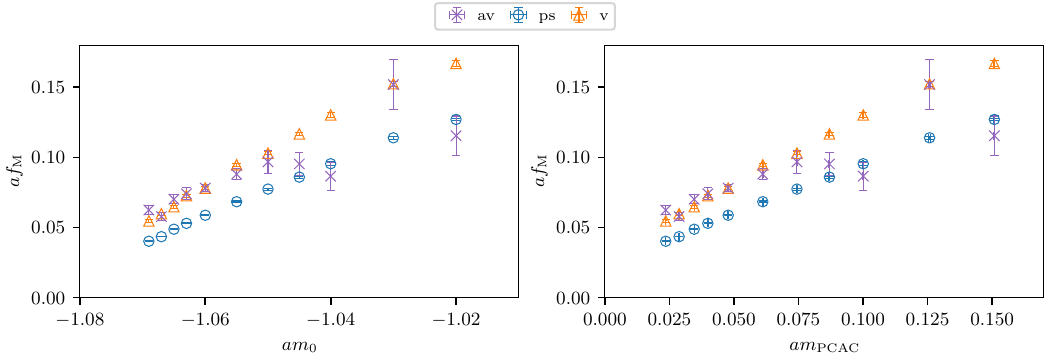}
\caption{%
\label{fig:decay_b6p7}%
Decay constants of flavored mesons, $a f_{M}$, extracted using $Z_2 \otimes Z_2$ stochastic wall sources in the definition of the correlation functions. We display only results obtained in ensembles with lattice coupling $\beta=\SpectrumPlotTargetBeta$, and vary the 
Wilson-Dirac mass of the hyperquarks, $a m_0$ (left panel), or, equivalently,  the PCAC hyperquark mass, $a m_{\rm PCAC}$ (right panel). 
The colour and marker in the legend denote measurements with $M={\rm ps,\, v,\, av}$ for pseudoscalar, vector, axial-vector mesons.
}
\end{center}
\end{figure*}

We have evidence  in our  numerical results suggesting  that in the limit of massless hyperquarks of the theory,  confinement is taking place, with spontaneous breaking of the (approximate) non-Abelian global symmetry.\footnote{As we have clear evidence of symmetry breaking, even in naive extrapolations to the massless limit, 
we do not discuss  symmetric mass generation---see Ref.~\cite{Butt:2024kxi} and references therein.}
The resulting low-energy spectrum contains light pseudoscalar meson states identified as the pseudo Nambu-Goldstone bosons. 
Yet, despite the wide range of hyperquark masses we explored, 
our results suggest that we expect to see significant deviations from the predictions of
 leading order (LO) chiral perturbation theory, as none of the ensembles are close
 enough to the massless limit. We devote the next paragraphs to presenting this evidence
and discussing its implications.

 In \Fig{mass_b6p7}, we present the masses of flavored mesons as a function both of the bare hyperquark mass, $a m_0$,
and of the alternative  PCAC mass.
As expected, over the range of masses considered, the lightest state in the spectrum is the pseudoscalar meson, followed by the vector meson. The tensor mesons are almost degenerate with the vectors, and thus we omit them from the plots, for simplicity. 
The scalar, axial-vector and axial-tensor flavored mesons are heavier, and 
the correlation functions from which they are extracted are affected by
significant amounts of statistical noise, which is reflected in larger errors in the measurements.
The overall trend displayed in \Fig{mass_b6p7} 
is that the masses grow approximately linearly with the hyperquark mass
for $a m_{\rm PCAC}\lesssim 0.07$, 
but show  deviations from such trend for larger fermion masses,
$a m_{\rm PCAC}\gtrsim 0.07$.

The decay constants for the pseudoscalar, vector and axial-vector mesons are displayed in \Fig{decay_b6p7}. 
The pseudoscalar decay constant, $f_{\rm ps}$, is consistently smaller than the vector one, $f_{\rm v}$. 
The measurements for
 the axial-vector decay constants are affected by larger uncertainties, yet their value is close to
 that of the vectors, when $a m_{\rm PCAC}\lesssim 0.07$. 
All  three decay constants grow approximately linearly with the hyperquark mass, and seem to approach a non-vanishing value if naively extrapolated towards the massless limit. This observation further disfavours the hypothesis that the continuum, massless theory is IR-conformal, and is consistent with the expectations of a confining theory with spontaneous breaking of approximate global symmetries.

\begin{figure}[t]
\begin{center}
\includegraphics{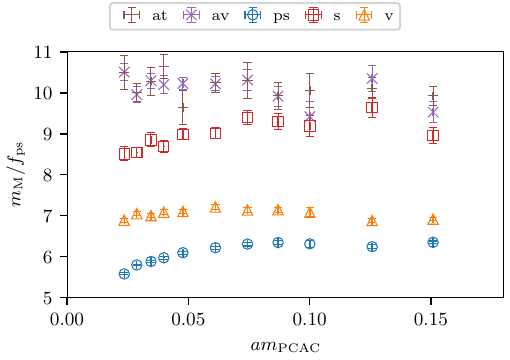}
\caption{%
\label{fig:mmfps_b6p7}%
Mesons masses, expressed in units of the pseudoscalar decay constant, $f_{\rm ps}$,  extracted using $Z_2 \otimes Z_2$ stochastic wall sources in the definition of the correlation functions.
 We display only results obtained in ensembles with lattice coupling $\beta=\SpectrumPlotTargetBeta$, and vary the 
Wilson-Dirac mass of the hyperquarks, $a m_0$, or, equivalently,  the PCAC hyperquark mass, $a m_{\rm PCAC}$.
The colour and marker in the legend denote measurements with $M={\rm ps,\, v,\, s,\, av,\,at}$ for pseudoscalar, vector, scalar, axial-vector, and axial-tensor mesons.
}
\end{center}
\end{figure}

In \Fig{mmfps_b6p7} we combine our measurements
of the masses of  ${\rm ps}$, ${\rm v}$, ${\rm s}$, ${\rm av}$, and ${\rm at}$  mesons, and present them
in units of the decay constant of the ${\rm ps}$ meson, $f_{\rm ps}$. 
We display  the results as a function of $a m_{\rm PCAC}$.
The figure highlights the contrast between the scaling properties of the ${\rm ps}$ meson, 
and all the other mesons. 
This observation further enforces the conclusion that the massless extrapolation of this theory is chirally broken, not IR conformal, and the ${\rm ps}$ mesons are indeed (light) Goldstone bosons. This therefore justifies the adoption of an analysis strategy that assumes confinement and spontaneous symmetry breaking of (approximate) global symmetry.

\begin{figure}
\begin{center}
\includegraphics{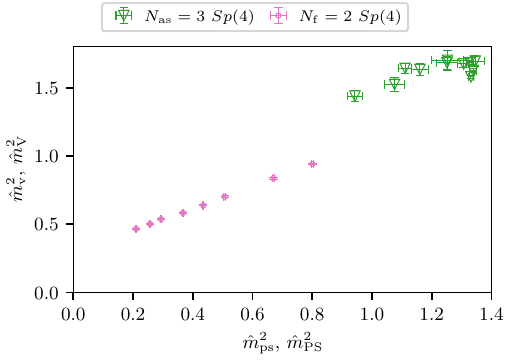}
\caption{%
\label{fig:m2vm2ps_b6p7}%
The square of the masses of vector mesons in the $Sp(4)$ theory coupled to $N_{\rm as}=3$ Dirac fermions transforming in the antisymmetric representation, $m_{\rm v}^2$, as a function of the pseudoscalar masses, $m_{\rm ps}^2$ (along the abscissa), and the $Sp(4)$ theory with $N_{\rm f}=2$ fermions transforming in the fundamental, 
$m_{\rm V}^2$,  as a function of the pseudoscalar masses, $m_{\rm PS}^2$ (along the abscissa).
Numerical data on the $N_{\rm f}=2$ theory with the fixed lattice coupling of $\beta=7.2$ is taken from Ref.~\cite{Bennett:2019jzz}, and the measurements displayed for the $N_{\rm as}=3$ theory have been performed on ensembles with $\beta=\SpectrumPlotTargetBeta$.
 Both vector and pseudoscalar masses are expressed in units of the Wilson flow scale, $w_0$.
}
\end{center}
\end{figure}

In \Fig{m2vm2ps_b6p7}, we show a comparison between measurements in the $Sp(4)$ theory with $N_{\rm as}=3$ antisymmetric fermions (for fixed $\beta=\SpectrumPlotTargetBeta$),
and those reported in Ref.~\cite{Bennett:2019jzz}, for the $Sp(4)$  theory with $N_{\rm f}=2$ fundamental fermions (for fixed $\beta=7.2$). We display the mass squared of 
 the vector mesons, in the two theories, denoted by $m_{\rm v}^2$ and $m_{\rm V}^2$, 
 as a function of the mass squared of the pseudoscalar mesons,  $m_{\rm ps}^2$ and $m_{\rm PS}^2$, respectively. The masses are expressed in units of the gradient flow scale, $w_0$. 
In the $N_{\rm f}=2$ theory, $\hat{m}_{\rm V}^2$ clearly displays a linear dependence on $\hat{m}_{\rm PS}^2$, 
over a wide range of different values for the masses. 
In the case of the $N_{\rm as}=3$ theory, on the other hand, the figure shows that 
both the pseudoscalar and vector mesons are heavier,
in respect to the $N_{\rm f}=2$ case, and hence the extrapolation towards the massless limit is less reliable. 
We note that, when we express the mass measurements in lattice units, both data sets span the similar mass ranges, $0.04 \lesssim (a m_{\rm ps,\,PS})^2 \lesssim 0.7$, and can be shown to be well described by polynomial functions with an appropriate massless limit, as discussed with our preliminary results in Ref.~\cite{Hsiao:2022gju}. 
However, care should be taken in performing the massless extrapolation, setting the scale with $w_0$ highlights that the physical scale of the masses is far away from the massless limit.

Furthermore, only for the lightest ensembles, those with $a m_{\rm PCAC}\lesssim 0.07$,
 does one see evidence of an increase of 
$\hat{m}_{\rm v}^2$ with $\hat{m}_{\rm ps}^2$.
Conversely, an unexpected qualitative behaviour appears for the measurements in heavier ensembles, those with $a m_{\rm PCAC}\gtrsim 0.07$, which cluster together in a small region of parameter space. We conservatively discard these heaviest ensembles from the rest of the subsequent analysis, attributing their behaviour to 
artefacts of the lattice theory, when used far from the continuum and massless limits.

The mass dependence of the Wilson flow governs the qualitative changes
observed in \Fig{m2vm2ps_b6p7}. Ultimately, this phenomenon descends from the dynamics of the theory.
We have presented convincing evidence supporting an interpretation of long distance data in terms of
confinement with spontaneous symmetry breaking.
Yet, the ratios $m_{\rm ps}/m_{\rm v}$ and $m_{\rm ps}/f_{\rm ps}$ vary at most by about $15\%$, 
while $a m_{\rm ps}$ spans a wide range, with the largest values
 about $3\div 4$ times larger than the smallest available ones. 
The slow evolution of dimensionless ratios,
visible even at smaller hyperquark masses, might be a consequence of the non-trivial 
properties associated with near-conformal dynamics. We relegate to  \App{largemass} a  more extensive set of results, in particular by including  measurements obtained at different
values of  $\beta$.

Despite all these cautionary remarks, the six lightest ensembles with $\beta=\SpectrumPlotTargetBeta$ do exhibit the expected behaviour, 
$\hat{m}_{\rm v}^2$ depending linearly on $\hat{m}_{\rm ps}^2$, as shown in \Fig{m2vm2ps_b6p7}. 
Similar results hold for the other available ensembles.
Based on our investigation of the ensembles 
with $\beta=6.65$ and $\beta=6.7$, we consistently find that the aforementioned linear behaviour appears when
$a m_{\rm ps}\lesssim \HeavyPSMassLimit$, and we apply this upper bound to restrict the available ensembles,
even those for which we do not have enough independent measurements to carry out a fixed-$\beta$ analysis.
With the resulting, restricted set of ensembles, we 
repeat and refine the measurements by implementing noise-reduction techniques (smearing), 
and perform a first  massless and continuum extrapolation.

We close this subsection with a technical, but important,  comment on the limitations of our calculations. 
As we saw in \Fig{mpsfps_b6p7}, there appears to be a maximum value 
 in the ratio of $m_{\rm ps}/f_{\rm ps}$. Similarly, the masses of composite states, expressed  in units of the gradient flow scale, $w_0$,
 are bounded from above, as shown in \Fig{m2vm2ps_b6p7}. 
 These upper bounds can be exceeded only if the lattice coupling increases, as shown in \App{largemass}. 
Alas, numerical studies of finer lattices are hindered by the difficulties related with topological freezing. 
 In the opposite direction, exploring small masses, $\hat{m}_{\rm ps}$, requires using larger lattice sizes,
 in order to overcome finite volume effects. This is not feasible, at the present time,  with realistic computing resources.
 In summary, the ensembles we reported here  provide
 a good representation of the region of parameter space that is accessible to us with the technology deployed for this study.
Compared to the $Sp(4)$ theory with $N_{\rm f}=2$, such a region is far more restricted,
and, to gain further numerical progress in the $N_{\rm as}=3$ theory, we will require
to improve  the action, possibly by adopting also a different formulation for the fermions.

  \begin{figure*}
\begin{center}
\includegraphics{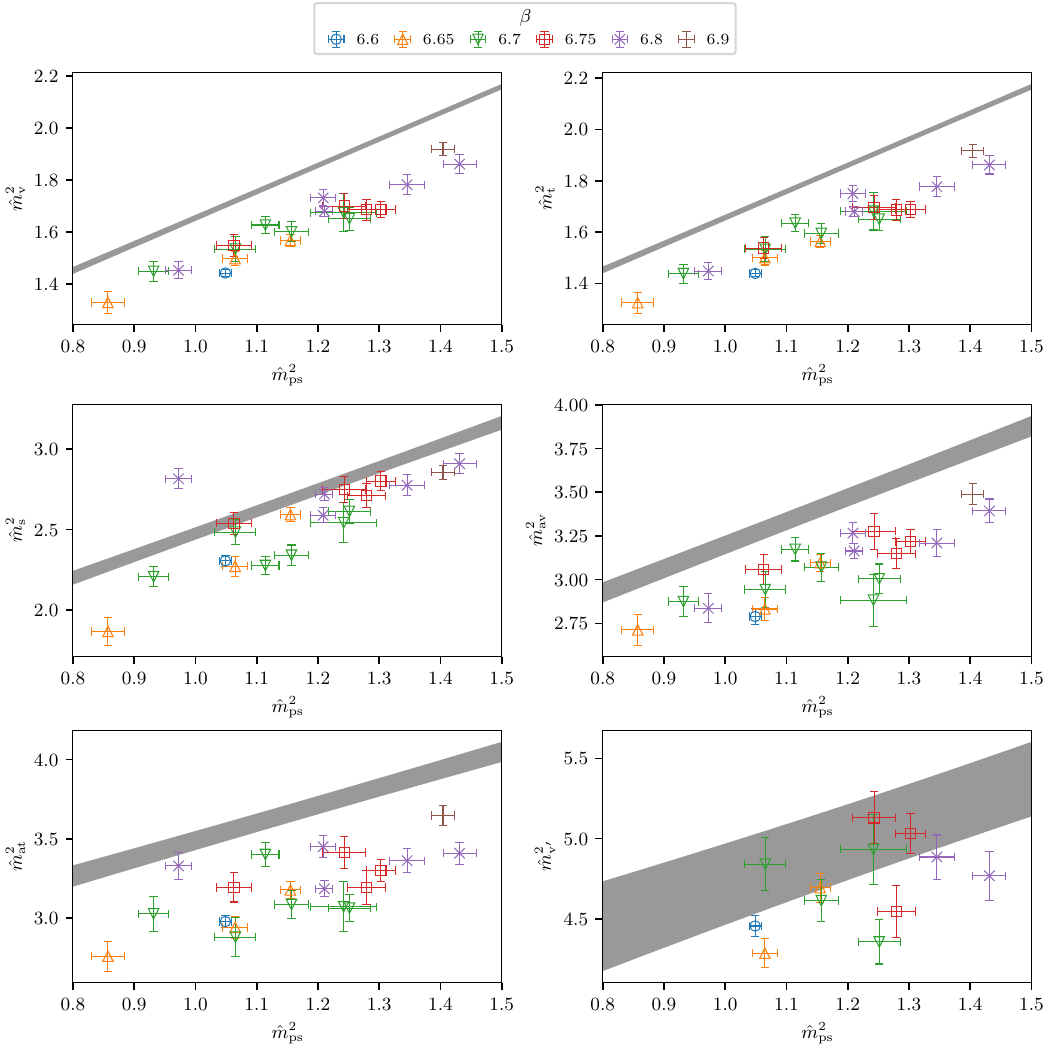}
\caption{%
\label{fig:mass_con}%
Square of the meson masses, $\hat{m}_M^2$, expressed in units of the Wilson flow scale, $w_0$,
plotted as a function of the mass squared  of the pseudoscalar meson,
$\hat{m}_{\rm ps}^2$, in
the $Sp(4)$ lattice gauge theory  coupled to $N_{\rm as}=3$ dynamical fermions.
These measurements  are obtained implementing Wuppertal and APE smearing,
and are restricted to ensembles satisfying the upper bound
$a m_{\rm ps}\lesssim \HeavyPSMassLimit$.
Top to bottom, and left to right, we display the individual measurements of $\hat{m}^2_{\rm v}$, $\hat{m}^2_{\rm t}$, $\hat{m}^2_{\rm av}$, 
$\hat{m}^2_{\rm at}$,  $\hat{m}^2_{\rm s}$, and, finally,  $\hat{m}^2_{{\rm v}^{\prime}}$, the first excited state sourced by the $({\rm v},\,{\rm t})$ operators. 
The value of $\beta$ is indicated by the colour and marker, as shown in the legend.
The grey bands denote the result of the extrapolation to the  continuum limit, as defined in Eq.~(\ref{eq:fit_m}). 
}
\end{center}
\end{figure*}

\begin{figure*}
\begin{center}
\includegraphics{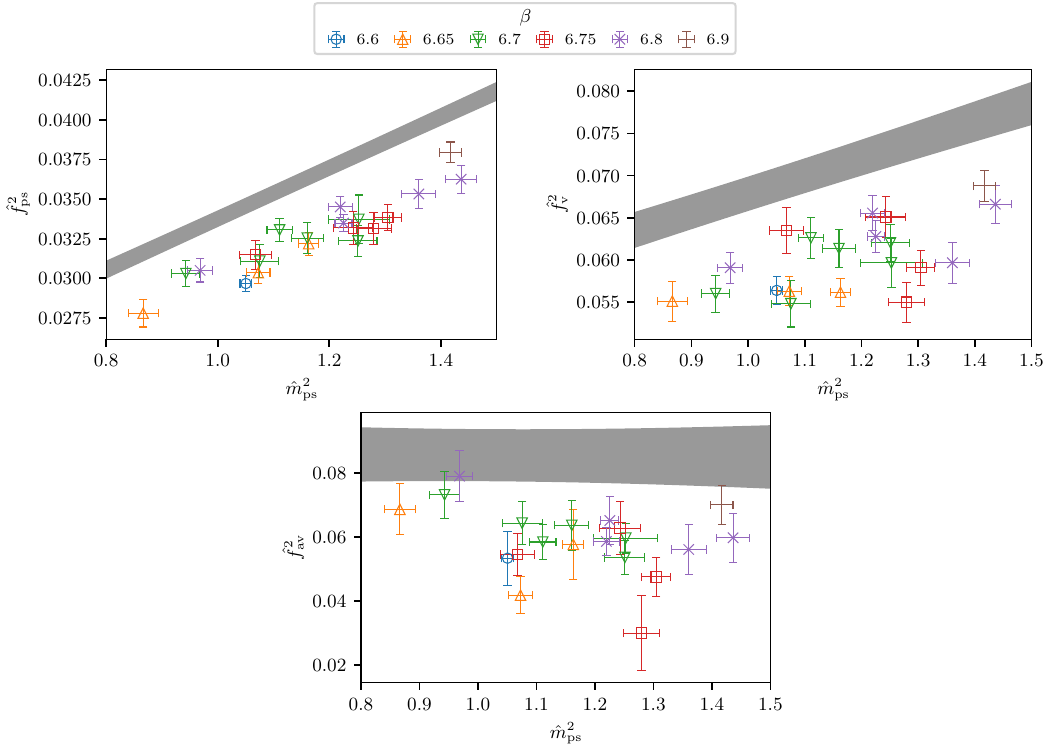}
\caption{%
\label{fig:decay_con}%
Square of the decay constants of the mesons, $\hat{f}_M^2$, expressed in units of the Wilson flow scale, $w_0$,
plotted as a function of the mass squared  of the pseudoscalar meson,
$\hat{m}_{\rm ps}^2$, in
the $Sp(4)$ lattice gauge theory  coupled to $N_{\rm as}=3$ dynamical fermions.
These measurements  of the masses are obtained implementing Wuppertal and APE smearing,
and are restricted to ensembles satisfying the upper bound
$a m_{\rm ps}\lesssim \HeavyPSMassLimit$.
The decay constants are extracted from correlation functions defined with $Z_2 \otimes Z_2$ stochastic wall sources.
Top to bottom, and left to right, we display the individual measurements of $\hat{f}^2_{\rm ps}$, $\hat{f}^2_{\rm v}$, and $\hat{f}^2_{\rm av}$, respectively. 
The value of $\beta$ is indicated by the colour and marker, as shown in the legend. 
The grey bands denote the result of the extrapolation to the  continuum limit, as defined in Eq.~(\ref{eq:fit_f}).
}
\end{center}
\end{figure*}

\subsection{\label{Sec:continuum} Continuum extrapolations}

We adopt fitting ansatzs inspired by the NLO Wilson chiral perturbation theory 
 to perform  continuum extrapolations in which we combine the available lattice data obtained at six different beta values, \ChiPTBetaValues. 
 We also
 assume a linear dependence on $\hat{m}_M^2$ of the measurements performed in ensembles with different masses. 
As discussed in Sect.~\ref{Sec:mesonN}, we restrict attention to ensembles in which
$a m_{\rm ps}\lesssim \HeavyPSMassLimit$. We refine the measurement of the masses of the mesons by
applying APE and Wuppertal smearing. We extract the masses of the first excited states of vector mesons using the GEVP method, as explained  in Sect.~\ref{Sec:measurement}.
For the decay constants, we retain the formulation of the correlation functions in terms of 
 $Z_2 \otimes Z_2$ stochastic wall sources.
The functional dependence we assume  for the observables is the following:
\beqs
\label{eq:fit_m}
\hat{m}^2_M &=& \hat{m}_{M,\,\chi}^2\left(1+L_{M}^m \hat{m}_{\rm PS}^2\right)+W_{M}^{m} \hat{a}\,,\\
\hat{f}^2_M &=& \hat{f}_{M,\,\chi}^2\left(1+L_{M}^f \hat{m}_{\rm PS}^2\right)+W_{M}^{f} \hat{a}\,,
\label{eq:fit_f}
\eeqs
where $\hat{m}_{M,\,\chi}$, $\hat{f}_{M,\,\chi}$, $L^{m}_{M}$, and $W^{m}_{M}$ are the low-energy constants (LECs) determined with the fits.

 \begin{table}[t]
\caption{%
\label{tab:fit_results}
Fit coefficients appearing
in NLO Wilson chiral perturbation theory,
Eqs.~(\ref{eq:fit_m}) and~(\ref{eq:fit_f}). 
We also present the normalised chi-square values, $\chi^2/{\rm N_{d.o.f}}$, associated with the individual fits. 
}
\begin{center}
\begin{tblr}{width=\textwidth,colspec=|c|c|c|c|c|}
\hline\hline
M & $\hat{f}_{M,\,\chi}^2$ & $L_{\rm M}^f$ & $W_M^f$ & $\chi^2/{\rm N_{d.o.f}}$ \\
\hline\hline
$\rm ps$ & $0.01769(97)$ & $0.913(88)$ & $-0.0130(17)$ & $0.41$ \\
$\rm v$ & $0.0464(37)$ & $0.47(10)$ & $-0.0374(68)$ & $1.55$ \\
$\rm av$ & $0.087(14)$ & $0.00(14)$ & $-0.089(28)$ & $1.70$ \\
\hline\hline
\end{tblr}

\begin{tblr}{width=\textwidth,colspec=|c|c|c|c|c|}
\hline\hline
M & $\hat{m}_{M,\,\chi}^2$ & $L_{\rm M}^m$ & $W_M^m$ & $\chi^2/{\rm N_{d.o.f}}$ \\
\hline\hline
$\rm v$ & $0.644(24)$ & $1.570(81)$ & $-0.676(33)$ & $0.28$ \\
$\rm t$ & $0.637(24)$ & $1.604(85)$ & $-0.692(33)$ & $0.38$ \\
$\rm av$ & $1.841(98)$ & $0.741(76)$ & $-1.14(17)$ & $1.36$ \\
$\rm at$ & $2.37(12)$ & $0.475(58)$ & $-1.54(19)$ & $2.66$ \\
$\rm s$ & $1.098(79)$ & $1.27(14)$ & $-0.51(13)$ & $7.98$ \\
$\rm v^\prime$ & $3.41(44)$ & $0.40(13)$ & $-0.90(68)$ & $4.36$ \\
\hline\hline
\end{tblr}

\end{center}
\end{table}

The resulting best-fit values of these LECs are presented in \Tab{fit_results}. 
The values of  $\hat{m}^2_{M,\,\chi}$ and $\hat{f}^2_{M,\,\chi}$ correspond to the massless and continuum extrapolations of meson masses and decay constants. We must utilise caution in using and interpreting these numbers, 
in view of the considerations exposed in Sect.~\ref{Sec:mesonN}, in particular the long extrapolation toward the massless limit they make use of.
Nevertheless, these quantities are finite and positive in the massless limit, $\hat{m}_{\rm ps}^2\rightarrow 0$,
and display qualitatively the expected features, in particular the fact that $\hat{m}_{{\rm v},\,\chi} = \hat{m}_{{\rm t},\,\chi} 
< \hat{m}_{{\rm av},\,\chi} $.

In Figs.~\ref{fig:mass_con} and~\ref{fig:decay_con}, we present our measurements of the meson masses and decay constants, expressed in units of the gradient flow scale.  We also display
 the continuum-limit extrapolations, obtained by using Eqs.~(\ref{eq:fit_m}) and~(\ref{eq:fit_f}), 
 together with the best-fit values in 
\Tab{fit_results}, to
 subtract finite lattice-spacing corrections. 
Mass ratios, such as $\hat{m}_{\rm v}^2/\hat{m}_{\rm ps}^2$, computed after 
this subtraction,
are larger,
 as all the $W_M$ coefficients in \Tab{fit_results} are negative.
 The masses of vector and tensor states are compatible with each other, as expected. 
The axial-tensor mesons are the heaviest states among the ground states, though no statistically significant difference with the axial-vector is 
present. 
The scalar meson lies between the vector (tensor) and axial-vector mesons.

\begin{figure}
\begin{center}
\includegraphics{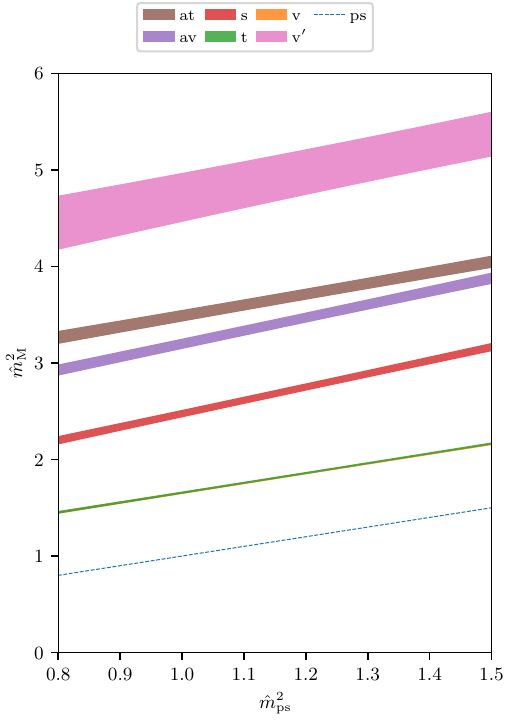}
\caption{%
\label{fig:meson_spectrum_con}%
Summary plot of the meson spectrum in the continuum limit, 
obtained by applying the ansatzs in Eq.~(\ref{eq:fit_m}), inspired by NLO Wilson chiral perturbation theory, to the $Sp(4)$
theory with $N_{\rm as}=3$ fermions. The masses have been measured using a combination of APE and Wuppertal smearing, 
on ensembles restricted to a comparatively low mass regime. The coloured bands represent the uncertainty of the extrapolation
of the fitting results. Lightest to heavier, we  display, as a function of the pseudoscalar mass squared, the 
mass squared of the pseudoscalar (ps), vector (v) and tensor (t), scalar (s), axial-vector (av), axial-tensor (at), and the first-excited state of vector (${\rm v}^{\prime}$) mesons. 
}
\end{center}
\end{figure}

We also present the results for the first excited states of the vector mesons. They are heavier than the ground state mesons in all  channels considered.
This measurement uses the subset of the available ensembles for which there is sufficient numerical data to perform the GEVP analysis successfully.
In \Fig{meson_spectrum_con}, we show a summary plot of 
the continuum extrapolations for the masses of all flavoured mesons.

Looking at the decay constants,  in Fig.~\ref{fig:decay_con}, we observe that of the pseudoscalar meson, $\hat{f}_{\rm ps}$, is the smallest of the three over the whole range of masses considered.
It further shows a strong mass dependence, yet the extrapolation 
towards massless and continuum limits is positive.
The decay constant of axial vector meson shows
 large statistical and systematical uncertainties, hence its extrapolation has to be taken with caution.

While our extrapolations to the massless and continuum limits have to be taken with a grain of salt,
especially because of  potentially large systematic uncertainties due to the long extrapolations,
it is legitimate to compare them to the results of the quenched calculations reported in 
Refs.~\cite{Bennett:2019cxd,Bennett:2023qwx}---see Table III of Ref.~\cite{Bennett:2023qwx}.
We find good agreement, within statistical errors,  for the masses of vector, tensor, axial-vector, and axial-tensor mesons. 
The noticeable exception is the scalar meson, our value of the mass squared being about $35\%$ smaller than in the quenched case. 
Our current results for the decay constant squared are about $30\%$ and $15\%$ larger than the in the quenched case, for the pseudoscalar and vector mesons, respectively. For the axial-vector meson we find that the dynamical result is statistically compatible with the quenched one.
In summary, the comparison with the quenched results shows that our calculations are robust, and indeed confirm that we are exploring a region of parameter space which is not close to the massless limit.
This is suitable for applications to composite Higgs or strongly coupled dark matter models.

\subsection{\label{Sec:vector}More about vector mesons}

The vector meson subsystem is particularly suitable to our numerical strategy,
as we study the region of parameter space in which the decay channels for the
decay of the ${\rm v}$ mesons are kinematically forbidden.
Thanks also to  the application of the GEVP method, we  both obtain
good control over statistical errors, and gain access to the first excited state, which we denote as ${\rm v}^{\prime}$. Details can be found in the tables in Appendix~\ref{Sec:data}. We devote this section to commenting on these results, and comparing them to those obtained in closely related theories.

\begin{figure}
\begin{center}
\includegraphics{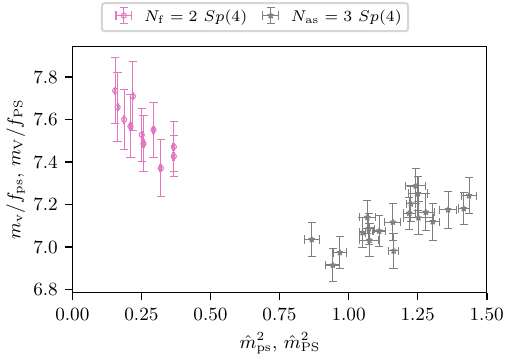}
\caption{%
\label{fig:mvfps_F_vs_AS}%
The ratio between vector meson mass and pseudoscalar decay constant,
$m_{\rm v}/f_{\rm ps}$, extrapolated to the continuum limit by means of our simplified W$\chi$PT
best-fit results, as a function of the mass squared of the pseudoscalar, $\hat{m}_{\rm ps}^2$, expressed in units of the Wilson flow, $w_0$.
For comparison purposes, besides the results for the $Sp(4)$ with $N_{\rm as}=3$ obtained in this work, 
we also present those for the $Sp(4)$ theory with $N_{\rm f}=2$ fundamental Dirac flavours, $m_{\rm V}/f_{\rm PS}$ as a function of $\hat{m}_{\rm PS}^2$, taken from Ref.~\cite{Bennett:2019jzz}. 
}
\end{center}
\end{figure}

A particularly interesting quantity is the ratio between the vector meson mass and the pseudoscalar decay constant, $m_{\rm v}/f_{\rm ps}$. 
As it shows only a mild dependence  on the pseudoscalar mass squared, $\hat{m}_{\rm ps}^2$, as well as the lattice coupling $\hat{a}$, as an exercise
we perform a linear fit to the values of $m_{\rm v}/f_{\rm ps}$ with a double expansion in $\hat{m}_{\rm ps}^2$ and $\hat{a}$, analogous to the continuum extrapolation discussed in the previous section, using the functional dependence
\beq
\frac{m_{\rm v}}{f_{\rm ps}} = \frac{m_{{\rm v},\,\chi}}{f_{{\rm ps},\,\chi}} \left(1+c_{\ell} \hat{m}_{\rm ps}^2\right)+c_{w} \hat{a},\\
\label{eq:mvfps_ext}
\eeq
and find
\beq
\frac{m_{{\rm v},\,\chi}}{f_{{\rm ps},\,\chi}} = \MVFPSExtrapolationRatio,~ c_{\ell} = \MVFPSExtrapolationCL,~c_{w} = \MVFPSExtrapolationCW.
\eeq
The resulting value of the ratio in the massless and continuum limits can be compared to the one obtained by taking the ratio between the individual extrapolations of the spectral observables, 
$\hat{m}_{{\rm v},\,\chi}$ and $\hat{f}_{{\rm ps},\,\chi}$, reported in \Tab{fit_results}. This exercise yields $\hat{m}_{{\rm v},\,\chi}/\hat{f}_{{\rm ps},\,\chi}=\MHatVFHatPSRatio$. 
The discrepancy between these two results points to the fact that the hyperquark masses are still large, hence long extrapolations towards the massless limit are being used, so that subleading terms truncated in the expansion in Eq.~(\ref{eq:mvfps_ext}) are not entirely negligible.

In \Fig{mvfps_F_vs_AS}, we present the continuum extrapolated values of this ratio, $m_{\rm v}/f_{\rm ps}$, which are obtained by subtracting the last term in \Eq{mvfps_ext} from the lattice measurements.  
For comparison purposes, we also present the corresponding results of the $Sp(4)$ theory with $N_f=2$ fundamental fermions, taken from Ref.~\cite{Bennett:2019jzz}.
As the  mass ranges considered are quite different, in the two studies, a direct comparison is not possible. 
Yet, the ratio for the case of $N_{\rm as}=3$ antisymmetric fermions is consistently smaller than in the $N_{\rm f}=2$ case.

Somewhat speculatively, we compare this finding to expectations arising from large-$N$ scaling arguments.
One expects the meson masses in $Sp(2N)$ theories to scale as $(2N)^0$,
while the decay constant in the fundamental and 2-index antisymmetric representation can be argued to
scale as the square root of the dimension of the representation, hence as
 $\sqrt{2N}$ and $\sqrt{(2N+1)(N-1)}$, respectively, by generalising 
the arguments in Ref.~\cite{Manohar:1998xv}. 
Furthermore, various lattice calculations in strongly coupled gauge theories 
with different gauge group and fermion content
suggest that $m_{\rm v}/f_{\rm ps}$ might be independent of the number of fermions, at least in the confined phase of the gauge theory~\cite{Nogradi:2019iek,Nogradi:2019auv,Kotov:2021mgp}.
If we take these scaling relations to hold also 
for $Sp(4)$, we arrive at the prediction that the ratio,
$m_{\rm v}/f_{\rm ps}=\sqrt{4/5}\, m_{\rm V}/f_{\rm PS}\simeq 
0.89\, m_{\rm V}/f_{\rm PS}$, which is consistent with what we observe numerically in \Fig{mvfps_F_vs_AS}.

Even more speculatively, this ratio, ${m_{{\rm v}}}/{f_{{\rm ps}}}$, 
 can be used to provide a rough
estimate of the on-shell coupling constant associated with the decay
of the vector 
 into two pseudoscalar mesons, $g_{\rm v\,ps\,ps}$, 
via the phenomenological KSRF relation, $g_{\rm v\,ps\,ps}=m_{\rm v}/\sqrt{2}f_{\rm ps}$ \cite{Kawarabayashi:1966kd,Riazuddin:1966sw}. 
In the continuum and massless limits, our numerical results yield $g_{\rm v\,ps\,ps}=\MVFPSExtrapolationROverRootTwo$, which is  smaller than 
the QCD value of $\sim 5.9$~\cite{ParticleDataGroup:2024cfk} and the one in the $Sp(4)$ gauge theory with two fundamental fermions, $m_{\rm V}/\sqrt{2}f_{\rm PS}=5.72(18)(13)$~\cite{Bennett:2019jzz}. 
It would be interesting to perform a direct measurement of the coupling $g_{\rm v\,ps\,ps}$,
which requires to study the theory with lower hyperquark masses.

\begin{figure}
\begin{center}
\includegraphics{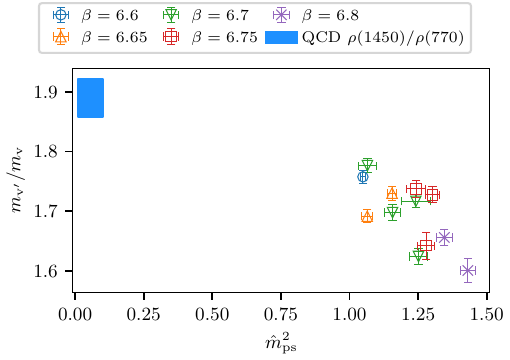}
\caption{%
\label{fig:mv_excited}%
The mass ratio between the ground and first excited states of the vector meson, $m_{{\rm v}^{\prime}}/m_{\rm v}$. 
The value of $\beta$ is indicated by the colour and marker, as shown in the legend. 
The blue rectangle denotes the QCD value for $\rho(1450)$, whose width represents the experimental uncertainty. 
}
\end{center}
\end{figure}

Finally, in \Fig{mv_excited}, we present our results for the ratio between the mass of the ground state and the first excited state of the vector meson, $m_{{\rm v}^{\prime}}/m_{\rm v}$, for all available values of the bare coupling and hyperquark mass for which the GEVP analysis yields statistically significant results.
We uncover clear evidence that this ratio depends on the mass of the pseudoscalar meson, increasing towards the massles limit.
Yet, this measurement is affected by non-negligible systematic effects, arising from the fact 
that the determination of the first-excited states, even with the GEVP method, 
 is  limited  by the short time extent for the plateau.
 We hence do not try to model the dependence of this ratio of the lattice parameter,
 but only compare to the analogous quantity in QCD.
For all the ensembles our measurements yield a smaller ratio than the QCD value obtained from experimental results, $m_{\rho(1440)}/m_{\rho(770)}\simeq 1.89(32)$~\cite{ParticleDataGroup:2024cfk}.

\section{\label{sec:conclusion}Summary and Outlook}

We reported the results of an extensive numerical study of the $Sp(4)$ (hypercolour) lattice gauge theory coupled to $N_{\rm as}=3$  Dirac fermions transforming  in the two-index antisymmetric representation of the gauge group. 
The enhanced global  $SU(6)$  symmetry acting on the fermions is broken to its $SO(6)$ subgroup because of the presence of non-zero (degenerate) hyperquark masses and the formation of hyperquark condensate. 
The corresponding continuum theory is of interest to the model building community, as it may  serve as an ultra-violet realisation of a class of  composite Higgs models with top partial compositeness. Alternatively, it can give rise to a candidate composite dark matter model.

The Euclidean, four dimensional lattice theory is formulated in terms of the (unimproved) plaquette  action, coupled to 
Wilson fermions.
We explored the two-dimensional  parameter space of the lattice theory, and identified
the presence of a line of first-order bulk phase transitions, terminating 
 at the critical lattice coupling $\beta^c \simeq 6.45$. 
 We conducted a preliminary, finite-volume study, and ascertained that
 finite-volume corrections to the mass of the lightest mesons have opposite sign in respect to that observed in the literature on $SU(3)$ theories coupled to fundamental matter fields.
We found that finite-volume effects are statistically negligible as long as the lattice size satisfies the empirical relation $m_{\rm ps}^{\rm inf} L \geq 7.5$.  
 We generated gauge ensembles with six different values of lattice coupling, $\beta>\beta^c$, and  a broad range of    hyperquark masses, $am_0$, with volumes sufficiently large to satisfy the relation $m_{\rm ps}^{\rm inf} L \geq 7.5$.

We adopted the gradient (Wilson) flow method for the purpose of setting the scale, and as a noise reduction technique in the computation of the topological charge. As observed in earlier lattice studies of new physics scenarios,  the Wilson flow scale displays strong dependence on the hyperquark mass. 
The choices of lattice parameters used for this study are good representatives of the portion of parameter space that
is accessible with the type of lattice action and algorithms we deployed. We saw hints of 
topological freezing in
the available ensembles with lightest masses and largest couplings.
The masses of the mesons,  the pseudoscalar ones in particular, are comparatively large. 
The  decay of the lightest vector meson
 into pairs of pseudoscalars is kinematically forbidden. 
 To be able to produce and analyse ensembles testing the lattice parameter space close
 to the massless and continuum limit would require a drastic change of simulation strategy.
 Nevertheless, this is the main regime of interest to composite Higgs models
and strongly interacting dark matter, hence our results have applicability to new physics models.

We reported our determination of masses and decay constants of composite states interpolated by flavored hyperquark bilinear operators (mesons).
In the lightest ensembles  available, we find clear evidence 
of mass-dependence in dimensionless combinations, such as the ratio of mass and decay constant of the lightest pseudoscalar particle, $m_{\rm ps}/f_{\rm ps}$.
This is  inconsistent with the hyperscaling hypothesis, motivated by IR conformal dynamics.
We have evidence that the theory displays confinement, accompanied by  spontaneous  breaking of its (approximately) global symmetries.
This is the main result of this paper.
We also performed a preliminary analysis of the available data in light of dEFT, as a test of near-conformal dynamics in the confining theory. Such exercise proved inconclusive, 
with available ensembles.

We find that the mass squared and the decay constant squared of composite states,
measured at fixed lattice coupling, $\beta$, exhibit a linear dependence on the pseudoscalar mass squared,  in all the ensembles in which $a m_{\rm ps}\lesssim \HeavyPSMassLimit$.
We  therefore perform continuum and massless extrapolations of our measurements, restricted to these (lighter) ensembles.
We first express our measurements in units of the Wilson-flow scale. Inspired by Wilson chiral perturbation theory, we then adopt a fit ansatz for the masses, $\hat{m}_{\rm M}^2$, and the decay constants, $\hat{f}_{\rm M}^2$, that is  linear in both $\hat{m}_{\rm ps}^2$ and $\hat{a}$. 
The resulting linear fit  well describes the numerical data. 
Yet,  the massless extrapolations are rather long, and hence the results of this analysis might be affected by significant systematic extrapolation effects.

Our results for the masses of mesons, extrapolated to  the continuum and massless limits,
 are overall comparable to the quenched results reported in Ref.~\cite{Bennett:2023qwx}. 
 Yet, the discrepancies are about $20\%$ for the scalar mesons. For pseudoscalar and vector meson decay constants, results obtained with dynamical fermions are larger than the quenched case.
We pay particular attention to the vector meson states,  for which we  develop a bespoke GEVP analysis that allows to extract the masses of both the ground and first excited states.

In the future, it would be interesting to explore the region of parameter space closer to the massless limit.
By doing so, it should be possible to improve significantly our control over the massless extrapolation, 
and hence  our spectroscopic results.
While doing so may not be a priority for the phenomenological purposes related to composite Higgs and dark  matter model building,
it would be valuable as a way to address the more general theoretical question of understanding
 whether this theory lies close to the lower edge of the conformal window.
To do so, it will be necessary to introduce more sophisticated versions of the lattice theory, involving improvement, and possibly modifying our treatment of  fermions.

Furthermore, it would be of interest to  measure  the properties of  mesons that do not carry flavor,
in particular the masses of the lightest scalar and pseudoscalar flavor singlets. The former might be a candidate dilaton,
with applicability to alternative models of electroweak symmetry breaking. 
 The latter might be of interest as an axion, which could play a role in a different type of dark matter models.
In either case, pursuing these goals would require significant new investment in software development, optimisation, and testing,
and in computing resources. We leave all these endeavours for the future.

\begin{table*}[t]
\caption{%
\label{tab:spin0_mass_wall}
Hyperquark mass, defined through the partially-conserved-axial-current (PCAC), $a m_{\rm PCAC}$,  
masses and decay constants of flavoured spin-$0$  mesons, in the pseudoscalar (ps) and scalar (s) channels, expressed
 in lattice units. For the ps mesons, we also express them in units of the spatial extent, as $m_{\rm ps} L$
 and $f_{\rm ps} L$. 
 All measurements use 
$Z_2 \otimes Z_2$ stochastic wall sources and sinks. 
The decay constants are renormalised using one-loop perturbative matching 
with  tadpole improvement.  }
\begin{center}
\begin{tblr}{width=\textwidth,colspec=|c|c|c|c|c|c|c|}
\hline\hline
Ensemble & $am_{\mathrm{PCAC}}$ & $am_{\mathrm{ps}}$ & $af_{\mathrm{ps}}$ & $am_{\mathrm{s}}$ & $m_{\mathrm{ps}}L$ & $f_{\mathrm{ps}}L$ \\
\hline
\hline
ASB0M1 & $0.08631(27)$ & $0.5640(15)$ & $0.0919(11)$ & $0.820(35)$ & $10.152(27)$ & $1.655(19)$ \\
ASB0M2 & $0.06943(13)$ & $0.48451(74)$ & $0.08092(56)$ & $0.714(20)$ & $11.628(18)$ & $1.942(13)$ \\
ASB0M3 & $0.052539(88)$ & $0.39808(64)$ & $0.06690(39)$ & $0.603(15)$ & $12.738(21)$ & $2.141(13)$ \\
\hline
ASB1M1 & $0.11558(25)$ & $0.6797(11)$ & $0.1083(10)$ & $1.006(25)$ & $12.234(19)$ & $1.950(18)$ \\
ASB1M2 & $0.08628(19)$ & $0.5527(11)$ & $0.08901(72)$ & $0.782(34)$ & $9.949(20)$ & $1.602(13)$ \\
ASB1M3 & $0.07723(19)$ & $0.5122(14)$ & $0.0828(12)$ & $0.769(13)$ & $9.219(25)$ & $1.490(21)$ \\
ASB1M4 & $0.05690(11)$ & $0.41476(80)$ & $0.06900(57)$ & $0.623(14)$ & $9.954(19)$ & $1.656(14)$ \\
ASB1M5 & $0.042270(77)$ & $0.33816(61)$ & $0.05689(34)$ & $0.5108(57)$ & $9.469(17)$ & $1.5928(95)$ \\
ASB1M6 & $0.027553(84)$ & $0.25472(66)$ & $0.04563(26)$ & $0.4087(67)$ & $8.151(21)$ & $1.4600(82)$ \\
\hline
ASB2M1 & $0.15086(17)$ & $0.80669(61)$ & $0.12706(69)$ & $1.138(24)$ & $12.9070(97)$ & $2.033(11)$ \\
ASB2M2 & $0.12577(25)$ & $0.7114(11)$ & $0.11397(94)$ & $1.098(25)$ & $11.383(17)$ & $1.823(15)$ \\
ASB2M3 & $0.10007(24)$ & $0.6026(15)$ & $0.0955(15)$ & $0.877(19)$ & $10.846(27)$ & $1.718(28)$ \\
ASB2M4 & $0.08697(17)$ & $0.54507(94)$ & $0.08593(73)$ & $0.799(15)$ & $13.082(23)$ & $2.062(18)$ \\
ASB2M5 & $0.07440(13)$ & $0.4880(10)$ & $0.07740(67)$ & $0.727(11)$ & $11.712(24)$ & $1.857(16)$ \\
ASB2M6 & $0.06117(11)$ & $0.42535(79)$ & $0.06840(50)$ & $0.6161(84)$ & $10.208(19)$ & $1.641(12)$ \\
ASB2M7 & $0.047757(77)$ & $0.35813(67)$ & $0.05875(38)$ & $0.5277(78)$ & $10.028(19)$ & $1.645(11)$ \\
ASB2M8 & $0.039866(85)$ & $0.31637(87)$ & $0.05298(43)$ & $0.4601(73)$ & $8.858(24)$ & $1.483(12)$ \\
ASB2M9 & $0.034595(65)$ & $0.28696(65)$ & $0.04878(31)$ & $0.4317(83)$ & $9.183(21)$ & $1.5611(99)$ \\
ASB2M10 & $0.028748(53)$ & $0.25200(54)$ & $0.04347(21)$ & $0.3711(52)$ & $9.072(20)$ & $1.5648(74)$ \\
ASB2M11 & $0.023573(50)$ & $0.22427(48)$ & $0.04020(20)$ & $0.3420(65)$ & $8.074(17)$ & $1.4473(73)$ \\
\hline
ASB3M1 & $0.08962(17)$ & $0.54658(92)$ & $0.08402(87)$ & $0.824(16)$ & $9.838(17)$ & $1.512(16)$ \\
ASB3M2 & $0.06421(11)$ & $0.43374(78)$ & $0.06985(53)$ & $0.6520(89)$ & $10.410(19)$ & $1.676(13)$ \\
ASB3M3 & $0.051517(99)$ & $0.36973(98)$ & $0.05950(56)$ & $0.499(12)$ & $8.874(24)$ & $1.428(14)$ \\
ASB3M4 & $0.039302(69)$ & $0.30733(65)$ & $0.05021(35)$ & $0.430(11)$ & $8.605(18)$ & $1.4059(98)$ \\
ASB3M5 & $0.029091(59)$ & $0.25332(61)$ & $0.04350(22)$ & $0.3971(46)$ & $8.106(20)$ & $1.3921(70)$ \\
\hline
ASB4M1 & $0.10354(17)$ & $0.59853(92)$ & $0.09217(72)$ & $0.880(10)$ & $9.577(15)$ & $1.475(11)$ \\
ASB4M2 & $0.08184(18)$ & $0.5055(11)$ & $0.07765(77)$ & $0.719(15)$ & $8.087(18)$ & $1.242(12)$ \\
ASB4M3 & $0.059877(85)$ & $0.40516(75)$ & $0.06437(50)$ & $0.5920(70)$ & $9.724(18)$ & $1.545(12)$ \\
ASB4M4 & $0.048377(62)$ & $0.34663(60)$ & $0.05585(36)$ & $0.5123(87)$ & $8.319(14)$ & $1.3404(86)$ \\
ASB4M5 & $0.036719(71)$ & $0.28917(60)$ & $0.04780(30)$ & $0.4414(58)$ & $9.254(19)$ & $1.5296(95)$ \\
ASB4M6 & $0.029649(53)$ & $0.25227(51)$ & $0.04243(21)$ & $0.3721(41)$ & $8.073(16)$ & $1.3577(67)$ \\
ASB4M7 & $0.022517(45)$ & $0.20975(66)$ & $0.03722(20)$ & $0.3641(31)$ & $7.551(24)$ & $1.3401(73)$ \\
\hline
ASB5M1 & $0.050683(52)$ & $0.34583(63)$ & $0.05424(34)$ & $0.4945(52)$ & $8.300(15)$ & $1.3017(82)$ \\
ASB5M2 & $0.036389(50)$ & $0.27657(59)$ & $0.04525(24)$ & $0.3906(56)$ & $8.850(19)$ & $1.4482(78)$ \\
\hline\hline
\end{tblr}

\end{center}
\end{table*}

\begin{table*}[t]
\caption{%
\label{tab:spin1_mass_wall}
Masses and decay constants of spin-$1$ flavoured vector (v) and axial-vector (av) mesons, 
and masses  in the tensor (t) and axial-tensor (at) mesons, in lattice units. 
All measurements use 
$Z_2 \otimes Z_2$ stochastic wall sources and sinks.
The decay constants are renormalised using the one-loop perturbative matching with tadpole improvement. 
}
\begin{center}
\begin{tblr}{width=\textwidth,colspec=|c|c|c|c|c|c|c|,colsep=1em}
\hline\hline
Ensemble & $am_{\mathrm{v}}$ & $af_{\mathrm{v}}$ & $am_{\mathrm{av}}$ & $af_{\mathrm{av}}$ & $am_{\mathrm{t}}$ & $am_{\mathrm{at}}$ \\
\hline
\hline
ASB0M1 & $0.6445(22)$ & $0.1265(18)$ & $0.898(30)$ & $0.101(13)$ & $0.6437(33)$ & $0.976(27)$ \\
ASB0M2 & $0.5608(16)$ & $0.1110(12)$ & $0.807(21)$ & $0.0972(91)$ & $0.5630(20)$ & $0.804(30)$ \\
ASB0M3 & $0.4701(16)$ & $0.0922(12)$ & $0.686(16)$ & $0.0897(69)$ & $0.4691(21)$ & $0.686(17)$ \\
\hline
ASB1M1 & $0.7558(19)$ & $0.1473(20)$ & $1.126(42)$ & $0.140(24)$ & $0.7542(25)$ & $1.161(39)$ \\
ASB1M2 & $0.6272(21)$ & $0.1211(17)$ & $0.916(22)$ & $0.1113(94)$ & $0.6287(26)$ & $0.932(23)$ \\
ASB1M3 & $0.5835(22)$ & $0.1136(15)$ & $0.821(21)$ & $0.0932(82)$ & $0.5807(38)$ & $0.873(18)$ \\
ASB1M4 & $0.4789(20)$ & $0.0912(12)$ & $0.703(19)$ & $0.0923(85)$ & $0.4749(30)$ & $0.712(41)$ \\
ASB1M5 & $0.4011(15)$ & $0.07748(90)$ & $0.553(12)$ & $0.0668(46)$ & $0.4025(17)$ & $0.558(21)$ \\
ASB1M6 & $0.3197(23)$ & $0.0643(11)$ & $0.484(11)$ & $0.0717(41)$ & $0.3209(37)$ & $0.474(22)$ \\
\hline
ASB2M1 & $0.8789(15)$ & $0.1670(18)$ & $1.211(31)$ & $0.115(14)$ & $0.8784(19)$ & $1.261(29)$ \\
ASB2M2 & $0.7843(20)$ & $0.1525(16)$ & $1.179(35)$ & $0.152(18)$ & $0.7825(26)$ & $1.152(26)$ \\
ASB2M3 & $0.6763(21)$ & $0.1303(16)$ & $0.899(28)$ & $0.087(10)$ & $0.6735(28)$ & $0.959(34)$ \\
ASB2M4 & $0.6141(13)$ & $0.11681(97)$ & $0.852(19)$ & $0.0953(83)$ & $0.6127(19)$ & $0.853(26)$ \\
ASB2M5 & $0.5525(17)$ & $0.1030(13)$ & $0.798(19)$ & $0.0968(81)$ & $0.5529(23)$ & $0.797(34)$ \\
ASB2M6 & $0.4933(16)$ & $0.0947(11)$ & $0.699(11)$ & $0.0880(41)$ & $0.4915(25)$ & $0.700(16)$ \\
ASB2M7 & $0.4174(16)$ & $0.07815(99)$ & $0.6008(78)$ & $0.0781(27)$ & $0.4195(21)$ & $0.566(24)$ \\
ASB2M8 & $0.3754(17)$ & $0.0728(10)$ & $0.540(11)$ & $0.0741(44)$ & $0.3752(24)$ & $0.564(15)$ \\
ASB2M9 & $0.3416(20)$ & $0.0648(12)$ & $0.5021(84)$ & $0.0702(32)$ & $0.3443(27)$ & $0.501(16)$ \\
ASB2M10 & $0.3064(17)$ & $0.05983(98)$ & $0.4325(79)$ & $0.0578(27)$ & $0.3054(28)$ & $0.4346(93)$ \\
ASB2M11 & $0.2770(16)$ & $0.05466(84)$ & $0.4223(79)$ & $0.0625(30)$ & $0.2803(26)$ & $0.422(16)$ \\
\hline
ASB3M1 & $0.6133(17)$ & $0.1151(14)$ & $0.870(31)$ & $0.107(18)$ & $0.6152(23)$ & $0.843(17)$ \\
ASB3M2 & $0.4945(19)$ & $0.0923(13)$ & $0.691(15)$ & $0.0827(51)$ & $0.4968(22)$ & $0.729(16)$ \\
ASB3M3 & $0.4242(23)$ & $0.0767(14)$ & $0.565(24)$ & $0.0566(97)$ & $0.4246(32)$ & $0.635(24)$ \\
ASB3M4 & $0.3645(17)$ & $0.07033(89)$ & $0.517(11)$ & $0.0690(44)$ & $0.3597(23)$ & $0.541(10)$ \\
ASB3M5 & $0.3095(18)$ & $0.06177(99)$ & $0.4310(92)$ & $0.0572(33)$ & $0.3074(21)$ & $0.441(16)$ \\
\hline
ASB4M1 & $0.6587(16)$ & $0.1200(14)$ & $0.934(16)$ & $0.1108(63)$ & $0.6592(21)$ & $0.941(18)$ \\
ASB4M2 & $0.5668(22)$ & $0.1042(15)$ & $0.773(23)$ & $0.084(10)$ & $0.5683(29)$ & $0.759(28)$ \\
ASB4M3 & $0.4639(19)$ & $0.0873(13)$ & $0.652(12)$ & $0.0826(52)$ & $0.4638(28)$ & $0.674(17)$ \\
ASB4M4 & $0.3990(19)$ & $0.0726(12)$ & $0.563(11)$ & $0.0704(46)$ & $0.4001(22)$ & $0.610(14)$ \\
ASB4M5 & $0.3430(16)$ & $0.06546(93)$ & $0.4936(97)$ & $0.0667(38)$ & $0.3426(27)$ & $0.493(13)$ \\
ASB4M6 & $0.3028(15)$ & $0.05849(78)$ & $0.4172(62)$ & $0.0553(21)$ & $0.3011(21)$ & $0.433(12)$ \\
ASB4M7 & $0.2588(16)$ & $0.05181(65)$ & $0.3925(77)$ & $0.0599(29)$ & $0.2583(25)$ & $0.4161(86)$ \\
\hline
ASB5M1 & $0.3911(15)$ & $0.0703(10)$ & $0.5481(86)$ & $0.0674(33)$ & $0.3927(16)$ & $0.557(11)$ \\
ASB5M2 & $0.3239(13)$ & $0.06093(71)$ & $0.4560(69)$ & $0.0615(25)$ & $0.3213(21)$ & $0.447(15)$ \\
\hline\hline
\end{tblr}

\end{center}
\end{table*}

\begin{table*}[t]
\caption{%
\label{tab:mass_in_fps_wall}
Ratio of the masses of flavoured mesons to the pseudoscalar decay constant,  $m_M/f_{\rm ps}$,  measured with   $Z_2 \otimes Z_2$ stochastic wall sources and sinks.
Decay constants are renormalised with tadpole improved one-loop perturbative matching.
}
\begin{center}
\begin{tblr}{width=\textwidth,colspec=|c|c|c|c|c|c|c|,colsep=1em}
\hline\hline
Ensemble & $m_{\mathrm{ps}} / f_{\mathrm{ps}}$ & $m_{\mathrm{s}}/ f_{\mathrm{ps}}$ & $m_{\mathrm{v}}/ f_{\mathrm{ps}}$ & $m_{\mathrm{t}}/ f_{\mathrm{ps}}$ & $m_{\mathrm{av}}/ f_{\mathrm{ps}}$ & $m_{\mathrm{at}}/ f_{\mathrm{ps}}$ \\
\hline
\hline
ASB0M1 & $6.134(60)$ & $8.92(41)$ & $7.010(75)$ & $7.001(86)$ & $9.77(36)$ & $10.61(34)$ \\
ASB0M2 & $5.988(37)$ & $8.83(25)$ & $6.930(46)$ & $6.958(48)$ & $9.97(27)$ & $9.93(37)$ \\
ASB0M3 & $5.950(29)$ & $9.01(22)$ & $7.027(41)$ & $7.011(43)$ & $10.25(25)$ & $10.26(26)$ \\
\hline
ASB1M1 & $6.274(53)$ & $9.29(24)$ & $6.976(62)$ & $6.962(63)$ & $10.39(39)$ & $10.72(38)$ \\
ASB1M2 & $6.210(41)$ & $8.79(38)$ & $7.046(53)$ & $7.064(58)$ & $10.29(26)$ & $10.47(29)$ \\
ASB1M3 & $6.188(74)$ & $9.30(20)$ & $7.05(10)$ & $7.016(99)$ & $9.93(28)$ & $10.55(28)$ \\
ASB1M4 & $6.011(46)$ & $9.02(21)$ & $6.941(58)$ & $6.883(62)$ & $10.19(29)$ & $10.31(59)$ \\
ASB1M5 & $5.945(29)$ & $8.98(12)$ & $7.052(44)$ & $7.076(51)$ & $9.71(23)$ & $9.80(37)$ \\
ASB1M6 & $5.583(28)$ & $8.96(16)$ & $7.007(56)$ & $7.033(92)$ & $10.60(24)$ & $10.39(47)$ \\
\hline
ASB2M1 & $6.349(32)$ & $8.95(20)$ & $6.917(35)$ & $6.913(35)$ & $9.53(25)$ & $9.92(23)$ \\
ASB2M2 & $6.242(46)$ & $9.64(25)$ & $6.882(56)$ & $6.866(56)$ & $10.34(32)$ & $10.11(25)$ \\
ASB2M3 & $6.312(93)$ & $9.19(26)$ & $7.08(12)$ & $7.06(12)$ & $9.42(36)$ & $10.05(42)$ \\
ASB2M4 & $6.343(47)$ & $9.30(19)$ & $7.146(60)$ & $7.130(59)$ & $9.91(24)$ & $9.92(33)$ \\
ASB2M5 & $6.305(46)$ & $9.40(17)$ & $7.139(61)$ & $7.143(67)$ & $10.31(25)$ & $10.30(44)$ \\
ASB2M6 & $6.219(38)$ & $9.01(13)$ & $7.212(57)$ & $7.187(63)$ & $10.22(17)$ & $10.24(24)$ \\
ASB2M7 & $6.096(33)$ & $8.98(13)$ & $7.104(47)$ & $7.140(49)$ & $10.23(14)$ & $9.64(41)$ \\
ASB2M8 & $5.972(38)$ & $8.68(15)$ & $7.087(62)$ & $7.082(66)$ & $10.20(22)$ & $10.64(29)$ \\
ASB2M9 & $5.882(30)$ & $8.85(17)$ & $7.003(53)$ & $7.059(69)$ & $10.29(18)$ & $10.28(34)$ \\
ASB2M10 & $5.798(21)$ & $8.54(13)$ & $7.050(48)$ & $7.027(66)$ & $9.95(19)$ & $10.00(22)$ \\
ASB2M11 & $5.578(25)$ & $8.51(17)$ & $6.891(49)$ & $6.972(70)$ & $10.50(20)$ & $10.50(41)$ \\
\hline
ASB3M1 & $6.505(61)$ & $9.80(23)$ & $7.300(74)$ & $7.322(77)$ & $10.35(40)$ & $10.04(24)$ \\
ASB3M2 & $6.209(42)$ & $9.33(15)$ & $7.079(56)$ & $7.112(57)$ & $9.89(22)$ & $10.44(25)$ \\
ASB3M3 & $6.214(47)$ & $8.39(20)$ & $7.130(61)$ & $7.135(68)$ & $9.50(42)$ & $10.66(41)$ \\
ASB3M4 & $6.121(36)$ & $8.56(23)$ & $7.259(57)$ & $7.165(68)$ & $10.29(23)$ & $10.77(22)$ \\
ASB3M5 & $5.823(23)$ & $9.13(11)$ & $7.114(50)$ & $7.067(56)$ & $9.91(21)$ & $10.13(37)$ \\
\hline
ASB4M1 & $6.494(45)$ & $9.55(13)$ & $7.147(52)$ & $7.151(54)$ & $10.13(19)$ & $10.21(22)$ \\
ASB4M2 & $6.509(56)$ & $9.26(21)$ & $7.300(70)$ & $7.318(71)$ & $9.95(30)$ & $9.77(36)$ \\
ASB4M3 & $6.294(42)$ & $9.20(12)$ & $7.206(60)$ & $7.205(66)$ & $10.13(20)$ & $10.47(26)$ \\
ASB4M4 & $6.206(35)$ & $9.17(17)$ & $7.144(54)$ & $7.164(59)$ & $10.08(21)$ & $10.92(25)$ \\
ASB4M5 & $6.050(30)$ & $9.23(12)$ & $7.176(57)$ & $7.167(71)$ & $10.33(21)$ & $10.32(28)$ \\
ASB4M6 & $5.946(24)$ & $8.77(10)$ & $7.136(47)$ & $7.097(57)$ & $9.83(16)$ & $10.20(29)$ \\
ASB4M7 & $5.635(24)$ & $9.780(95)$ & $6.951(55)$ & $6.939(74)$ & $10.54(21)$ & $11.18(24)$ \\
\hline
ASB5M1 & $6.376(33)$ & $9.12(11)$ & $7.210(44)$ & $7.241(50)$ & $10.10(17)$ & $10.26(23)$ \\
ASB5M2 & $6.111(26)$ & $8.63(13)$ & $7.156(46)$ & $7.099(52)$ & $10.08(16)$ & $9.88(35)$ \\
\hline\hline
\end{tblr}

\end{center}
\end{table*}

\begin{table*}[t]
\caption{%
\label{tab:spin0_mass_smearing}
Masses, $a m_M$, of the flavored mesons included in the continuum and massless extrapolations, measured by applying APE and Wuppertal smearing in the definition of the correlation functions. The mass of the first excited state of vector meson,  ${\rm v}^{\prime}$, is determined through a GEVP analysis that includes the two, inequivalent,  vector and tensor meson operators.
}
\begin{center}
\begin{tblr}{width=\textwidth,colspec=|c|c|c|c|c|c|c|c|,colsep=0.5em}
\hline\hline
Ensemble & $am_{\mathrm{ps}}$ & $am_{\mathrm{s}}$ & $m_{\mathrm{v}}$ & $am_{\mathrm{t}}$ & $m_{\mathrm{av}}$ & $m_{\mathrm{at}}$ & $m_{\mathrm{v}^\prime}$ \\
\hline
\hline
ASB0M3 & $0.39776(34)$ & $0.5897(36)$ & $0.46643(59)$ & $0.46591(63)$ & $0.6486(45)$ & $0.6702(26)$ & $0.8198(51)$ \\
\hline
ASB1M4 & $0.41334(35)$ & $0.6193(20)$ & $0.48158(48)$ & $0.48095(56)$ & $0.6768(23)$ & $0.6856(30)$ & $0.8333(58)$ \\
ASB1M5 & $0.33690(33)$ & $0.4921(51)$ & $0.39960(62)$ & $0.39990(55)$ & $0.5495(37)$ & $0.5600(38)$ & $0.6759(42)$ \\
ASB1M6 & $0.25334(34)$ & $0.3737(65)$ & $0.31548(59)$ & $0.31502(70)$ & $0.4507(21)$ & $0.4546(41)$ & $\cdots$ \\
\hline
ASB2M6 & $0.42536(34)$ & $0.6147(21)$ & $0.48865(50)$ & $0.48851(52)$ & $0.6591(39)$ & $0.6655(53)$ & $0.7936(68)$ \\
ASB2M7 & $0.35657(43)$ & $0.5102(56)$ & $0.41402(71)$ & $0.41489(73)$ & $0.5432(79)$ & $0.5608(96)$ & $0.7107(45)$ \\
ASB2M8 & $0.31587(41)$ & $0.4493(33)$ & $0.37160(70)$ & $0.37091(86)$ & $0.5147(29)$ & $0.5159(44)$ & $0.6308(46)$ \\
ASB2M9 & $0.28559(29)$ & $0.4361(14)$ & $0.34276(42)$ & $0.34264(51)$ & $0.4747(28)$ & $0.4696(60)$ & $0.6088(38)$ \\
ASB2M10 & $0.25237(27)$ & $0.3607(25)$ & $0.30498(45)$ & $0.30571(43)$ & $0.4259(19)$ & $0.4410(26)$ & $\cdots$ \\
ASB2M11 & $0.22293(24)$ & $0.3431(18)$ & $0.27802(56)$ & $0.27696(58)$ & $0.3916(22)$ & $0.4018(39)$ & $\cdots$ \\
\hline
ASB3M2 & $0.43338(36)$ & $0.6356(38)$ & $0.49310(72)$ & $0.49329(78)$ & $0.6812(40)$ & $0.6898(47)$ & $0.8517(66)$ \\
ASB3M3 & $0.36971(36)$ & $0.5382(28)$ & $0.42442(59)$ & $0.42449(71)$ & $0.5800(33)$ & $0.5842(60)$ & $0.6969(96)$ \\
ASB3M4 & $0.30733(31)$ & $0.4571(21)$ & $0.35938(52)$ & $0.35904(61)$ & $0.4990(32)$ & $0.5093(23)$ & $0.6244(51)$ \\
ASB3M5 & $0.25275(38)$ & $0.3906(15)$ & $0.30510(65)$ & $0.30402(58)$ & $0.4288(23)$ & $0.4381(25)$ & $\cdots$ \\
\hline
ASB4M3 & $0.40446(27)$ & $0.5767(22)$ & $0.46126(38)$ & $0.46130(43)$ & $0.6228(19)$ & $0.6243(25)$ & $0.7383(93)$ \\
ASB4M4 & $0.34472(33)$ & $0.4951(18)$ & $0.39669(56)$ & $0.39619(57)$ & $0.5324(23)$ & $0.5449(23)$ & $0.6569(55)$ \\
ASB4M5 & $0.28740(27)$ & $0.4309(16)$ & $0.33855(42)$ & $0.33856(50)$ & $0.4646(20)$ & $0.4662(28)$ & $\cdots$ \\
ASB4M6 & $0.25112(28)$ & $0.3675(13)$ & $0.30058(45)$ & $0.30219(43)$ & $0.4129(21)$ & $0.4243(21)$ & $\cdots$ \\
ASB4M7 & $0.21017(31)$ & $0.3577(11)$ & $0.25691(49)$ & $0.25652(52)$ & $0.3590(34)$ & $0.3889(22)$ & $\cdots$ \\
\hline
ASB5M1 & $0.34496(50)$ & $0.4881(25)$ & $0.39154(72)$ & $0.39107(77)$ & $0.5333(28)$ & $0.5343(28)$ & $0.6501(40)$ \\
ASB5M2 & $0.27531(36)$ & $0.3927(17)$ & $0.32183(49)$ & $0.32167(56)$ & $0.4340(28)$ & $0.4437(26)$ & $\cdots$ \\
\hline\hline
\end{tblr}

\end{center}
\end{table*}

\begin{table*}[t]
\caption{%
\label{tab:mass_in_fps_smearing}
Ratio of the masses of the flavored mesons included in the continuum and massless extrapolations, to
the pseudoscalar decay constant, $m_M/f_{\rm ps}$.  The masses are measured by applying APE and Wuppertal smearing in the definition of the correlation functions. The mass of the first excited state of vector mesons,  ${\rm v}'$, is determined through a GEVP analysis that includes the two, inequivalent,  vector and tensor meson operators.
We also report the ratio of the masses of the first-excited and ground states of the vector meson, $m_{{\rm v}'}/m_{\rm v}$.}
\begin{center}
\begin{tblr}{width=\textwidth,colspec=|c|c|c|c|c|c|c|c|c|,colsep=0.5em}
\hline\hline
Ensemble  & $m_{\mathrm{ps}} / f_{\mathrm{ps}}$ & $m_{\mathrm{s}}/ f_{\mathrm{ps}}$ & $m_{\mathrm{v}}/ f_{\mathrm{ps}}$ & $m_{\mathrm{t}}/ f_{\mathrm{ps}}$ & $m_{\mathrm{av}}/ f_{\mathrm{ps}}$ & $m_{\mathrm{at}}/ f_{\mathrm{ps}}$ & $m_{\mathrm{v}^\prime} / f_{\mathrm{ps}} $ & $m_{\rm v^\prime} / m_{\rm v}$ \\
\hline
\hline
ASB0M3 & $5.946(35)$ & $8.814(73)$ & $6.972(42)$ & $6.964(42)$ & $9.696(86)$ & $10.018(70)$ & $12.25(11)$ & $1.758(11)$ \\
\hline
ASB1M4 & $5.991(50)$ & $8.975(83)$ & $6.980(59)$ & $6.971(59)$ & $9.809(88)$ & $9.937(95)$ & $12.08(13)$ & $1.730(12)$ \\
ASB1M5 & $5.922(36)$ & $8.650(96)$ & $7.025(44)$ & $7.030(44)$ & $9.659(86)$ & $9.844(88)$ & $11.88(11)$ & $1.691(11)$ \\
ASB1M6 & $5.553(32)$ & $8.19(15)$ & $6.915(40)$ & $6.905(41)$ & $9.879(74)$ & $9.96(11)$ & $\cdots$ & $\cdots$ \\
\hline
ASB2M6 & $6.219(46)$ & $8.987(71)$ & $7.145(53)$ & $7.142(53)$ & $9.637(93)$ & $9.73(10)$ & $11.60(13)$ & $1.624(14)$ \\
ASB2M7 & $6.069(39)$ & $8.68(11)$ & $7.047(47)$ & $7.062(46)$ & $9.25(14)$ & $9.54(17)$ & $12.10(11)$ & $1.716(11)$ \\
ASB2M8 & $5.962(49)$ & $8.480(96)$ & $7.014(57)$ & $7.001(59)$ & $9.716(95)$ & $9.74(11)$ & $11.91(13)$ & $1.698(13)$ \\
ASB2M9 & $5.854(37)$ & $8.939(62)$ & $7.026(46)$ & $7.024(45)$ & $9.732(85)$ & $9.63(13)$ & $12.48(11)$ & $1.776(11)$ \\
ASB2M10 & $5.806(28)$ & $8.299(71)$ & $7.017(36)$ & $7.033(35)$ & $9.799(62)$ & $10.145(68)$ & $\cdots$ & $\cdots$ \\
ASB2M11 & $5.545(29)$ & $8.533(66)$ & $6.915(37)$ & $6.889(38)$ & $9.740(74)$ & $9.99(11)$ & $\cdots$ & $\cdots$ \\
\hline
ASB3M2 & $6.204(47)$ & $9.099(88)$ & $7.059(55)$ & $7.062(55)$ & $9.753(93)$ & $9.87(10)$ & $12.19(13)$ & $1.727(13)$ \\
ASB3M3 & $6.213(59)$ & $9.045(96)$ & $7.133(69)$ & $7.134(69)$ & $9.75(11)$ & $9.82(14)$ & $11.71(19)$ & $1.642(23)$ \\
ASB3M4 & $6.121(42)$ & $9.104(70)$ & $7.158(49)$ & $7.151(50)$ & $9.938(86)$ & $10.143(83)$ & $12.44(14)$ & $1.737(14)$ \\
ASB3M5 & $5.810(32)$ & $8.980(58)$ & $7.013(40)$ & $6.989(39)$ & $9.856(75)$ & $10.072(81)$ & $\cdots$ & $\cdots$ \\
\hline
ASB4M3 & $6.283(49)$ & $8.959(79)$ & $7.166(55)$ & $7.166(55)$ & $9.675(82)$ & $9.698(81)$ & $11.47(17)$ & $1.601(20)$ \\
ASB4M4 & $6.172(40)$ & $8.865(63)$ & $7.103(47)$ & $7.094(47)$ & $9.532(73)$ & $9.757(77)$ & $11.76(13)$ & $1.656(14)$ \\
ASB4M5 & $6.012(38)$ & $9.014(64)$ & $7.083(45)$ & $7.083(45)$ & $9.720(69)$ & $9.753(80)$ & $\cdots$ & $\cdots$ \\
ASB4M6 & $5.919(30)$ & $8.662(54)$ & $7.084(38)$ & $7.122(37)$ & $9.730(70)$ & $10.000(73)$ & $\cdots$ & $\cdots$ \\
ASB4M7 & $5.646(33)$ & $9.609(61)$ & $6.902(41)$ & $6.891(41)$ & $9.64(10)$ & $10.447(81)$ & $\cdots$ & $\cdots$ \\
\hline
ASB5M1 & $6.360(40)$ & $8.999(66)$ & $7.219(46)$ & $7.211(46)$ & $9.833(76)$ & $9.852(76)$ & $11.99(10)$ & $1.660(11)$ \\
ASB5M2 & $6.084(34)$ & $8.677(60)$ & $7.111(40)$ & $7.108(40)$ & $9.590(80)$ & $9.804(76)$ & $\cdots$ & $\cdots$ \\
\hline\hline
\end{tblr}

\end{center}

\end{table*}

Finally, we envision a complementary, equally ambitious, research programme of study of the phase space of the theory.
There is not a sign problem, given the real nature of the fermion representations,
hence its behaviour in the presence of a (isospin) 
chemical potential can be studied numerically, from first principles.
Even at zero chemical potential, as this might be used as a model of a new dark sector
 it is of general interest to understand whether this theory undergoes
 a first-order deconfinement phase transition at high temperature. 
If this is the case, 
bubble dynamics may leave a 
 relic stochastic background of gravitational 
 waves, as mentioned in the introduction---see also Ref.~\cite{afzal2023nanograv}.

The highly challenging goal of measuring,  in strongly coupled theories,
 those parameters that enter into
the calculation of the power spectrum  of gravitation waves
is currently being pursued via several 
 complementary investigation strategies~\cite{Huang:2020crf,Halverson:2020xpg,Kang:2021epo,Reichert:2021cvs,Reichert:2022naa,Pasechnik:2023hwv},
that use, for example, of effective tools as
 the Polyakov-loop~\cite{Pisarski:2000eq, Pisarski:2001pe,Pisarski:2002ji,Sannino:2002wb,
Ratti:2005jh,Fukushima:2013rx,Fukushima:2017csk,Lo:2013hla,Hansen:2019lnf}
or the
matrix models~\cite{
Meisinger:2001cq,Dumitru:2010mj,
Dumitru:2012fw, Kondo:2015noa,Pisarski:2016ixt,Nishimura:2017crr,Guo:2018scp,
KorthalsAltes:2020ryu,Hidaka:2020vna}.
The Logarithmic Linear Relaxation (LLR) algorithm~\cite{Langfeld:2012ah,Langfeld:2013xbf,Langfeld:2015fua,Cossu:2021bgn} offers a
new, alternative opportunity to perform such high precision numerical studies.
Finite-temperature studies of Yang-Mills theories using the LLR  exist for  $Sp(4)$~\cite{Bennett:2024bhy},  $SU(3)$~\cite{Mason:2022trc,Mason:2022aka,Lucini:2023irm}, $SU(4)$~\cite{Springer:2021liy,
   Springer:2022qos},  and general $SU(N_c)$~\cite{Springer:2023wok,Springer:2023hcc}. It would be interesting to generalise these studies  to
   theories with matter content, such as the one proposed here.

\begin{acknowledgments}

We would like to thank Giacomo Cacciapaglia, Gabriele Ferretti, Thomas Flacke, Anna Hasenfratz, Chulwoo Jung, and Sarada Rajeev, 
for very helpful discussions during the ``PNU Workshop on Composite Higgs: Lattice study and all'', at Haeundae, Busan, in February 2024, 
where preliminary results of this study were presented. We also thank Will Detmold, Alberto Ramos, and Andr\'{e} Walker-Loud for useful discussions. 

The work of EB and BL is supported in part by the EPSRC ExCALIBUR programme ExaTEPP (project EP/X017168/1). 
The work of EB, BL, and MP has been supported by the STFC Consolidated Grant No. ST/X000648/1.
The work of EB has also been supported by the UKRI Science and Technology Facilities Council (STFC) Research Software Engineering Fellowship EP/V052489/1.
The work of DKH was supported by Basic Science Research Program through the National Research Foundation of Korea (NRF) funded by the Ministry of Education (NRF-2017R1D1A1B06033701). 
The work of DKH was further supported by the National Research Foundation of Korea (NRF) grant funded by the Korea government (MSIT) (2021R1A4A5031460).
The work of JWL is supported by IBS under the project code, IBS-R018-D1. 
The work of HH and CJDL is supported by the Taiwanese MoST grant 109-2112-M-009-006-MY3 and NSTC grant 112-2112-M-A49-021-MY3. 
The work of CJDL is also supported by Grants No. 112-2639-M-002-006-ASP and No. 113-2119-M-007-013-.
The work of BL and MP has been further supported in part by the STFC  Consolidated Grant No. ST/T000813/1.
BL and MP received funding from the European Research Council (ERC) under the European Union's Horizon 2020 research and innovation program under Grant Agreement No.~813942. 
The work of DV is supported by STFC under Consolidated Grant No. ST/X000680/1.

Numerical simulations have been performed on the 
Swansea University SUNBIRD cluster (part of the Supercomputing Wales project) and AccelerateAI A100 GPU system,
on the local HPC
clusters in Pusan National University (PNU), in Institute for Basic Science (IBS) and in National Yang Ming Chiao Tung University (NYCU),
and on the DiRAC Data Intensive service at Leicester.
The Swansea University SUNBIRD system and AccelerateAI are part funded by the European Regional Development Fund (ERDF) via Welsh Government.
The DiRAC Data Intensive service at Leicester is operated by 
the University of Leicester IT Services, which forms part of 
the STFC DiRAC HPC Facility (www.dirac.ac.uk). The DiRAC 
Data Intensive service equipment at Leicester was funded 
by BEIS capital funding via STFC capital grants ST/K000373/1 
and ST/R002363/1 and STFC DiRAC Operations grant ST/R001014/1. 
DiRAC is part of the National e-Infrastructure.

{\bf Research Data Access Statement}---The data generated for this manuscript can be downloaded from  
Ref.~\cite{data_release} and the analysis code from Ref.~\cite{analysis_release}. 

{\bf Open Access Statement}---For the purpose of open access, the authors have applied a Creative Commons  Attribution (CC BY) licence  to any Author Accepted Manuscript version arising.

\end{acknowledgments}

\appendix

\section{\label{Sec:data}More numerical results}

In this appendix, we summarise all the new numerical results on lattice measurements of
 meson masses and decay constants 
which have been carried out for this work. We adopted  two different strategies for the analysis of the correlation function.
Measurement extracted from correlation functions defined with
 $Z_2 \otimes Z_2$ stochastic wall sources and sinks,
for all the ensembles listed in \Tab{ensemble},
are tabulated in Tables ~\ref{tab:spin0_mass_wall} 
and ~\ref{tab:spin1_mass_wall}, and 
are discussed in \Sec{mesonN} and \App{largemass}. 
We also report the results for the PCAC (unrenormalised) hyperquark mass, 
$a m_{\rm PCAC}$, and for the quantities, $m_{\rm ps} L$ and $f_{\rm ps} L$. 
To facilitate comparison with phenomenological models and applications to model building, 
we find it useful to express the measurements also in units of the decay constant of the pseudoscalar 
meson, $f_{\rm ps}$, in \Tab{mass_in_fps_wall}.

As explained in the body of the paper, we 
select a subset of the ensembles, 
which we use for the massless and continuum extrapolation,
and apply a combination of  APE and Wuppertal smearing, to repeat the mass measurements,
performing also the extraction of the first-excited states of vector mesons with the GEVP method.
The results are tabulated
in Table~\ref{tab:spin0_mass_smearing}, and 
 discussed in Sects. \ref{Sec:continuum} and \ref{Sec:vector}. 
We express them in units of $f_{\rm ps}$, in \Tab{mass_in_fps_smearing},
in which we also report the ratio between the  first excited and ground states of vector meson.

\begin{table*}[t]
\caption{%
\label{tab:autocorr}
Autocorrelation lengths of the average plaquette, $\tau^{\cal P}_{\rm exp}$, of the $2$-point correlation function of ps meson, $\tau^{\rm ps}_{\rm exp}$, of the energy density at the flow time $t=(w_0/a)^2$, $\tau^{w_0}_{\rm exp}$, and of the topological charge at the flow time $t=L^2/32$, $\tau^{\cal Q}_{\rm exp}$, in units of HMC trajectories. 
The separations in trajectories between adjacent configurations used for the measurements of spectral quantities and the gradient flow scale are denoted by $\delta_{\rm traj}$ and $\delta_{\rm traj}^{\rm GF}$, respectively. 
}
\begin{center}
\begin{tblr}{width=\textwidth,colspec=|c|c|c|c|c|c|c|}
\hline\hline
Ensemble & $\delta_{\mathrm{traj}}$ & $\tau_{\mathrm{exp}}^{\mathcal{P}}$ & $\tau_{\mathrm{exp}}^{\mathrm{ps}}$  & $\delta_{\mathrm{traj}}^{\mathrm{GF}}$  & $\tau_{\mathrm{exp}}^{w_0}$ & $\tau_{\mathrm{exp}}^{\mathcal{Q}}$ \\
\hline
\hline
ASB0M1 & 24 & 15.3(1.8) & 8.7(2.1) & 72 & 74.1(8.2) & 69.5(2.9) \\
ASB0M2 & 16 & 10.4(1.2) & 4.35(83) & 48 & 43.4(3.9) & 130.2(7.8) \\
ASB0M3 & 16 & 10.6(1.3) & 8.5(1.7) & 112 & 119(12) & 105(11) \\
\hline
ASB1M1 & 28 & 15.6(2.0) & 6.0(1.3) & 84 & 64.7(4.5) & 32.4(3.0) \\
ASB1M2 & 20 & 7.4(1.2) & 4.05(79) & 100 & 94.7(5.0) & 81.1(5.3) \\
ASB1M3 & 16 & 6.93(79) & 7.5(1.6) & 80 & 76.2(3.6) & 64.6(2.2) \\
ASB1M4 & 16 & 8.08(67) & 3.87(96) & 64 & 77.8(2.6) & 273(20) \\
ASB1M5 & 12 & 5.98(45) & 4.31(68) & 144 & 120.2(1.0) & 115.6(4.4) \\
ASB1M6 & 12 & 5.63(62) & 3.06(58) & 240 & 259.0(6.4) & 96.4(1.7) \\
\hline
ASB2M1 & 12 & 12.5(2.3) & 4.98(68) & 36 & 31.4(2.2) & 22.9(1.1) \\
ASB2M2 & 20 & 18.3(5.3) & 7.3(4.7) & 60 & 62.6(5.5) & 54.8(5.5) \\
ASB2M3 & 20 & 8.2(1.5) & 8.4(1.7) & 120 & 107.2(5.4) & 63.3(5.1) \\
ASB2M4 & 20 & 12.5(1.9) & 8.3(1.6) & 60 & 65.4(7.5) & 63.5(4.3) \\
ASB2M5 & 20 & 8.30(99) & 5.8(1.1) & 80 & 69.5(6.1) & 159.2(7.1) \\
ASB2M6 & 12 & 6.09(68) & 5.60(91) & 144 & 124.43(93) & 170.4(5.6) \\
ASB2M7 & 12 & 6.19(99) & 6.5(1.7) & 240 & 324(14) & 225.6(7.9) \\
ASB2M8 & 20 & 6.35(52) & 5.0(1.0) & 160 & 191.8(5.9) & 299.2(8.4) \\
ASB2M9 & 12 & 4.28(46) & 4.77(93) & 144 & 127.8(2.7) & 61.5(8.3) \\
ASB2M10 & 12 & 5.38(29) & 3.96(75) & 108 & 89.3(3.1) & 207(13) \\
ASB2M11 & 12 & 5.50(32) & 3.80(48) & 120 & 87.1(3.9) & 47.5(6.8) \\
\hline
ASB3M1 & 12 & 6.36(65) & 6.0(1.3) & 84 & 88.7(2.1) & 94.4(7.7) \\
ASB3M2 & 20 & 5.38(50) & 3.53(85) & 120 & 223(21) & 118.3(5.1) \\
ASB3M3 & 12 & 3.64(36) & 4.00(78) & 96 & 80.5(1.9) & 140.6(8.9) \\
ASB3M4 & 8 & 2.98(19) & 4.23(93) & 160 & 155.9(7.4) & 348(12) \\
ASB3M5 & 8 & 3.40(43) & 2.83(18) & 144 & 125.9(1.6) & 569(12) \\
\hline
ASB4M1 & 12 & 5.52(75) & 2.61(44) & 60 & 60.5(1.0) & 30.00(93) \\
ASB4M2 & 12 & 3.92(41) & 1.21(47) & 84 & 82.5(2.2) & 156.1(9.2) \\
ASB4M3 & 12 & 3.82(21) & 4.16(84) & 108 & 75.57(59) & 88.1(4.0) \\
ASB4M4 & 8 & 3.44(35) & 2.33(38) & 104 & 88.6(1.1) & 153.5(6.7) \\
ASB4M5 & 12 & 4.69(25) & 4.06(80) & 72 & 60.4(3.2) & 491(30) \\
ASB4M6 & 12 & 4.33(31) & 5.5(1.0) & 84 & 63.1(1.0) & 345(25) \\
ASB4M7 & 12 & 3.37(35) & 5.19(97) & 192 & 174(10) & 814(52) \\
\hline
ASB5M1 & 8 & 3.89(15) & 5.1(1.3) & $\cdots$ & $\cdots$ & 368(18) \\
ASB5M2 & 12 & 5.95(80) & 3.06(53) & 72 & 74.4(2.9) & 193(18) \\
\hline\hline
\end{tblr}

\end{center}
\end{table*}

\section{\label{Sec:autocorrelation} Autocorrelation}

The local updates used by the (R)HMC algorithm necessarily lead to residual autocorrelation between the configurations retained in an ensemble. We devote this appendix to estimating the autocorrelation length of such effect, and to explaining the strategy adopted in our data analysis to ensure that our statistical analysis is not comprised by autocorrelation effects. As we shall see this is dependent on the observable. 
For a given observable, $\mathcal{O}$, we define the autocorrelation function, $C_{\cal \mathcal{O}}(\tau)$, as follows:
\beq
C_{\cal O}(\tau) \equiv 
\sum_{i=1}^{N_{\rm traj}-\tau} \frac{({\cal O}_i-\langle {\cal O}\rangle)({\cal O}_{i+\tau}-\langle {\cal O}\rangle)}{{\rm var}({\cal O})},
\label{eq:autocorr}
\eeq
where $\tau$ labels the individual trajectory. In \Eq{autocorr}, ${\rm var}({\cal O})$ is the ensemble variance of $\cal O$,  and $N_{\rm traj}$ is the total number of trajectories. 
After visually checking the behaviour of $C_{\cal O}(\tau)$, the exponential autocorrelation time, $\tau^{\cal O}_{\rm exp}$, is estimated by fitting  $C_{\cal O}(\tau)/C_{\cal O}(0) = {\rm exp} (-\tau/\tau^{\cal O}_{\rm exp})$.

The first observable we consider is the average plaquette, $\cal P$, defined in \Eq{plaquette}, measured on gauge configuration. 
The corresponding autocorrelation time, $\tau^{\cal P}_{\rm exp}$, is then used to determine the separation of trajectories, $\delta_{\rm traj}$, between two neighboring configurations selected for all the measurements of $2$-point mesonic correlation functions such that $\delta_{\rm traj} \gtrsim \tau^{\cal P}_{\rm exp}$.  
We report our results in \Tab{autocorr}, by tabulating both the values of $\delta_{\rm traj}$ and $\tau_{\rm exp}^{\cal{P}}$. 
We additionally compute the autocorrelation length of the $2$-point correlation function of the ps meson, taken at the choice of the Euclidean time at which a plateau appears in the effective mass. 
We denote this estimate of the autocorrelation as $\tau_{\rm exp}^{\rm ps}$. This measurement provides an alternative way to estimate the effect of autocorrelation in $2$-pint functions involving mesons. 
The resulting values of $\tau^{\cal P}_{\rm exp}$ and $\tau^{\rm ps}_{\rm exp}$ are reported in \Tab{autocorr}, from which one can conclude that the choice of $\delta_{\rm traj}$ is sufficient to retain independent configurations for all the spectral measurements we performed in this work.

As discussed in Sects.~\ref{Sec:scale} and \ref{Sec:topology}, the observables built from flowed fields through the gradient flow, including the action density ${\cal E}(t)$ and the topological charge $Q$, exhibit a long autocorrelation along the HMC trajectories. 
We calculate autocorrelation functions for the action density at $t=w_0^2$ and the topological charge at $t=L^2/32$, which we denote as $C_{\cal E}(\tau)$ and $C_{Q}(\tau)$, by replacing $\cal O$ in \Eq{autocorr} with ${\cal E}(t=w_0^2)$ and $Q(t=L^2/32)$, respectively. 
We then estimate the corresponding autocorrelation lengths, $\tau^{w_0}_{\rm exp}$ and $\tau^Q_{\rm exp}$, from exponential fits to $C_{\cal E}(\tau)$ and $C_{Q}(\tau)$. We present the results for $\tau^{w_0}_{\rm exp}$ and $\tau^Q_{\rm exp}$ in \Tab{autocorr}. 
The measurements of the flowed observables have been carried out on the configurations already separated by $\delta_{\rm traj}$ trajectories, leaving a few hundred samples much smaller than $N_{\rm traj}$. 
We notice that some correlation lengths obtained in this way are much larger than $\delta_{\rm traj}$. 
In these cases, the exponential function may be insufficient to capture the detailed magnitude of autocorrelation effects. The errors quoted in our results have been estimated through standard bootstrapping technique. 
Nevertheless, the clear conclusion of this exercise is that the autocorrelation length of flowed observables is substantially larger than that of the average plaquette and the $2$-point correlation functions of ps mesons. We use the estimated autocorrelation length, $\tau^{w_0}_{\rm exp}$, for guidance in the selection of the subset for a given ensemble, containing configurations separated by $\delta_{\rm traj}^{\rm GF}\simeq \tau^{w_0}_{\rm exp}$ trajectories, for the measurements of the gradient flow scale, $w_0/a$.

 \begin{figure}[h!]
\begin{center}
\includegraphics{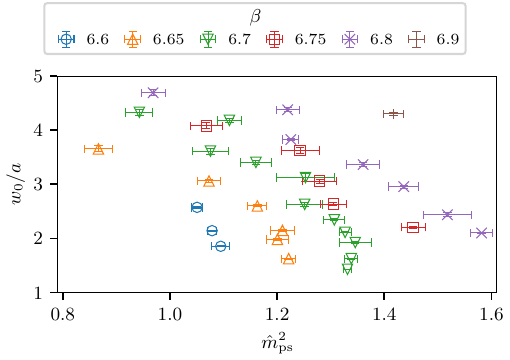}
\caption{%
\label{fig:w0_vs_mps}%
The Wilson flow scale, $w_0/a$, as a function of the pseudoscalar mass squared, expressed in units of the Wilson scale, $w_0$, and denoted as $\hat{m}_{\rm ps}^2$, for all the available choices of lattice coupling, $\beta$. 
The measurement of the pseudoscalar mass is extracted from correlation functions defined with 
$Z_2 \otimes Z_2$ stochastic wall sources and sinks. 
The value of $\beta$ is indicated by the colour and marker, as shown in the legend.
}
\end{center}
\end{figure}

\begin{figure}[h!]
\begin{center}
\includegraphics{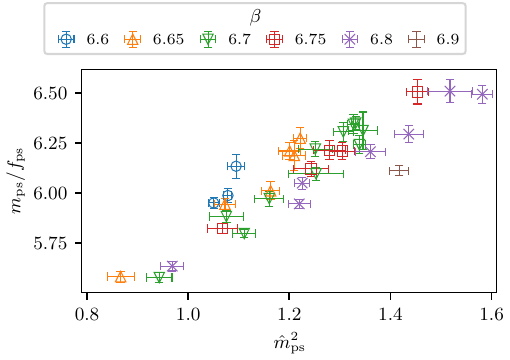}
\caption{%
\label{fig:ratio_mpsfps}%
The ratio between mass and decay constant of the flavored, pseudoscalar mesons, 
 $m_{\rm ps}/f_{\rm ps}$, 
as a function of the square of the mass of the pseudoscalar, expressed in units of Wilson flow scale.
The measurement are extracted from correlation functions defined with 
$Z_2 \otimes Z_2$ stochastic wall sources and sinks.
The value of $\beta$ is indicated by the colour and marker, as shown in the legend.
}
\end{center}
\end{figure}

\section{\label{Sec:largemass} Hyperquark mass dependence}

In this appendix we provide additional information  on the dependence of physical observables
on the 
hyperquark mass. 
We  present our results for the Wilson flow scale, $w_0/a$,  in \Fig{w0_vs_mps}. 
As expected, larger values of the  lattice coupling, $\beta$, generically translate into larger values of $w_0/a$,
 for any  fixed value of the pseudoscalar mass squared, $\hat{m}_{\rm ps}^2$. 
But we also find that  $w_0/a$ changes substantially as the pseudoscalar mass varies, for fixed $\beta$.
With a few exceptions (the three heaviest ensembles with $\beta=6.65$ and the five heaviest with $\beta=6.7$),
we find approximately linear dependence of $w_0/a$ on $\hat{m}_{\rm ps}^2$. 
In the eight exceptions, all characterised by large values of the hyperquark mass, 
we find that $w_0/a$ changes significantly, despite the fact that the psedoscalar mass, $\hat{m}_{\rm ps}$, is approximately constant.
In particular, there appears to be an upper bound on the pseudoscalar mass, expressed in units of the Wilson flow scale,
at fixed $\beta$.

The ratio between  mass and decay constant of the flavored  pseudoscalar meson, $m_{\rm ps}/f_{\rm ps}$, 
 is shown in \Fig{ratio_mpsfps}, as a function of $\hat{m}_{\rm ps}^2$, for all the ensembles.
As discussed above, the measured values of $\hat{m}_{\rm ps}^2$ at fixed $\beta$ value are bounded from above. 
We see that in the case of the heavy ensembles with
 $\beta=6.65$ and $6.7$, also the ratio $m_{\rm ps}/f_{\rm ps}$ converges, so that these heavier ensembles are,
 in practical terms, testing the same physical scales.

\section{\label{Sec:mpcac}PCAC  mass}

In the Wilson-Dirac fermionic action,  the hyperquark mass receives additive renormalisation.
To provide a  definition of hyperquark mass that vanishes when the global symmetry is restored,
 we use a  Ward-Takahashi identity, and introduce
the partially-conserved-axial-current (PCAC) mass. We devote this appendix
to  providing such definition, and fixing conventions.

Using the  PCAC relation, the variation of 
 an operator,  $\mathcal{O}$,  under the action of an infinitesimal chiral transformation,
can be written as 
 \begin{widetext}
\beqs
\label{Eq:pcac}
\langle 0 | \delta_x \mathcal{O}(y) | 0 \rangle =
\langle 0 | \{ \partial^\mu \mathcal{O}_{\rm av}^\mu (x) + 2 m_{\rm PCAC} \mathcal{O}_{\rm ps}(x) \} \mathcal{O}(y) | 0 \rangle\,,
\eeqs
 \end{widetext}
where $m_{\rm PCAC}$ is the PCAC mass, while
 $\mathcal{O}_{\rm av}^\mu(x)$ and $\mathcal{O}_{\rm ps}(x)$ are the axial-vector current and the
  pseudoscalar density, respectively, defined as
\beqs
\mathcal{O}_{\rm av}^\mu(x)&=&\overline{\Psi}(x)\gamma^5\gamma^\mu\Psi\,,
\quad
\mathcal{O}_{\rm ps}(x)\,=\,\overline{\Psi}(x) \gamma^5 \Psi\,.
\eeqs

If one specifies $\mathcal{O}(x)=\mathcal{O}_{\rm ps}(x)$ and $\mu=0$, and integrates over the spatial coordinates, 
the left hand side of Eq.~(\ref{Eq:pcac}) vanishes, and one finds
\beqs
m_{\rm PCAC} = -\frac{\partial_0 \langle \mathcal{O}_{\rm av}^0 (t) \mathcal{O}_{\rm ps}(0) \rangle }{2\langle \mathcal{O}_{\rm ps}(t) \mathcal{O}_{\rm ps}(0) \rangle} 
= -\frac{\partial_0 C_{\rm av,\, ps}(t)}{2 C_{\rm ps,\, ps}(t)}\,,
\eeqs

In the case of discretised, Euclidean time, with  finite extent, $T$, as in lattice calculations,
after taking into account  the asymptotic behaviour of the correlation functions, 
one can define the effective PCAC mass as
\beq
m^{\rm eff}_{\rm PCAC} \equiv -\frac{m_{\rm ps}^{\rm eff}}{\sinh m_{\rm ps}^{\rm eff}}
\frac{C_{\rm av,\,ps}(t+1) - C_{\rm av,\,ps}(t-1)}{4 C_{\rm ps,\,ps}(t)}\,.
\label{eq:eff_mpcac}
\eeq

 \begin{figure*}[t]
\begin{center}
\includegraphics{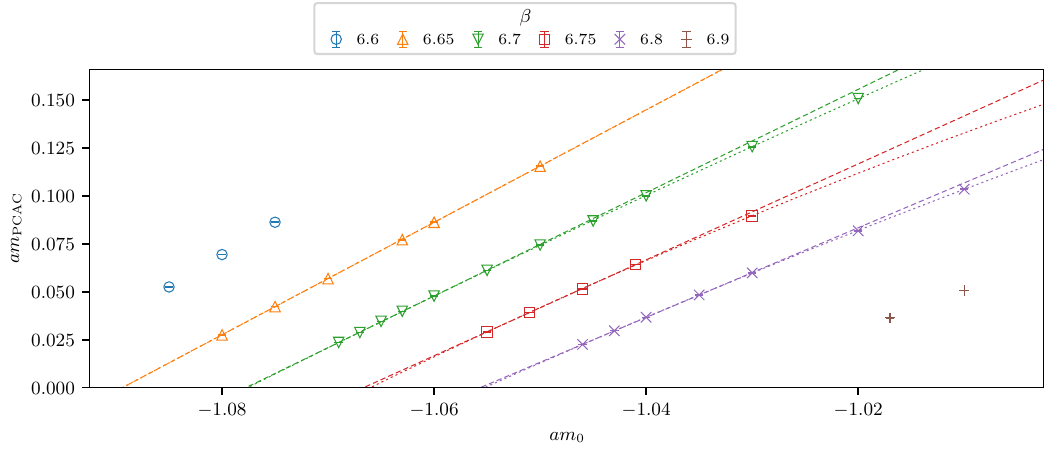}
\caption{%
\label{fig:mpcac}%
The partially-conserved-axial-current mass, $a m_{\rm PCAC}$, as a function of the Wilson-Dirac mass, $a m_0$,
measured in all the ensembles used in this work, together with linear (dashed) and polynomial (dotted) fits
 obtained for fixed  lattice coupling, $\beta$.  The value of $\beta$ is indicated by the colour and marker, as shown in the legend. 
}
\end{center}
\end{figure*}

We plot the resulting values for $a m_{\rm PCAC}$, as a function of the bare mass, $am_0$, in \Fig{mpcac}. 
We perform polynomial fits to the data, for each choice of lattice coupling,  $\beta$. 
We find that a linear fit describes well the results obtained from the ensembles used for the continuum extrapolations, 
while a quadratic term is required to extend the fits to all the ensembles, including the heavier ones.
Because of the smallness of the uncertainties (which include only statistical errors), some of the 
values of the chi-square per degrees of freedom, $\chi^2/N_{\rm d.o.f}$, are larger than unity. 
Yet, for the qualitative purposes of this Appendix, this is sufficient to provide 
support for the usage of $m_{\rm PCAC}$ as a replacement of bare hyperquark mass $m_0$ and for our choice of excluding the heavier ensembles from the continuum and massless extrapolations.

\section{\label{Sec:deft}dEFT, simplified}

We devote this appendix to examining an alternative approach, based upon dEFT~\cite{Matsuzaki:2013eva,Golterman:2016lsd,Kasai:2016ifi,Hansen:2016fri,Golterman:2016cdd,Appelquist:2017wcg,Appelquist:2017vyy,Golterman:2018mfm,Cata:2019edh,Cata:2018wzl,Appelquist:2019lgk,Golterman:2020tdq,Golterman:2020utm,Appelquist:2020bqj,Appelquist:2022qgl,Appelquist:2022mjb},
to the description of the long distance behaviour of the theory.
We restrict our attention to the mass and decay constant of the pseudoscalar mesons, ps, measured in lattice units, at fixed lattice couplings. 
We notice that the presence in the spectrum of 
a light scalar particle, the dilaton,  is an essential feature of dEFT,
which we cannot test with the available data and measurements.
Nevertheless, the 
 scaling relations  built into dEFT can be tested without a priori knowledge of the
 properties (nor the existence) of such a particle, and it is hence legitimate to carry out this exercise.
To this purpose, we borrow notation from Ref.~\cite{Appelquist:2019lgk},
and  write the scaling relation
\beq
m_{\rm ps}^2 f_{\rm ps}^{(2-y)}=C m_f\,,
\eeq
where the low-energy constant, $C$, is related to the fermion condensate, while $y$ is associated with its
scaling dimension~\cite{Leung:1989hw}.
We find it convenient to take the logarithm of this expression, and consider instead the expression:
\beq
\log \left[
\frac{a m_{\rm ps}^2}{m_f}
\right]
= \tilde{C}+Y \log \left[ a^2f_{\rm ps}^2 \right]\,,
\label{eq:deft1}
\eeq
where $Y=\frac{y}{2}-1$, $\tilde{C}=\log C$, and we identify $m_f=m_{\rm PCAC}$.

\begin{figure}[t]
\begin{center}
\includegraphics{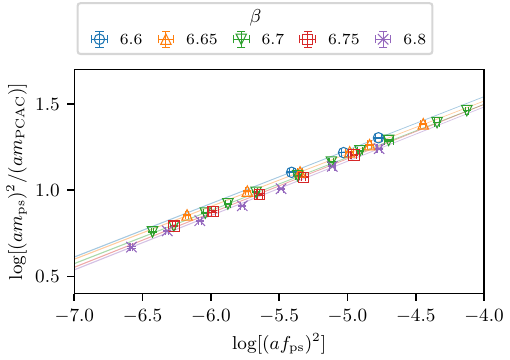}
\caption{%
\label{fig:deft1}%
Logarithmic plot of the scaling relation between fermion mass, 
 $am_f=am_{\rm PCAC}$, pseudoscalar mass, $a m_{\rm ps}$, and  decay constant, $a f_{\rm ps}$,  for all the available values of the lattice coupling, $\beta$. 
 The value of $\beta$ is indicated by the colour and marker, as shown in the legend. 
 The solid lines are the results of dEFT fits.
}
\end{center}
\end{figure}

In \Fig{deft1}, we plot the left-hand side of \Eq{deft1} as a function of  $\log (a^2f_{\rm ps}^2)$,
for all the available values of the coupling, $\beta$.
As shown in the figure, \Eq{deft1} describes qualitatively well the results, over the whole mass range, while we find that the quality of the fits worsens as $\beta$ increases. 
Surprisingly, the resulting values of the scaling dimension, $y$, are independent of the lattice coupling, $y\simeq \DEFTCommonYValue$.
This is lower than the naive, engineering dimension of the fermion condensate, $y_0=3$, but larger than the expectations
for a theory in close proximity of edge of the conformal window, $y_{\ast}=2$---see Ref.~\cite{Cohen:1988sq}, the recent Ref.~\cite{Zwicky:2023krx} and references therein.

We conclude by repeating that, even if the dEFT fit to the data yields intriguing results, it remains a somewhat speculative exercise.
Besides the aforementioned fact that we do not know whether a light scalar singlet is present in the spectrum,  this analysis  relies on the use of continuum relations,
while we know that the values of $\beta$ available are such that significant corrections are expected to be for this analysis due to
discretisation effects. Another concern might originate from the observation that the quality of the dEFT fit worsens as the 
$\beta$ value increases. Nevertheless,  the results are encouraging, and it would be interesting to redo such analysis on lattice data that is closer to both the continuum and the massless limits.

\bibliographystyle{JHEP} 
\bibliography{sp4as}

\end{document}